\documentclass[12pt,letterpaper]{article} 

\usepackage{hyperref} 
\usepackage{graphicx}
\usepackage{cite}

\begin{document}


\begin{center}
{\LARGE\bf Taste-split staggered actions: eigenvalues,\\[4pt]
chiralities and Symanzik improvement}
\end{center}
\vspace{5pt}

\begin{center}
{\large\bf Stephan D\"urr$\,^{a,b}$}
\\[10pt]
${}^a${\sl Bergische Universit\"at Wuppertal,
Gau{\ss}stra{\ss}e\,20, 42119 Wuppertal, Germany}\\
${}^{b\!}${\sl J\"ulich Supercomputing Center,
Forschungszentrum J\"ulich, 52425 J\"ulich, Germany}
\end{center}
\vspace{5pt}

\begin{abstract}
\noindent
The eigenvalue spectra of staggered fermions with an Adams and/or Hoelbling
mass term are studied.
The chiralities of the eigenmodes reflect whether the chirality linked to the
unflavored approximate ($\gamma_5\!\times\!1$) or the flavored exact
($\gamma_5\!\times\!\xi_5$) staggered symmetry is considered, and which one
of the RR, LR, RL, LL eigenmode definitions is used.
In either case a sensitivity to the topological charge of the gauge
background is found.
We discuss how to remove the leading cut-off effects of these actions by means
of a properly tuned improvement term and/or the overlap procedure.
The combination of Symanzik improvement and link smearing radically improves
the properties of the physical branch.
\end{abstract}
\vspace{5pt}



\newcommand{\pad}{\partial}
\newcommand{\hqu}{\hbar}
\newcommand{\ovr}{\over}
\newcommand{\til}{\tilde}
\newcommand{\pri}{^\prime}
\renewcommand{\dag}{^\dagger}
\newcommand{\<}{\langle}
\renewcommand{\>}{\rangle}
\newcommand{\gaf}{\gamma_5}
\newcommand{\xif}{\xi_5}
\newcommand{\nab}{\nabla}
\newcommand{\lap}{\triangle}
\newcommand{\dal}{{\sqcap\!\!\!\!\sqcup}}
\newcommand{\tim}{{\!\times\!}}
\newcommand{\trc}{\mathrm{tr}}
\newcommand{\Trc}{\mathrm{Tr}}
\newcommand{\Mpi}{M_\pi}
\newcommand{\Fpi}{F_\pi}
\newcommand{\Mka}{M_K}
\newcommand{\Fka}{F_K}
\newcommand{\Met}{M_\et}
\newcommand{\Fet}{F_\et}
\newcommand{\Mss}{M_{\bar{s}s}}
\newcommand{\Fss}{F_{\bar{s}s}}
\newcommand{\Mcc}{M_{\bar{c}c}}
\newcommand{\Fcc}{F_{\bar{c}c}}

\newcommand{\al}{\alpha}
\newcommand{\be}{\beta}
\newcommand{\ga}{\gamma}
\newcommand{\de}{\delta}
\newcommand{\ep}{\epsilon}
\newcommand{\ve}{\varepsilon}
\newcommand{\ze}{\zeta}
\newcommand{\et}{\eta}
\renewcommand{\th}{\theta}
\newcommand{\vt}{\vartheta}
\newcommand{\io}{\iota}
\newcommand{\ka}{\kappa}
\newcommand{\la}{\lambda}
\newcommand{\rh}{\rho}
\newcommand{\vr}{\varrho}
\newcommand{\si}{\sigma}
\newcommand{\ta}{\tau}
\newcommand{\ph}{\phi}
\newcommand{\vp}{\varphi}
\newcommand{\ch}{\chi}
\newcommand{\ps}{\psi}
\newcommand{\om}{\omega}

\newcommand{\psb}{\bar{\psi}}
\newcommand{\etb}{\bar{\eta}}
\newcommand{\psh}{\hat{\psi}}
\newcommand{\eth}{\hat{\eta}}
\newcommand{\psd}{\psi^{\dagger}}
\newcommand{\etd}{\eta^{\dagger}}
\newcommand{\qh}{\hat{q}}
\newcommand{\kh}{\hat{k}}

\newcommand{\bdm}{\begin{displaymath}}
\newcommand{\edm}{\end{displaymath}}
\newcommand{\bea}{\begin{eqnarray}}
\newcommand{\eea}{\end{eqnarray}}
\newcommand{\beq}{\begin{equation}}
\newcommand{\eeq}{\end{equation}}

\newcommand{\mr}{\mathrm}
\newcommand{\mb}{\mathbf}
\newcommand{\ri}{\mr{i}}
\newcommand{\Nf}{N_{\!f}}
\newcommand{\Nc}{N_{ c }}
\newcommand{\Nt}{N_{ t }}
\newcommand{\DW}{D_\mr{W}}
\newcommand{\DB}{D_\mr{B}}
\newcommand{\Dst}{D_\mr{st}}
\newcommand{\Dov}{D_\mr{ov}}
\newcommand{\Dke}{D_\mr{ke}}
\newcommand{\Dstm}{D_{\mr{st},m}}
\newcommand{\Dovm}{D_{\mr{ov},m}}
\newcommand{\Dkem}{D_{\mr{ke},m}}
\newcommand{\Dker}{D_{\mr{ke},-\rh/a}}
\newcommand{\MeV}{\,\mr{MeV}}
\newcommand{\GeV}{\,\mr{GeV}}
\newcommand{\fm}{\,\mr{fm}}
\newcommand{\MSbar}{\overline{\mr{MS}}}

\newcommand{\Ga}{\Gamma}
\newcommand{\half}{\frac{1}{2}}
\newcommand{\Dslash}{D\!\!\!\!\slash\,}

\hyphenation{topo-lo-gi-cal simu-la-tion theo-re-ti-cal mini-mum con-tinu-um}


\section{Introduction}


The so-called fermion doubling problem in lattice field theory has been
addressed in many ways; the two most popular approaches are known as Wilson
fermions \cite{Wilson:1974sk} and staggered fermions \cite{Susskind:1976jm},
respectively.
A problem being cured in several ways usually indicates that none of the
solutions is completely satisfactory in all respects, and the lattice is no
exception to this rule.

Wilson fermions represent the spinor components as internal degrees of freedom
(i.e.\ with explicit $\ga$-matrices), such that on a lattice with $N$ grid
points the Wilson formulation entails a matrix of size $4\Nc N\times4\Nc N$,
where $\Nc$ is the number of colors.
Susskind ``staggered'' fermions are based on the observation that the eigenvalue
spectra of naive fermions on interacting backgrounds are 4-fold degenerate
(i.e.\ not 16-fold in 4 space-time dimensions as one might have naively
guessed), and hence a reduction to 1-component spinors is possible (at the
price of distributing the spinor degrees of freedom over space-time), such that
one ends up with a staggered fermion matrix of size $\Nc N\times\Nc N$ that
corresponds to 4 species in the continuum.

Wilson fermions are convenient for their conceptual simplicity (there is a
1-to-1 correspondence between lattice and continuum flavor).
Their main technical disadvantage is the breaking of chiral symmetry through
the dimension 5 operator $\psb(-\lap)\ps$ that is added to lift 15 out of the
16 naive species to a mass of order $1/a$, where $\lap$ is the gauge covariant
Laplacian and $a$ is the lattice spacing.
Staggered fermions offer the advantage of a truncated form of chiral symmetry
and of more speedy simulations (the size of the matrix is smaller).
Their main technical disadvantage is that the taste symmetry (among the 4
species) is not exact and that the reduction to a single species proceeds in
different ways for sea-quarks (those which come from the functional
determinant) and valence-quarks (those which stem from interpolating fields).
This renders the rooted staggered formulation (at finite $a$) non-local and/or
non-unitary which, in principle, could affect the universality class of the
theory (though there is plenty of analytical and numerical evidence that this
does not happen), see e.g.\ the review \cite{Bazavov:2009bb}.

In either approach the main technical disadvantage can be mitigated.
Replacing the gauge links $U_\mu(x)$ of the covariant derivative
$(\nab_\mu\ph)(x)=[U_\mu(x)\ph(x\!+\!\hat\mu)-U_\mu\dag(x\!-\!\hat\mu)\ph(x\!-\!\hat\mu)]/2$
in the operator by smeared links $V_\mu(x)$ (see below for details) reduces the
amount of chiral symmetry breaking with Wilson fermions (particularly
effectively when combined with a clover term) and of taste symmetry breaking
with staggered fermions (see \cite{Bazavov:2009bb} for a guide to the
literature).
In the Wilson case chiral symmetry breaking can be completely removed through
the overlap procedure \cite{Neuberger:1997fp,Neuberger:1998wv}; unfortunately
this increases the computational requirements by a factor $O(100)$.

Recently, two modifications of the staggered action
$S_\mr{S}=\bar\ch[D_\mr{S}\!+\!m]\ch$ with $D_\mr{S}=\et_\mu\!\nab_\mu$ and
$\et_\mu(x)$ given below were proposed which go by the
somewhat confusing name ``staggered Wilson fermions''.
In this article we call them ``taste-split staggered actions''.
In essence the proposal is to replace/augment the usual staggered mass term
$m(1\!\otimes\!1)$ in $S_\mr{S}$, where the notation is spinor$\otimes$taste
\cite{Sharatchandra:1981si,KlubergStern:1983dg,Golterman:1984cy}, by taste
non-singlet mass terms $\propto\!(1\!\otimes\!\xi)$ which are designed to
(partly or fully) lift the staggered near-degeneracy, i.e.\ some species get
masses $2/a$ or $4/a$ such that they decouple in the continuum limit.
Such non-standard staggered mass terms were first considered in
\cite{Golterman:1984cy} where it was noticed that they break the remnant chiral
symmetry $D_\mr{S}(x,y)\to e^{\ri\ep(x)\th}D_\mr{S}(x,y)e^{\ri\ep(y)\th}$ of
the massless staggered action [with $\ep(x)$ defined below] more severely than
the usual mass term does.
The Adams proposal \cite{Adams:2010gx}
\beq
S_\mr{A}     =\bar\ch[\et_\mu\!\nab_\mu+r(M_\mr{A}\!+\!1)]\ch
\label{def_A}
\eeq
builds on a mass term $M_\mr{A}\simeq1\!\otimes\xif$, such that in the naive
continuum limit two species stay massless, while two have mass $2r/a$ (where
$0<r<2$ is akin to the Wilson parameter); the precise form of $M_\mr{A}$ will
be given below.
Similarly, Hoelbling proposes the two operators \cite{Hoelbling:2010jw}
\beq
S_\mr{Hori}=\bar\ch[\et_\mu\!\nab_\mu+r(M_\mr{A}\!+\!M_\mr{Hori}\!+\!2)]\ch
\label{def_Hori}
\eeq
\beq
S_\mr{Hsym}=\bar\ch[\et_\mu\!\nab_\mu+r(M_\mr{Hsym}\!+\!2)]\ch
\label{def_Hsym}
\eeq
which fully lift the staggered near-degeneracy, i.e.\ one species stays
massless, while two receive a mass $2r/a$ and the last one is brought up to
$4r/a$.
The two versions differ on how they break the rotational symmetry group
$R_{\mu\nu}$ of the massless staggered action; the precise form of
$M_\mr{Hori}$ and $M_\mr{Hsym}$ will be given below.
It is straight forward to write down the linear combination
\beq
S_\mr{Hmix}=\bar\ch[\et_\mu\!\nab_\mu+\frac{r}{2}(M_\mr{A}\!+\!M_\mr{Hsym}\!+\!3)]\ch
=\frac{1}{2}\Big[S_\mr{A}+S_\mr{Hsym}\Big]
\label{def_Hmix}
\eeq
which lifts all three non-continuum species to the same doubler point $2r/a$.

Evidently, the attractive feature of these operators is that the matrices are
just of size $\Nc N\times\Nc N$, and still only 1 or 2 species survive in the
continuum.
The question has been raised whether the remaining symmetries of (\ref{def_A})
or (\ref{def_Hori}, \ref{def_Hsym}, \ref{def_Hmix}) are sufficient for
taking the continuum limit (at a fixed pion mass) by tuning only the standard
(relevant) mass term or whether other (relevant, marginal or irrelevant)
operators need to be tuned simultaneously
\cite{SharpeNara,deForcrand:2012bm,Misumi:2012sp,Misumi:2012eh}.
In this paper we investigate the eigenvalue spectra of these operators on
interacting backgrounds and the chiralities (with respect to $\gaf\!\otimes\!1$,
$1\!\otimes\!\xif$ and $\gaf\!\otimes\!\xif$) of their eigenmodes.
The idea behind is that the willingness or reluctance of the ``bellies'' to
open up (and hence separate the branches) is indicative of how severe the
fermionic operator mixing is -- in the Wilson case both link smearing and
including the clover term tend to clear the first eigenvalue belly, since they
suppress mixing between the dimension 5 Laplacian and the dimension 3 mass
operator.

In the proposals \cite{Adams:2010gx,Hoelbling:2010jw} (and in
\cite{deForcrand:2012bm,Creutz:2010bm,Creutz:2011cd,Kimura:2011ik,Misumi:2012sp,
Misumi:2012eh}) a notation is used which obscures the link to the original work
\cite{Golterman:1984cy}.
This is why Sec.\,\ref{sec:review} contains a review of taste non-singlet mass
terms using standard staggered notation.
The core of the investigation, a look at the staggered eigenvalues and
chiralities, is presented in Sec.\,\ref{sec:eigenvalues}.
A brief discussion of some of the peculiarities of the Symanzik improvement
program, when applied to these partly or fully undoubled staggered actions, is
given in Sec.\,\ref{sec:symanzik}.
Results where these taste-split staggered actions serve as kernel to the
overlap procedure are shown in Sec.\,\ref{sec:overlap}.
A comparison to the eigenvalue spectra of Wilson-type actions is found in
Sec.\,\ref{sec:comparison}.
Comments on the breaking of rotational symmetry by the Hoelbling operators
(\ref{def_Hori}, \ref{def_Hsym}, \ref{def_Hmix}) are arranged in
Sec.\,\ref{sec:rotation}.
Finally, Sec.\,\ref{sec:summary} contains a summary.


\section{Review of staggered mass terms \label{sec:review}}


Define the Golterman-Smit staggered phase factors \cite{Golterman:1984cy}
\beq
\et_\mu(x)=(-1)^{\sum_{\nu<\mu}x_\nu}
\;,\qquad
\ze_\mu(x)=(-1)^{\sum_{\mu<\nu}x_\nu}
\eeq
which multiply to give
\beq
\et_5(x)=\et_1\et_2\et_3\et_4=(-1)^{x_1+x_3}
\;,\qquad
\ze_5(x)=\ze_1\ze_2\ze_3\ze_4=(-1)^{x_2+x_4}
\eeq
respectively.
The product of the latter two phase factors reads
(throughout the spinor$\otimes$taste identification holds up to $O(a)$
corrections \cite{Sharatchandra:1981si,KlubergStern:1983dg,Golterman:1984cy})
\beq
\ep(x)=\et_5\ze_5=(-1)^{x_1+x_2+x_3+x_4}
\quad\longleftrightarrow\quad
\ga_5\otimes\xi_5
\eeq
and induces the $U(1)_\ep$-symmetry and $\ep$-hermiticity
$\ep(x)D_\mr{S}(x,y)\ep(y)=D_\mr{S}\dag(x,y)=[D_\mr{S}(y,x)]\dag$ of the
massless staggered action.
After being lifted to a site-diagonal operator
\beq
\ep(x,y)=\Ga_{55}(x,y)=(-1)^{x_1+x_2+x_3+x_4}\de_{x,y}
\label{def_55}
\eeq
it couples to the $\gaf\!\otimes\!\xif$ ``Goldstone'' state when used as an
interpolating field \cite{Golterman:1984cy}.

The $(\ga_\mu\!\otimes\!1)$ and $(\gaf\!\otimes\!1)$ ``taste singlet''
operators are defined by
\bea
\Ga_{\mu0}(x,y)\;\equiv\;\Ga_\mu(x,y)&=&\frac{1}{2}\et_\mu(x)
\Big[U_\mu(x)\de_{x+\hat\mu,y}+U_\mu\dag(x\!-\!\hat\mu)\de_{x-\hat\mu,y}\Big]
\\
\Ga_{50}(x,y)\;\equiv\;\Ga_5(x,y)&=&
\frac{1}{4!}\sum_\mr{perm}\ep_\mr{perm}\Ga_1\Ga_2\Ga_3\Ga_4
\eea
and the $(1\!\otimes\!\xi_\mu)$ and $(1\!\otimes\!\xif)$ ``spinor
singlet'' operators are defined by
\bea
\Ga_{0\mu}(x,y)\;\equiv\;\Xi_\mu(x,y)&=&\frac{1}{2}\ze_\mu(x)
\Big[U_\mu(x)\de_{x+\hat\mu,y}+U_\mu\dag(x\!-\!\hat\mu)\de_{x-\hat\mu,y}\Big]
\\
\Ga_{05}(x,y)\;\equiv\;\Xi_5(x,y)&=&
\frac{1}{4!}\sum_\mr{perm}\ep_\mr{perm}\Xi_1\Xi_2\Xi_3\Xi_4
\eea
with the consequence that both $\Ga_{50}$ and $\Ga_{05}$ are 4-hop
operators.
Furthermore, the latter two operators relate to each other by a simple
$\Ga_{55}$ operation (either from the left or from the right).

In practice it can be advantageous to introduce the commutators
in spinor and taste space
\bea
\Ga_{\mu\nu}(x,y)&\equiv&
\frac{\ri}{2}\Big(\Ga_\mu\Ga_\nu-\Ga_\nu\Ga_\mu\Big)
\qquad\longleftrightarrow\qquad\ga_{\mu\nu}\!\otimes\!1
\label{def_Sigma_munu}
\\
\Xi_{\mu\nu}(x,y)&\equiv&
\frac{\ri}{2}\Big(\Xi_\mu\Xi_\nu-\Xi_\nu\Xi_\mu\Big)
\qquad\longleftrightarrow\qquad1\!\otimes\!\xi_{\mu\nu}
\label{def_Theta_munu}
\eea
respectively, with $\ga_{\mu\nu}\equiv\frac{\ri}{2}[\ga_\mu,\ga_\nu]$ known as
$\si_{\mu\nu}$ and $\xi_{\mu\nu}\equiv\frac{\ri}{2}[\xi_\mu,\xi_\nu]$.
Based on these one finds
\bea
\Ga_{50}(x,y)&\simeq&-\frac{1}{6}\Big(
\Ga_{12}\Ga_{34}-\Ga_{13}\Ga_{24}+\Ga_{14}\Ga_{23}+
\Ga_{23}\Ga_{14}-\Ga_{24}\Ga_{13}+\Ga_{34}\Ga_{12}
\Big)
\label{def_Ga50}
\\
\Ga_{05}(x,y)&\simeq&-\frac{1}{6}\Big(
\Xi_{12}\Xi_{34}-\Xi_{13}\Xi_{24}+\Xi_{14}\Xi_{23}+
\Xi_{23}\Xi_{14}-\Xi_{24}\Xi_{13}+\Xi_{34}\Xi_{12}
\Big)
\label{def_Ga05}
\eea
where the near-equality is exact if no SU(3) backprojection in intermediate
steps is applied.
The main advantage of the form (\ref{def_Ga50},\ref{def_Ga05}) is that one can
apply SU(3) backprojection twice, in each case the argument being a sum over
products of just two color matrices.

Following Golterman and Smit \cite{Golterman:1984cy}, everything which is a
singlet in spinor space is called a ``mass term''.
This leads to the categorization of potential mass terms as
\bea
1\!\otimes\!1\quad\longleftrightarrow&1&
\mbox{(0-hop, taste-scalar)}
\\
1\!\otimes\!\xi_\mu\quad\longleftrightarrow&\Xi_\mu&
\mbox{(1-hop, taste-vector)}
\\
1\!\otimes\!\xi_{\mu\nu}\quad\longleftrightarrow&\Xi_{\mu\nu}&
\mbox{(2-hop, taste-tensor)}
\\
1\!\otimes\!\ri\xi_\mu\xif\quad\longleftrightarrow&\ri\Xi_\mu\Xi_5&
\mbox{(3-hop, taste-pseudovector)}
\\
1\!\otimes\!\xif\quad\longleftrightarrow&\Xi_5&
\mbox{(4-hop, taste-pseudoscalar)}
\eea
in terms of hermitian operators.
As was pointed out in \cite{deForcrand:2012bm}, an acceptable mass term is
$1\!\otimes\!1$, $1\!\otimes\!\xi_{\mu\nu}$, $1\!\otimes\!\xif$ or a
combination thereof.
In this terminology the Adams term in (\ref{def_A}) is the taste-pseudoscalar
mass $\Gamma_{05}=\Xi_5$.
And the Hoelbling terms in (\ref{def_Hori}, \ref{def_Hsym}) represent
unsymmetrized and symmetrized versions of the taste-tensor mass $\Xi_{\mu\nu}$,
respectively, i.e.\ $M_\mr{Hori}=\Xi_{12}$ [to be used together with
$M_\mr{A}$, see (\ref{def_Hori})] and 
$M_\mr{Hsym}=(\Xi_{12}+\Xi_{34}+\Xi_{13}-\Xi_{24}+\Xi_{14}+\Xi_{23})/\sqrt{3}$
[to be used alone, see (\ref{def_Hsym})].
The operator (\ref{def_Hmix}), finally, involves all three types of valid
mass terms.

Note that there is some freedom in the definition of the operators
(\ref{def_A}--\ref{def_Hmix}) and of the taste non-singlet mass terms.
For instance with the Adams operator as defined in (\ref{def_A}) and the choice
$M_A=\gaf\!\otimes\!\xif$ it follows that the physical branch has $\xif=-1$.
Hence, with this convention the actions of $\gaf\!\otimes\!1$ and
$\gaf\!\otimes\!\xif$, in the physical branch, on a topological mode will
differ in sign.


\section{Eigenvalues and chiralities \label{sec:eigenvalues}}


The eigenvalue spectrum on a thermalized gauge background is a key to get an
impression of how suitable a given fermion discretization is in practical terms.
An issue with staggered fermions is their sensitivity to the topological charge
of the gauge configuration \cite{Karsten:1980wd,Smit:1986fn,Hands:1990wc,
Donald:2011if}.

The $\ep$-hermiticity ensures that the eigenvalues of the massless operator
occur in pairs $\pm\ri\la$ on the imaginary axis.
There are no exact zero-modes of $D_\mr{S}$; instead $4|q|$ would-be zero-modes
show up.
Both their separation from the non-topological modes and their chiralities (see
below) depend on how much the staggered action is ``improved'' through link
smearing.

\begin{figure}[!tb]
\includegraphics[width=0.5\textwidth]{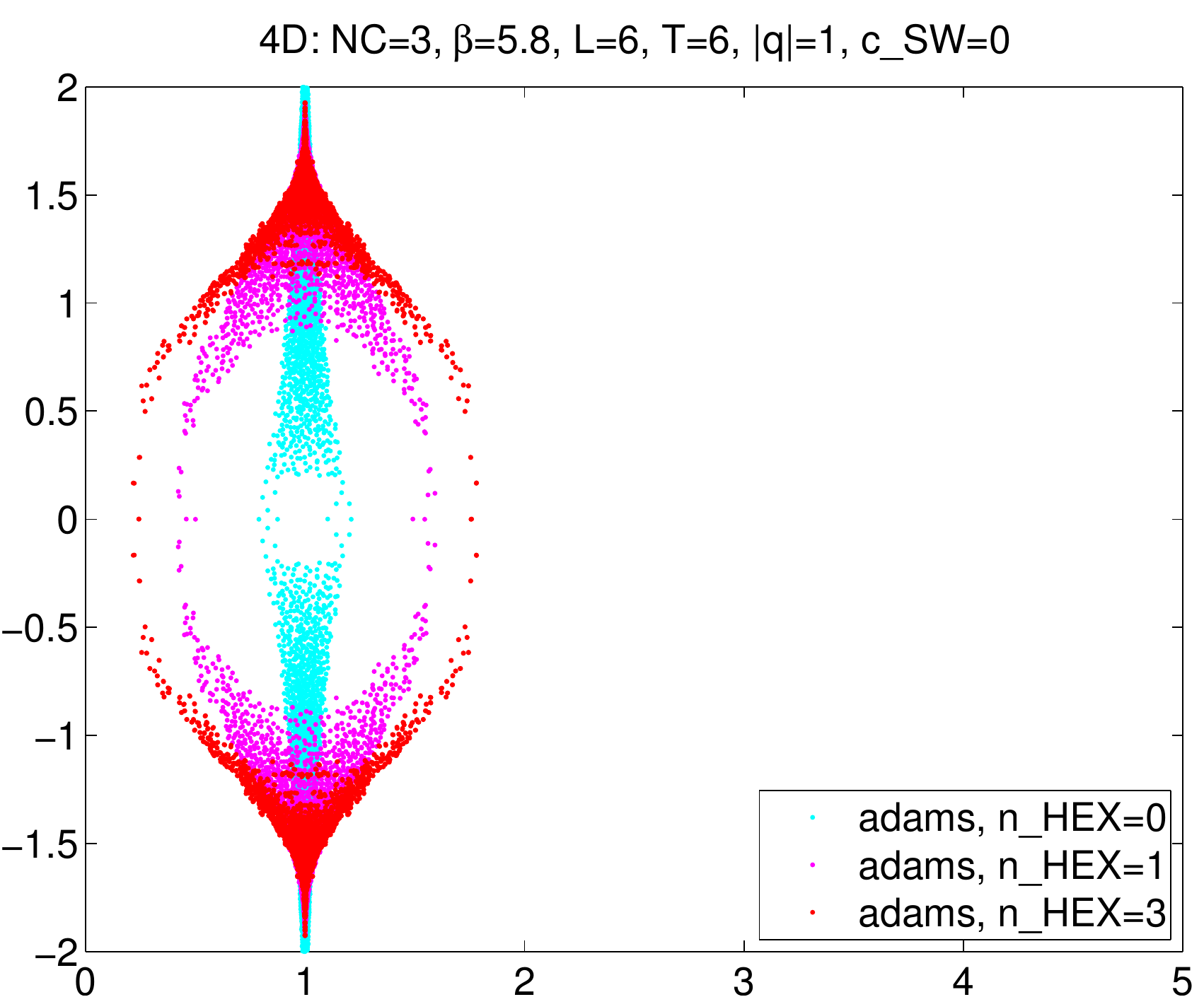}%
\includegraphics[width=0.5\textwidth]{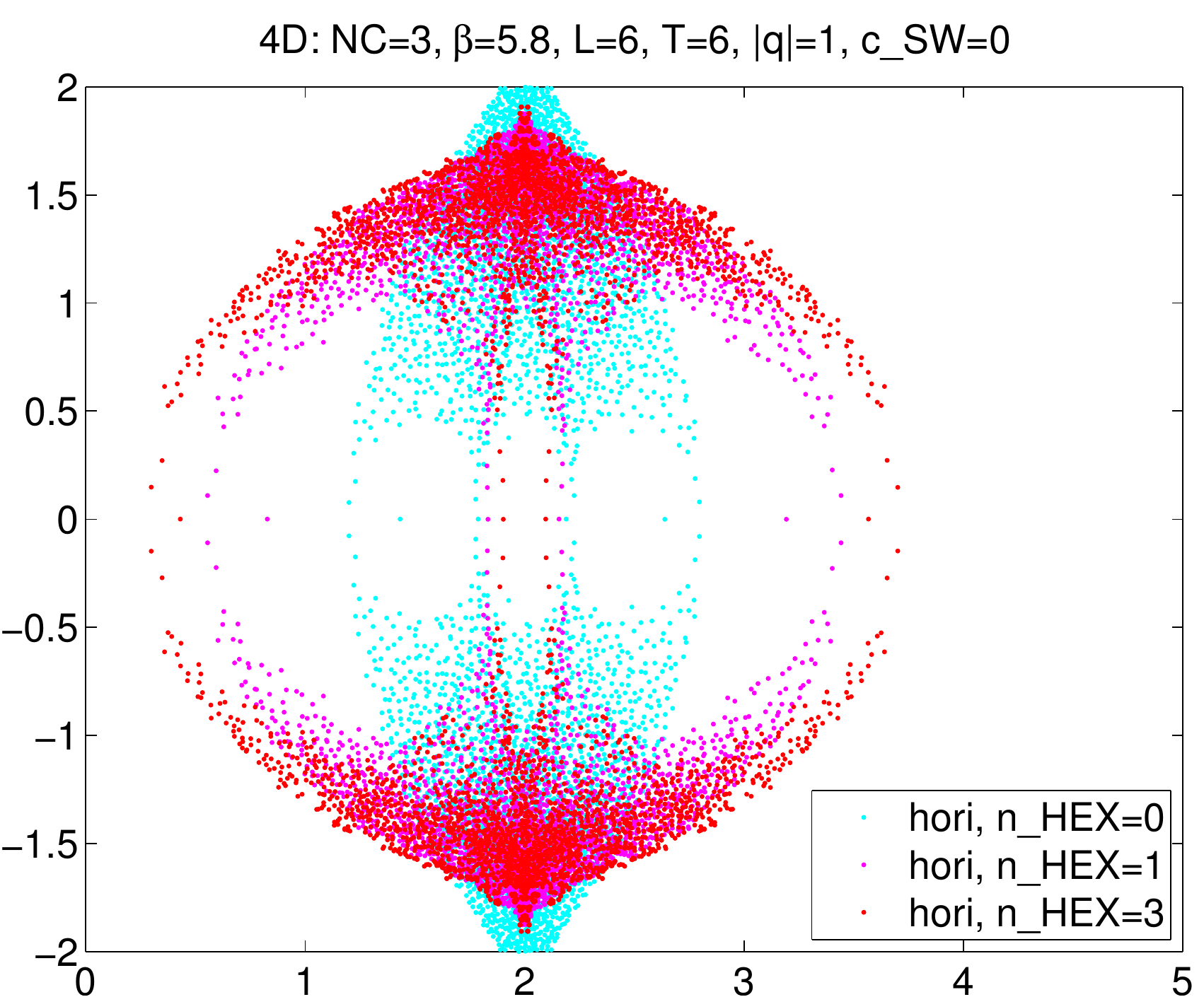}%
\\[2mm]
\includegraphics[width=0.5\textwidth]{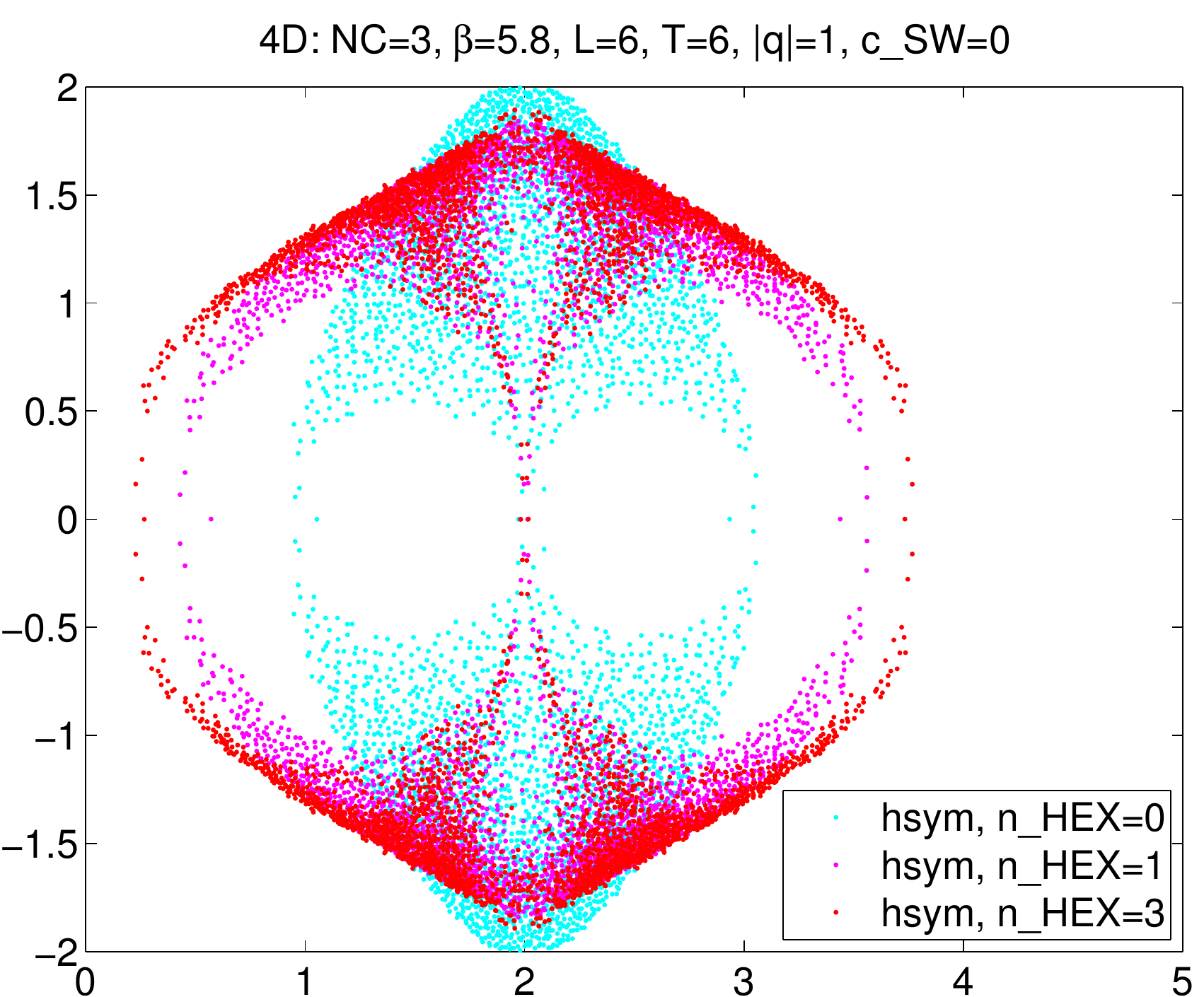}%
\includegraphics[width=0.5\textwidth]{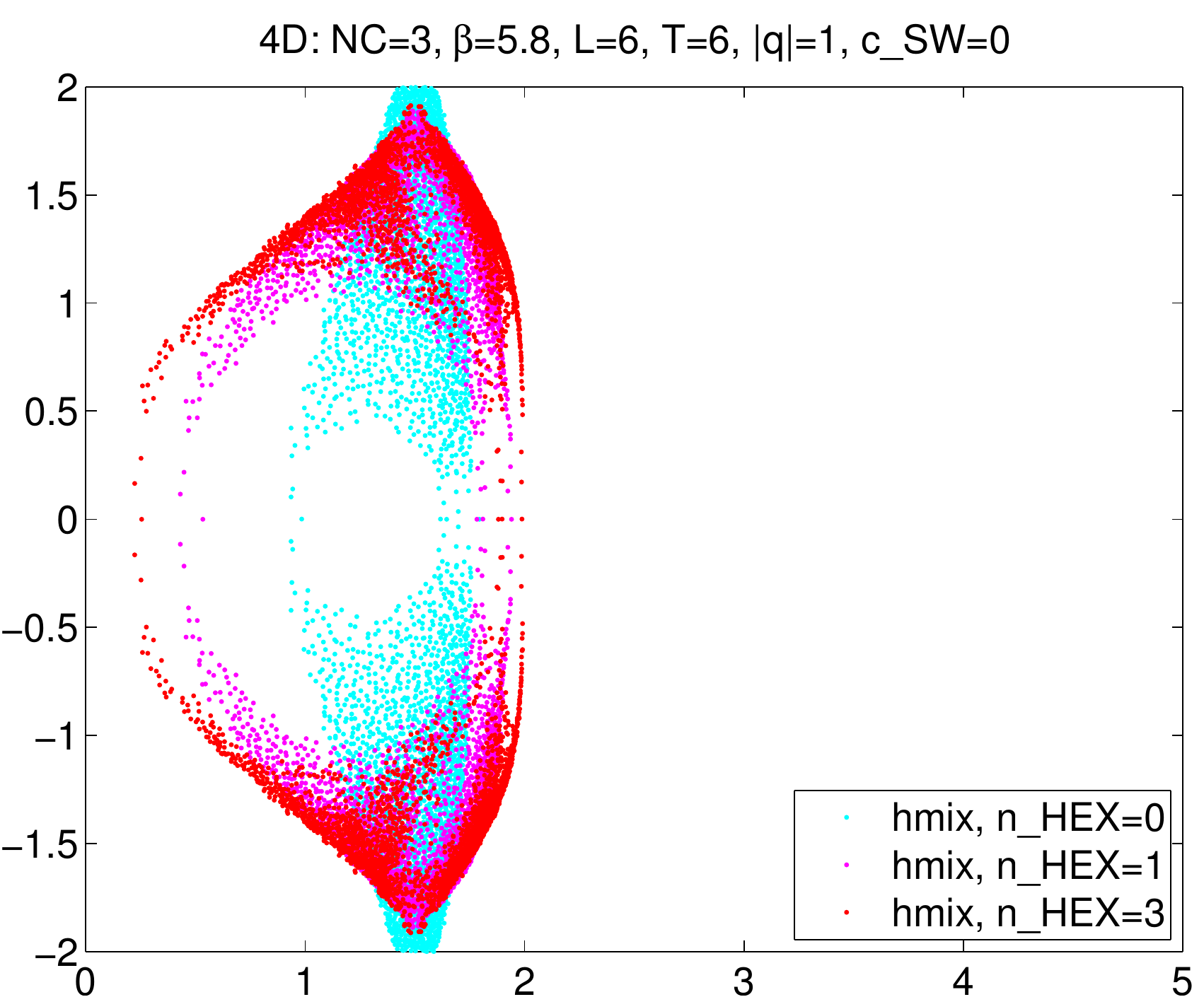}%
\caption{\label{fig1}\sl
Eigenvalue spectra of the four unimproved operators $(c_\mr{SW}\!=\!0)$ with
$r\!=\!1$ at $m\!=\!0$ for up to three levels of HEX smearing -- the
smearing seems essential for the bellies to form.}
\end{figure}

Fig.\,\ref{fig1} displays the eigenvalue spectra of the operators
(\ref{def_A}--\ref{def_Hmix}) on a $6^4$ lattice (generated with $\be=5.8$
Wilson glue) of unit topological charge, $|q|=1$.
The fermion boundary conditions are periodic in all directions; with
antiperiodic boundary conditions in the 4-direction mild changes would
occur in regions of large (positive or negative) imaginary part.
Throughout, we use gauge links $V_\mu(x)$ which have undergone 0, 1, or 3
levels of HEX smearing \cite{Capitani:2006ni}.
Backprojection of the mean of the $n$-hop paths to SU(3) could be applied,
but in the present work this is not done.

Our first observation in these plots is that without link-smearing the
``bellies'' have a hard time opening up (as was already observed in
\cite{deForcrand:2012bm}), but the situation improves considerably upon
applying 1 or 3 steps of smearing.
All operators have a bare (taste-scalar) mass $m=0$, but their physical
(left-most) branches cross the real axis at distinctively non-zero values.
This is a sign of additive mass renormalization, and the plots show that the
smearing reduces this effect significantly.
Last but not least the physical branches have one [or two for (\ref{def_A}),
since this operator is doubled] exactly real modes which tend to get
``soaked into the belly'' even for the higher smearing levels.
And of course all spectra are symmetric with respect to the real axis, which
implies that the determinant is real (this is a consequence of the requirement
of $\ep$-hermiticity which ruled out two possible mass terms,
cf.\ Sec.\,\ref{sec:review}).

The next step is to look at the chiralities of the pertinent eigenmodes,
i.e.\ the values of the operators $\Ga_{50}$, $\Ga_{05}$ and $\Ga_{55}$ when
sandwiched between the staggered eigenmodes.
The standard operator $D_\mr{S}$ or $D_\mr{S}+m$ is a normal operator,
i.e.\ $[D_\mr{S},D_\mr{S}\dag]=0$, and only $\Ga_{50}$ is sensitive to the
topological charge of the background (see below).
The taste-split staggered operators (\ref{def_A}--\ref{def_Hmix}) are
non-normal operators, i.e.\ $[D,D\dag]\neq0$.
This implies that the left-eigenvectors, which satisfy $\<L_i|D=\<L_i|\la_i$,
are not just the hermitian conjugates of the right-eigenvectors, which satisfy
$D|R_i\>=\la_i|R_i\>$, though they share the eigenvalue $\la_i$.
The situation is now analogous to that of the Wilson operator which is also
non-normal \cite{Hip:2001hc}.
This implies that for a given sandwich operator $\Gamma$ there are four
chiralities, $\<L_i|\Gamma|L_i\>$, $\<L_i|\Gamma|R_i\>$, $\<R_i|\Gamma|L_i\>$,
$\<R_i|\Gamma|R_i\>$, where $|L_i\>$ is the hermitian conjugate of the
(normalized) left-eigenvector $\<L_i|$, and $\<R_i|$ is the hermitian conjugate
of the (normalized) right-eigenvector $|R_i\>$, with $i$ running over all modes.

\begin{figure}[!tb]
\includegraphics[width=0.5\textwidth]{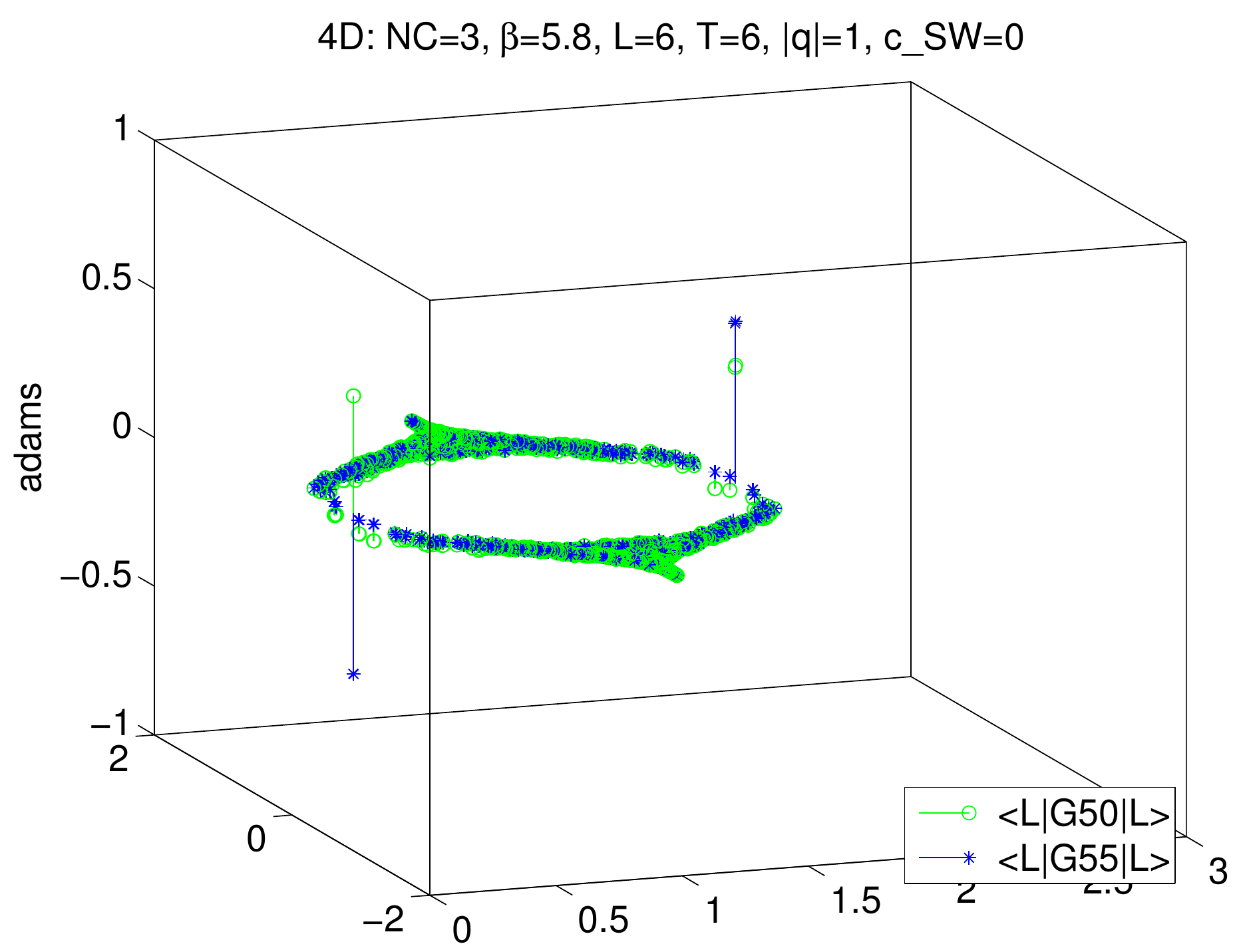}%
\includegraphics[width=0.5\textwidth]{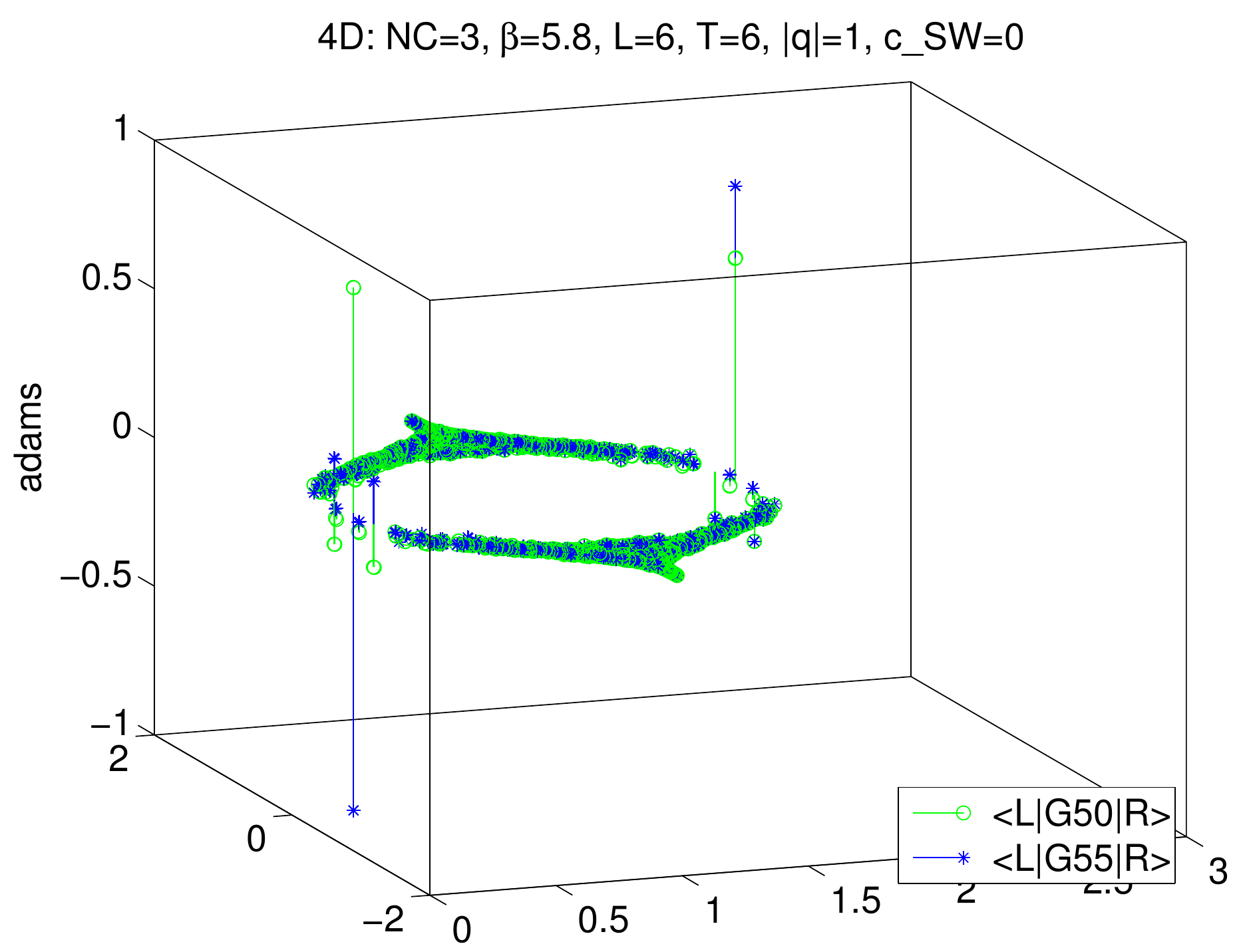}%
\\[2mm]
\includegraphics[width=0.5\textwidth]{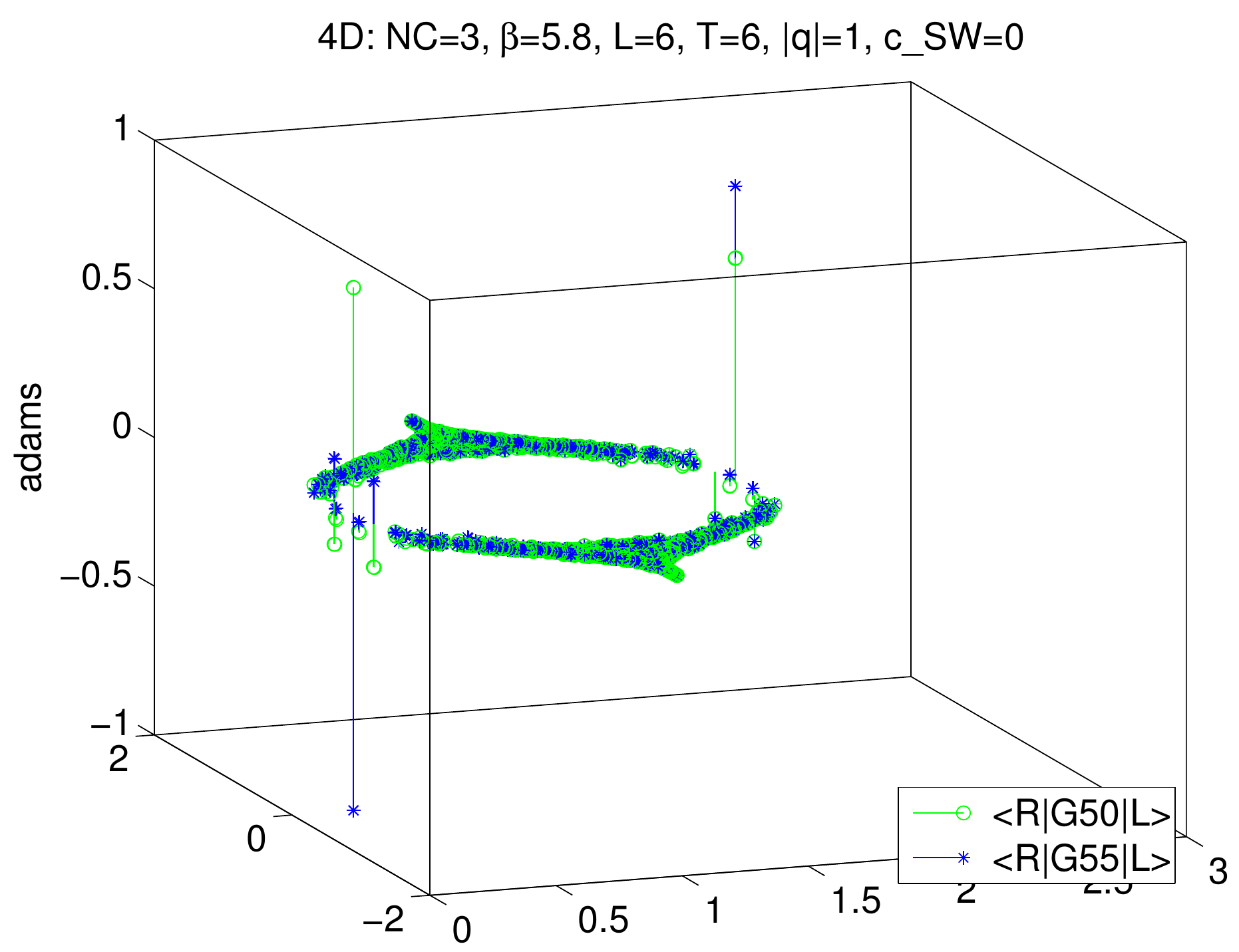}%
\includegraphics[width=0.5\textwidth]{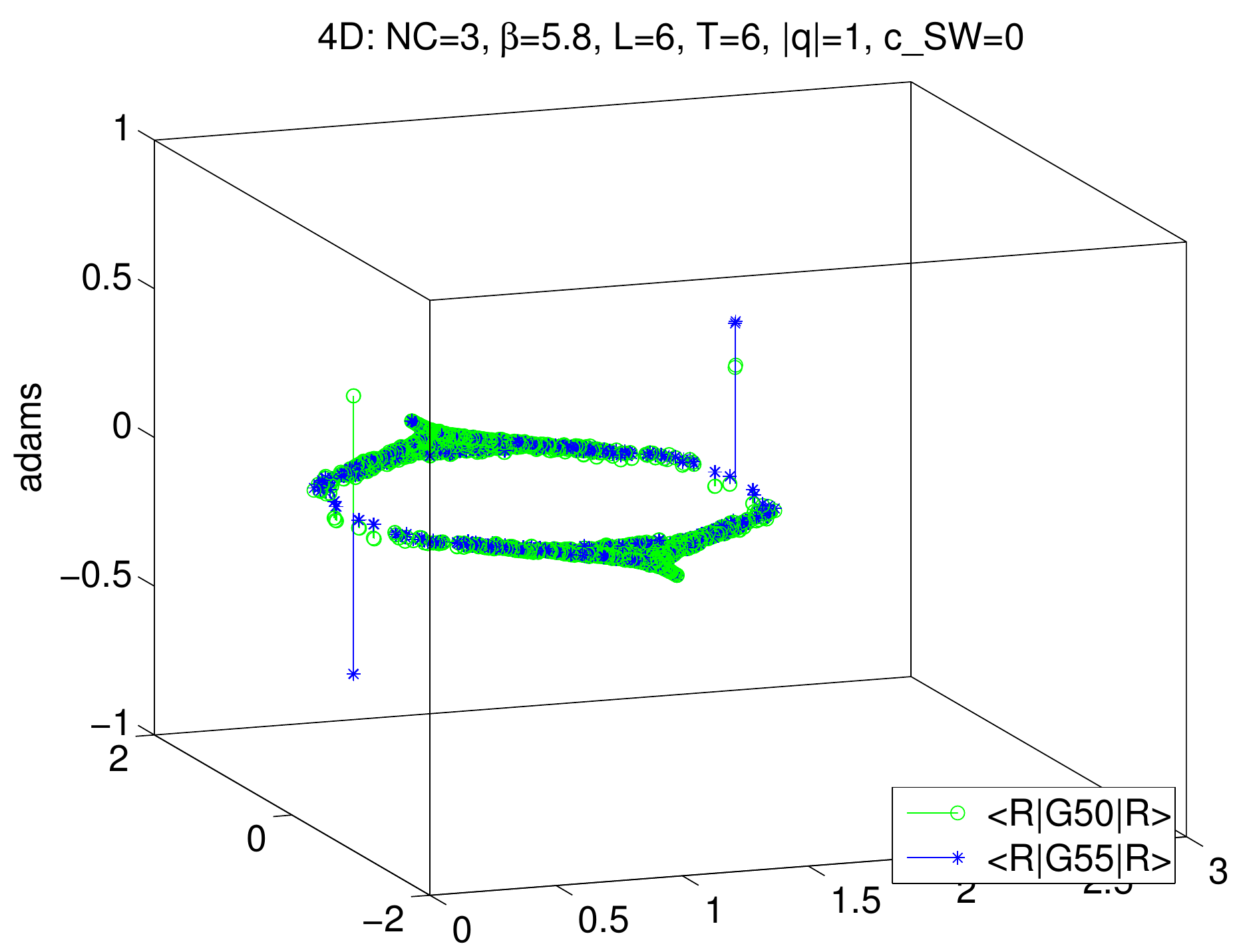}%
\caption{\label{fig2}\sl
Needle plots of the four L/R-chiralities of the unimproved Adams operator
$(c_\mr{SW}\!=\!0)$ with respect to $\Ga_{50}$ (green circles) or
$\Ga_{55}\!=\!\ep$ (blue stars), with three HEX smearings.}
\end{figure}

Fig.\,\ref{fig2} displays these four options with the sandwich operators
$\Gamma_{50}$ and $\Gamma_{55}$ for the eigenmodes of the Adams operator
(\ref{def_A}).
The chiralities are plotted in the $z$-direction above the respective
eigenvalue $\la=x+\ri y$ in the complex plane.
The four options collapse effectively into two, as it must be, due to the
$\ep$-hermiticity of the operator.
The $\<L|\Gamma|L\>$ or $\<R|\Gamma|R\>$ option shows two distinctively
non-zero physical modes (on a configuration with $|q|=1$), depicted at the
position of the two exactly real eigenvalues, both for $\Gamma=\Gamma_{50}$ and
$\Gamma=\Gamma_{55}$ (peeking into the $+z$ and $-z$ directions, respectively).
With the $\<L|\Gamma|R\>$ or $\<R|\Gamma|L\>$ option the chiralities tend
to be even more pronounced, and in either case it holds that the magnitude of
$\<.|\Gamma_{55}|.\>$ exceeds the magnitude of $\<.|\Gamma_{50}|.\>$.
Finally there is an equal number of chiral modes in the unphysical branch (near
$x=2$) with identical orientation for $\Gamma_{50}$ but opposite orientation
for $\Gamma_{55}$ (in accordance with Karsten-Smit \cite{Karsten:1980wd}).
The most surprising lesson is that both $\Gamma_{50}$ and $\Gamma_{55}$ prove
sensitive to the topology of the gauge background; this happens in sharp
contrast to the standard staggered case (cf.\ next paragraph).
And the recommendation of \cite{Hip:2001hc} that for any $\Gamma$ one should
focus on the $\<L|\Gamma|R\>$ option continues to be valid.

\begin{figure}[!tb]
\includegraphics[width=0.5\textwidth]{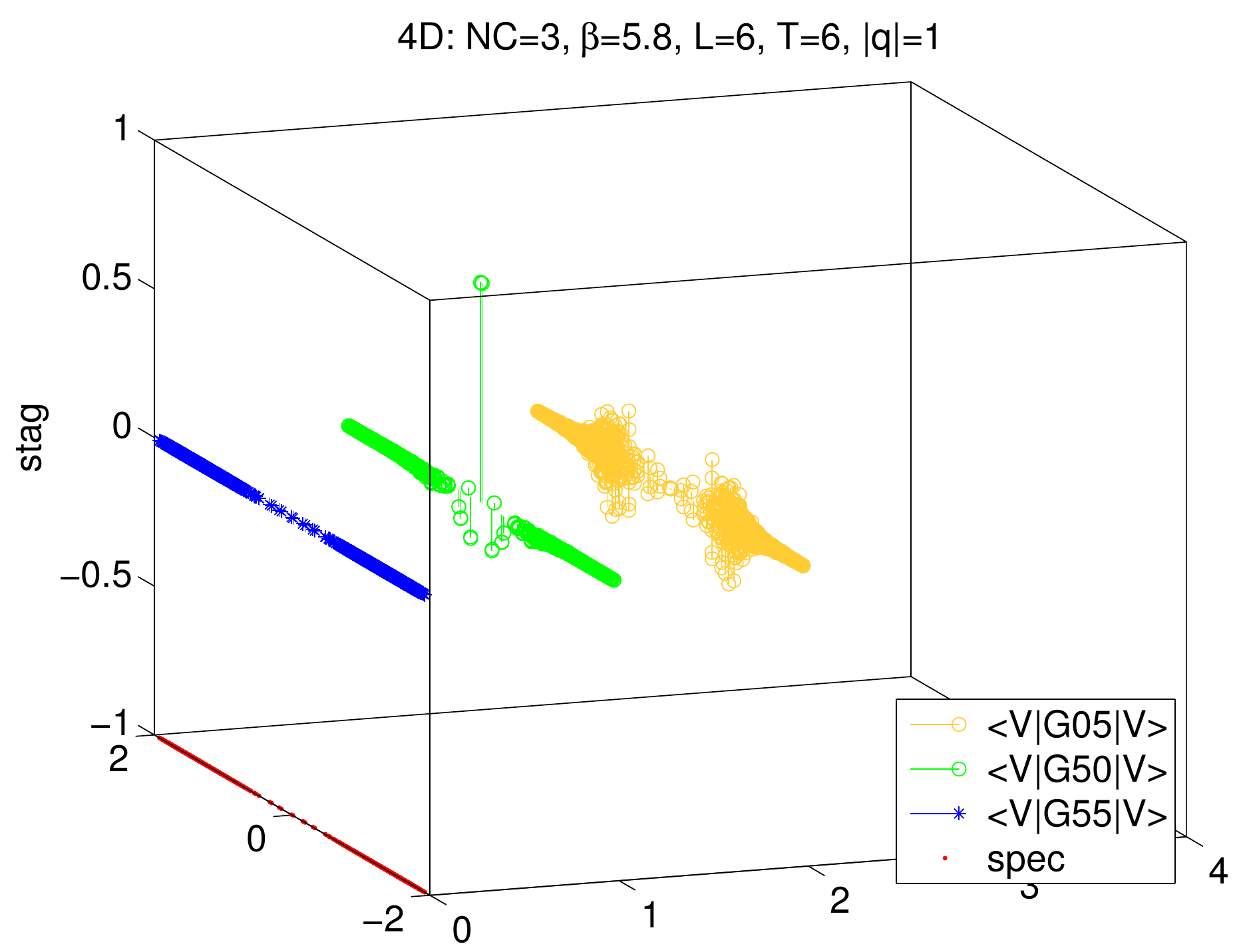}%
\includegraphics[width=0.5\textwidth]{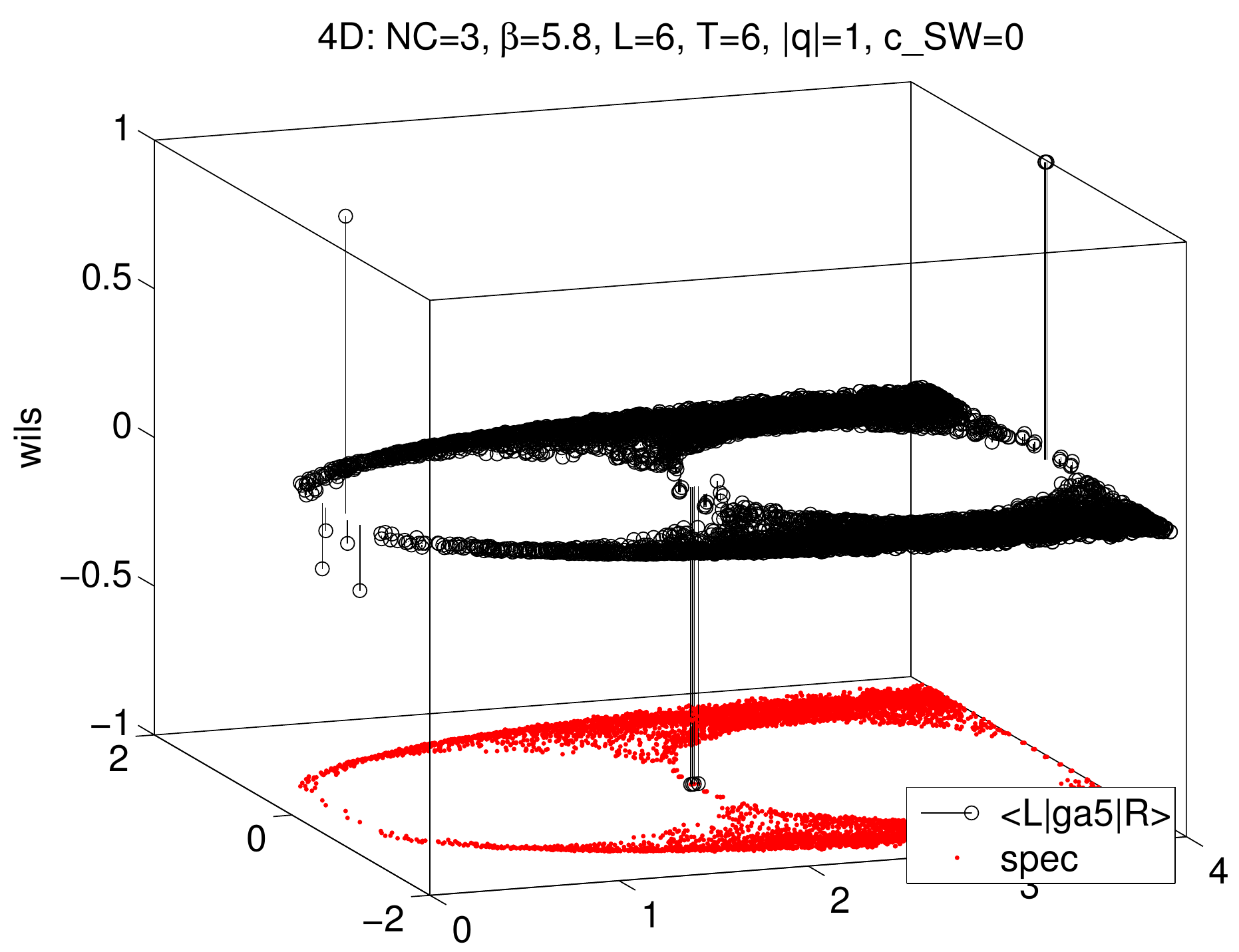}%
\caption{\label{fig3}\sl
Chiralities of the standard staggered action and of the unimproved Wilson
operator. For better visibility the results for $\<.|\Ga_{50}|.\>$ and
$\<.|\Ga_{05}|.\>$ are displaced in the real direction.}
\end{figure}

Fig.\,\ref{fig3} displays -- for comparison -- the chiralities of the standard
staggered action and of the Wilson action on the same $|q|=1$ gauge background
(only half of the spectrum is shown in the latter case).
In the staggered case there is only one type of eigenmode, due to
$\<V_i|=|V_i\>\dag$, while in the Wilson case only $\<L_i|\gaf|R_i\>$ is shown.
In the staggered case $\<V_i|\Gamma_{55}|V_i\>$ is absolutely flat (as required
by $\{\ep,D_\mr{S}\}=0$), while $\<V_i|\Gamma_{50}|V_i\>$ (plotted at $x=1$)
has 4 modes which peek upwards, and $\<V_i|\Gamma_{05}|V_i\>$ (plotted at
$x=2$) is wiggly but not sensitive to topology.
The Wilson operator shows the expected behavior -- one chiral mode in the
physical branch, four oppositely oriented modes near $x=2$, six originally
oriented modes near $x=4$, four oppositely oriented modes near $x=6$ (not
shown) and one originally oriented mode near $x=8$ (not shown).

\begin{figure}[!tb]
\includegraphics[width=0.5\textwidth]{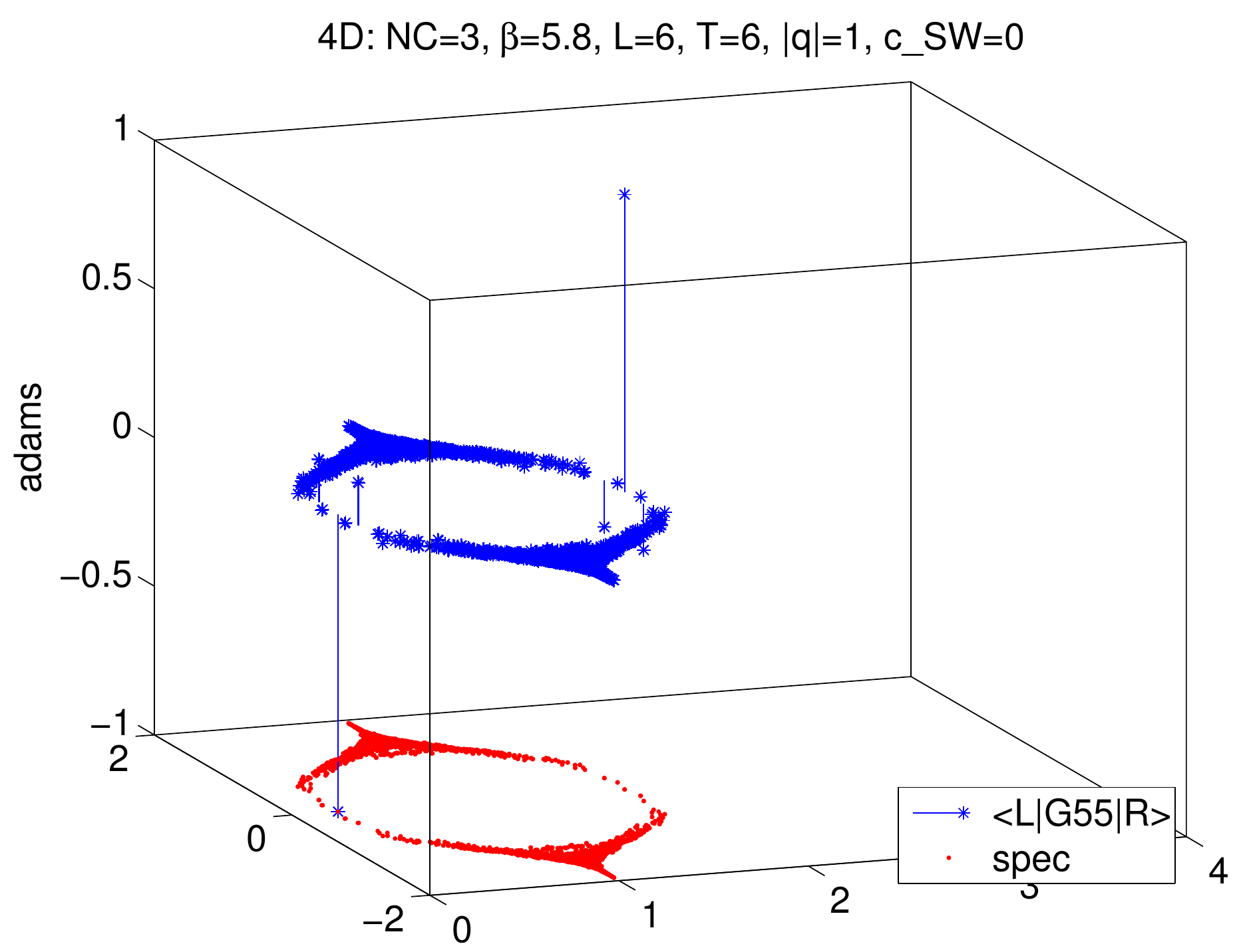}%
\includegraphics[width=0.5\textwidth]{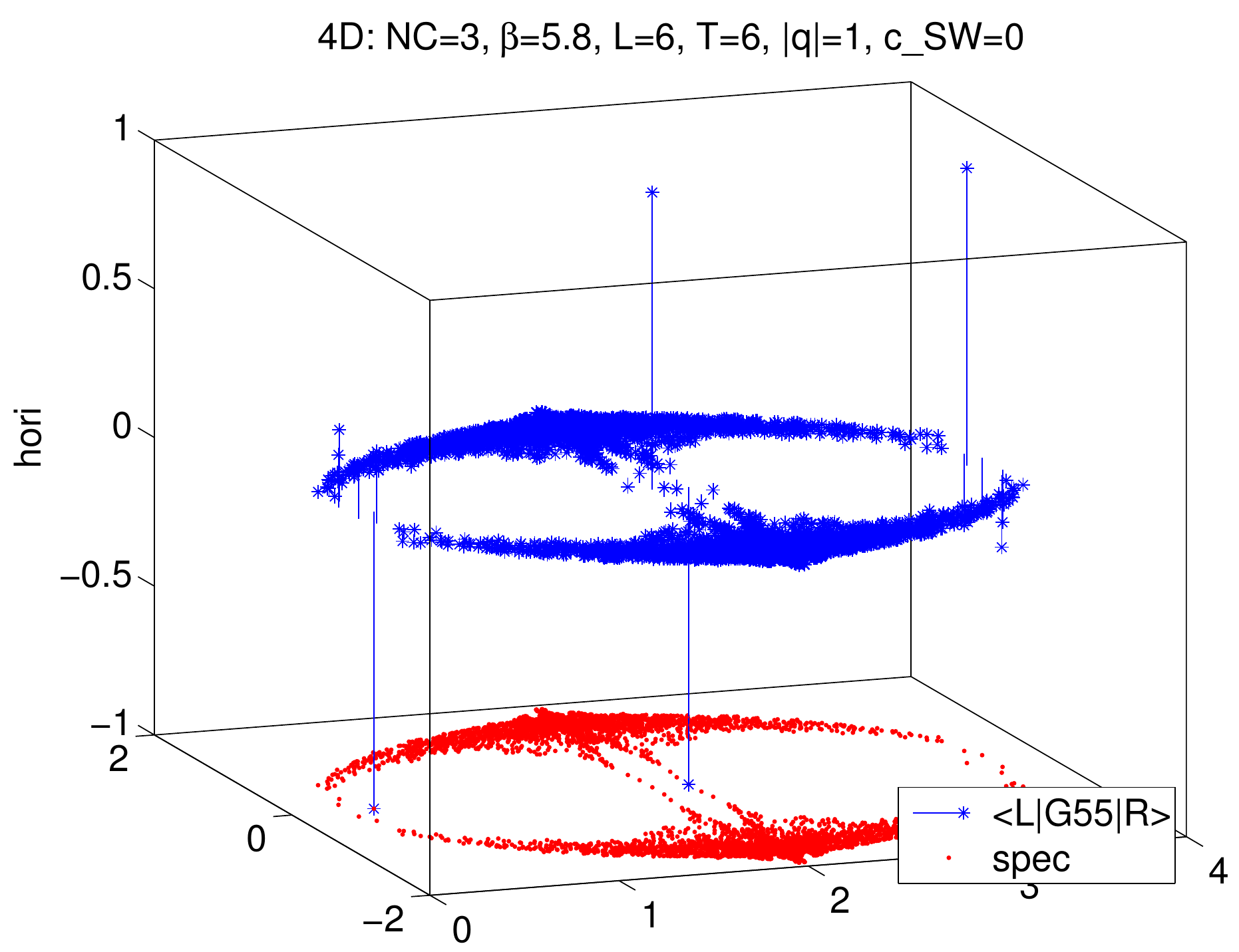}%
\\[2mm]
\includegraphics[width=0.5\textwidth]{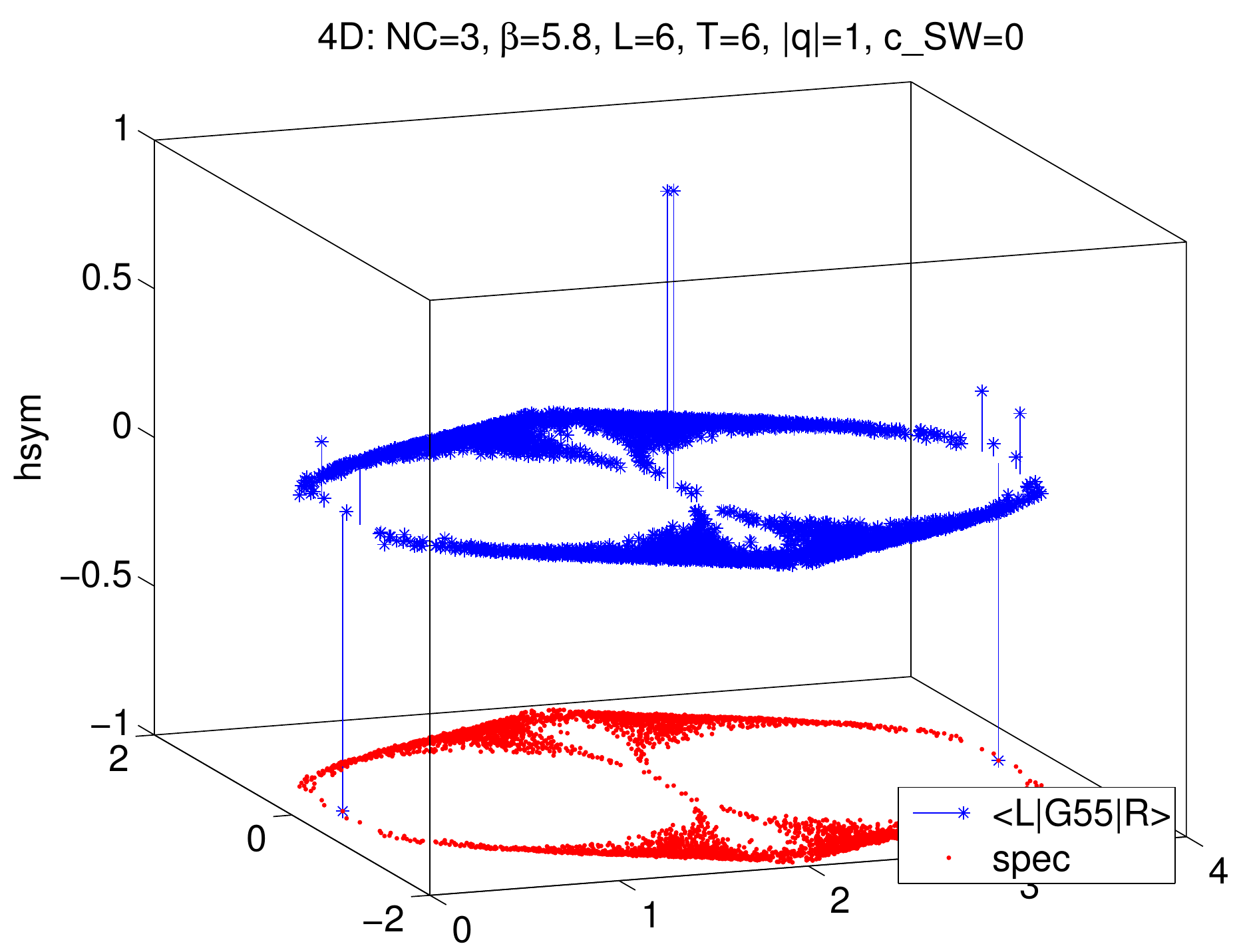}%
\includegraphics[width=0.5\textwidth]{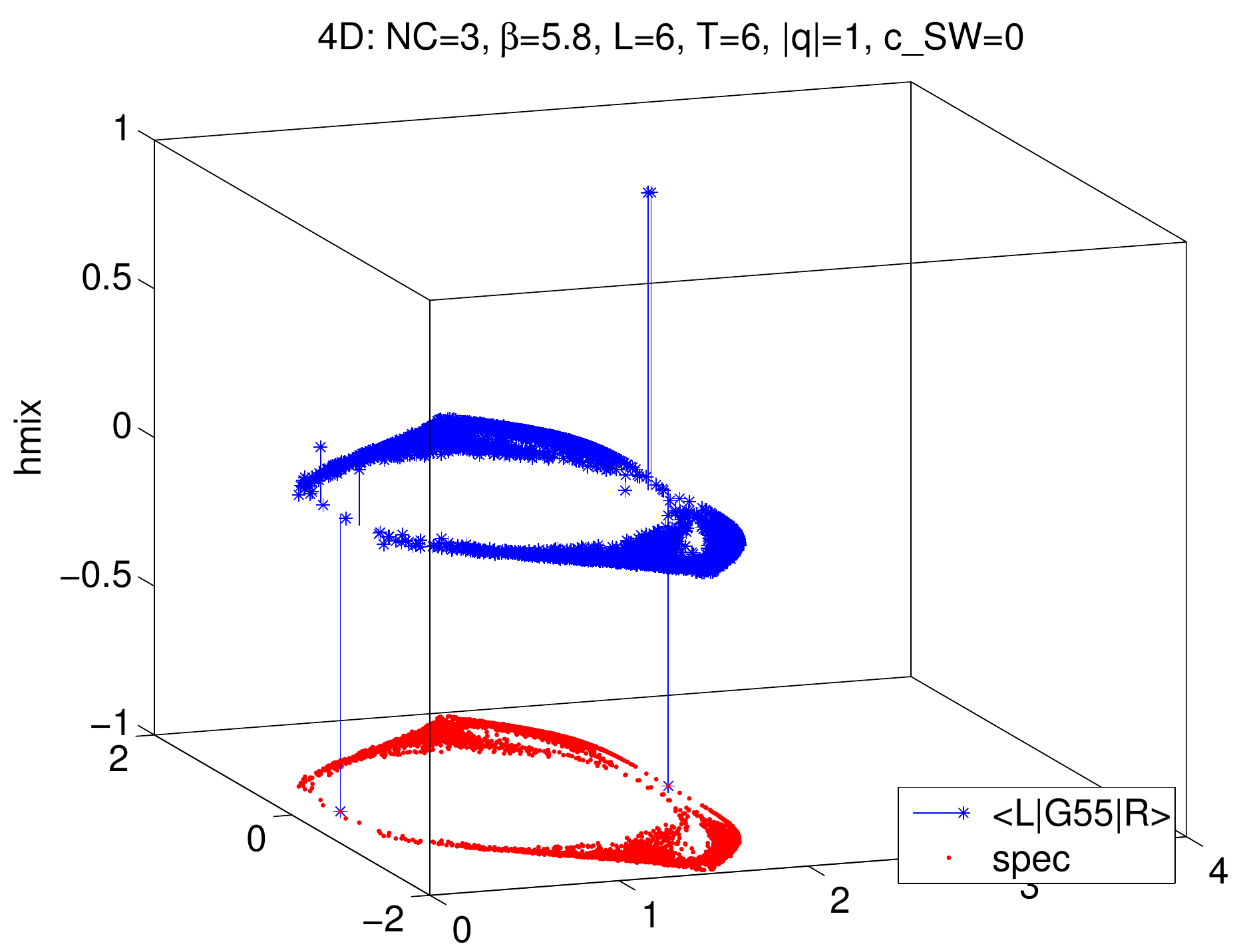}%
\caption{\label{fig4}\sl
Needle plots of the chiralities with respect to $\Ga_{55}$ of the unimproved
operators $(c_\mr{SW}\!=\!0)$ above their eigenvalues. The LR-definition of
$\<.|\Ga_{55}|.\>$ and three HEX smearings are used.}
\end{figure}

Fig.\,\ref{fig4} displays the $\<L|\Gamma_{55}|R\>$ chiralities for the four
operators (\ref{def_A}--\ref{def_Hmix}) above their eigenvalues.
The Adams operator has two chiral modes in the physical branch and two
unphysical modes with opposite orientation.
The unsymmetrized Hoelbling operator has one chiral mode which is physical,
two modes of unequal orientation near $x=2$ and one oppositely oriented mode
near $x=4$.
The symmetrized version has one chiral mode which is physical, two oppositely
oriented modes near $x=2$ and one mode with original orientation near $x=4$.
The mixed operator (\ref{def_Hmix}), finally, has one chiral mode which is
physical and three unphysical modes near $x=2$ (two of which are antiparallel,
one of which is parallel to the chirality of the physical mode).
In each case it was checked that the choice $\Gamma_{50}$ (instead of
$\Gamma_{55}$) or the naive option $\<R|.|R\>$ (instead of $\<L|.|R\>$) bring
exactly the kind of change that one would anticipate from Fig.\,\ref{fig2}.


\section{Results with Symanzik improvement \label{sec:symanzik}}


So far the investigation of the eigenvalues and eigenmode chiralities of the
taste-split staggered actions (\ref{def_A}--\ref{def_Hmix}) leaves us with
the impression that they are technically rather close to the usual Wilson
action -- with chiral symmetry breaking and all the consequences of
non-normality, e.g.\ $\<L|$ and $|R\>$ eigenmodes.
This raises the question whether some of the remedies which have proven useful
in taming these effects with the Wilson action might be taken over and/or
adapted to these taste-split staggered actions.
The most successful remedies were link smearing, clover improvement and the
overlap procedure -- they all mitigate (in the last case eliminate) the effects
of chiral symmetry breaking, and they can be used in various combinations.

\begin{figure}[!tb]
\includegraphics[width=0.5\textwidth]{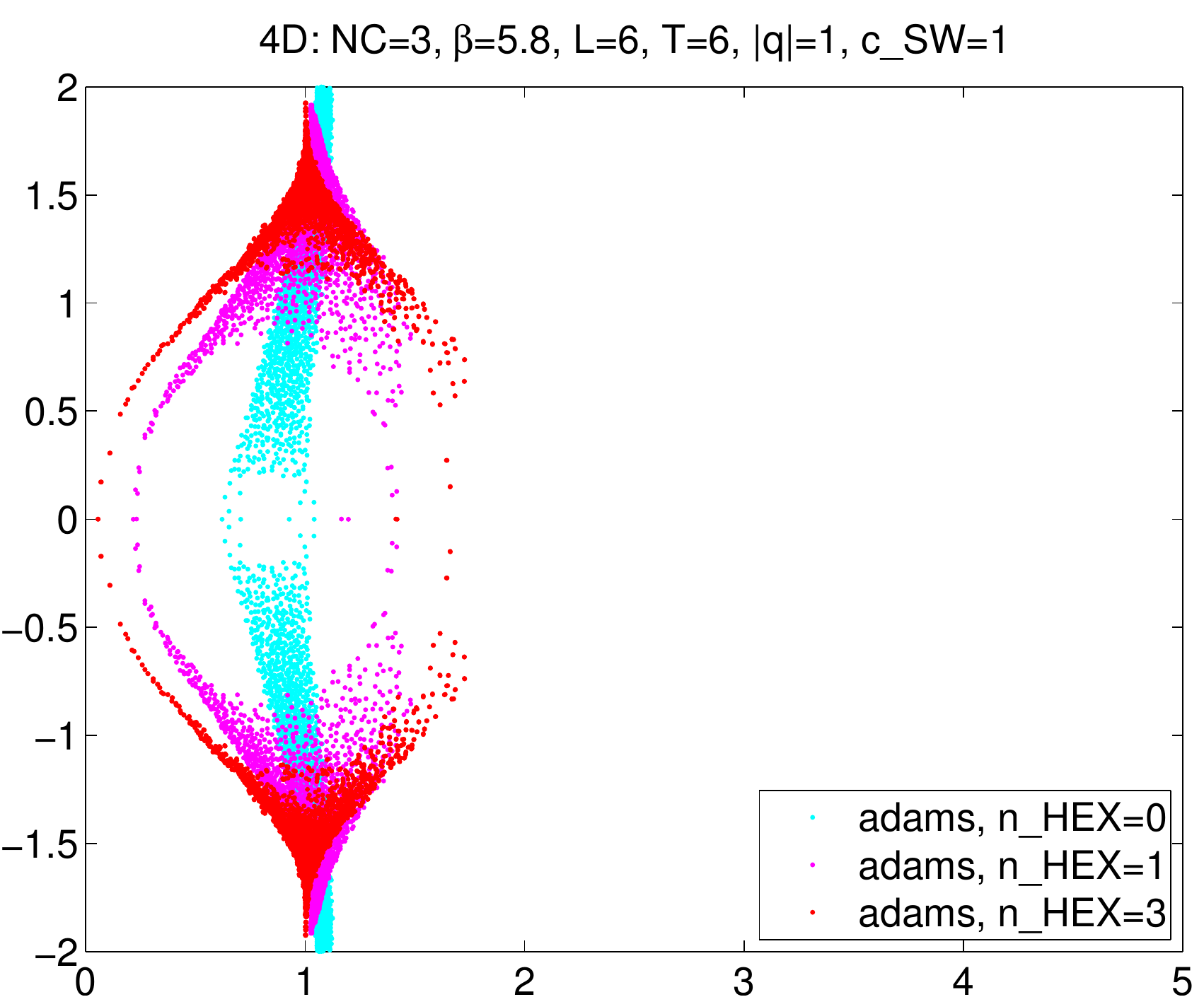}%
\includegraphics[width=0.5\textwidth]{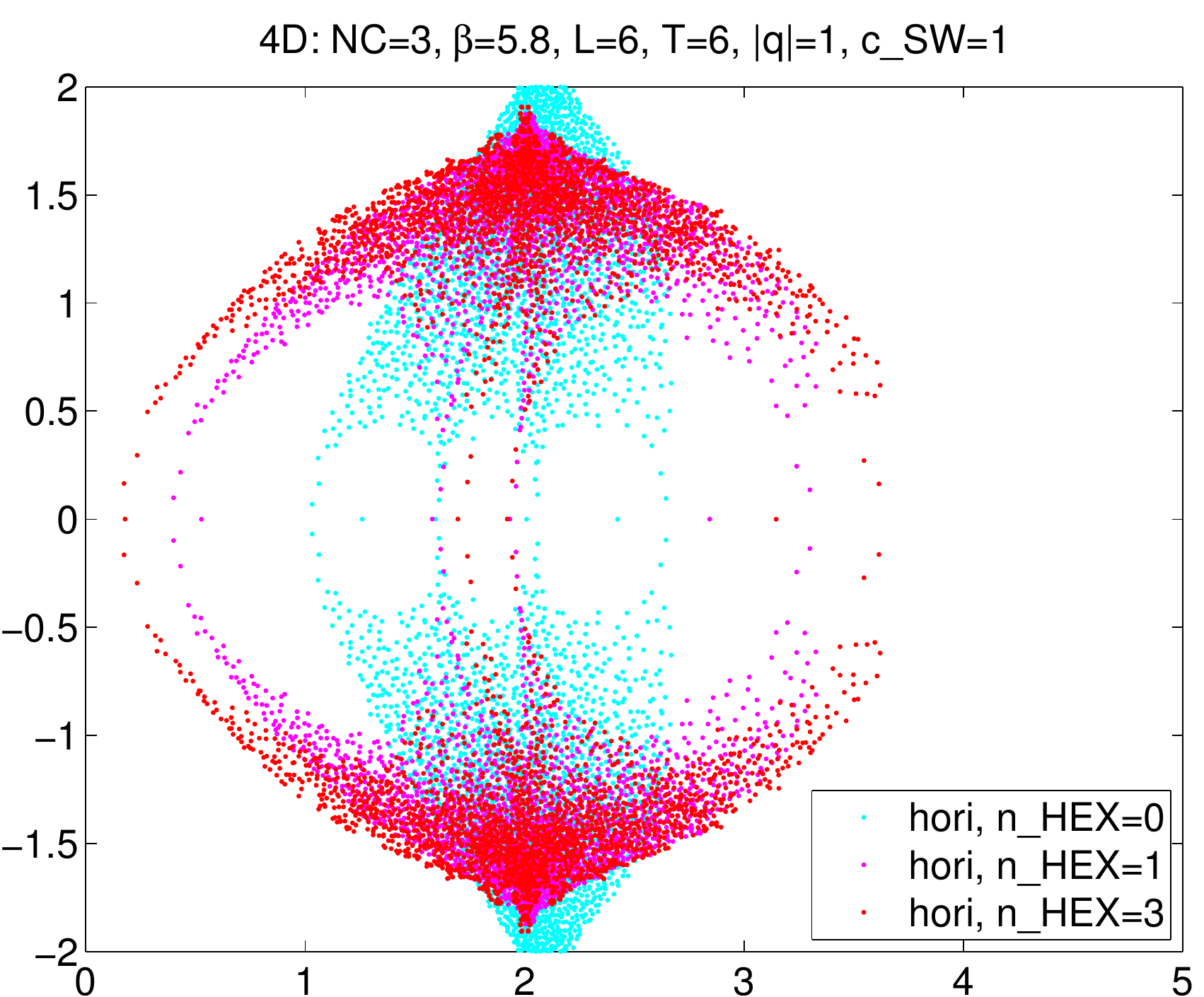}%
\\[2mm]
\includegraphics[width=0.5\textwidth]{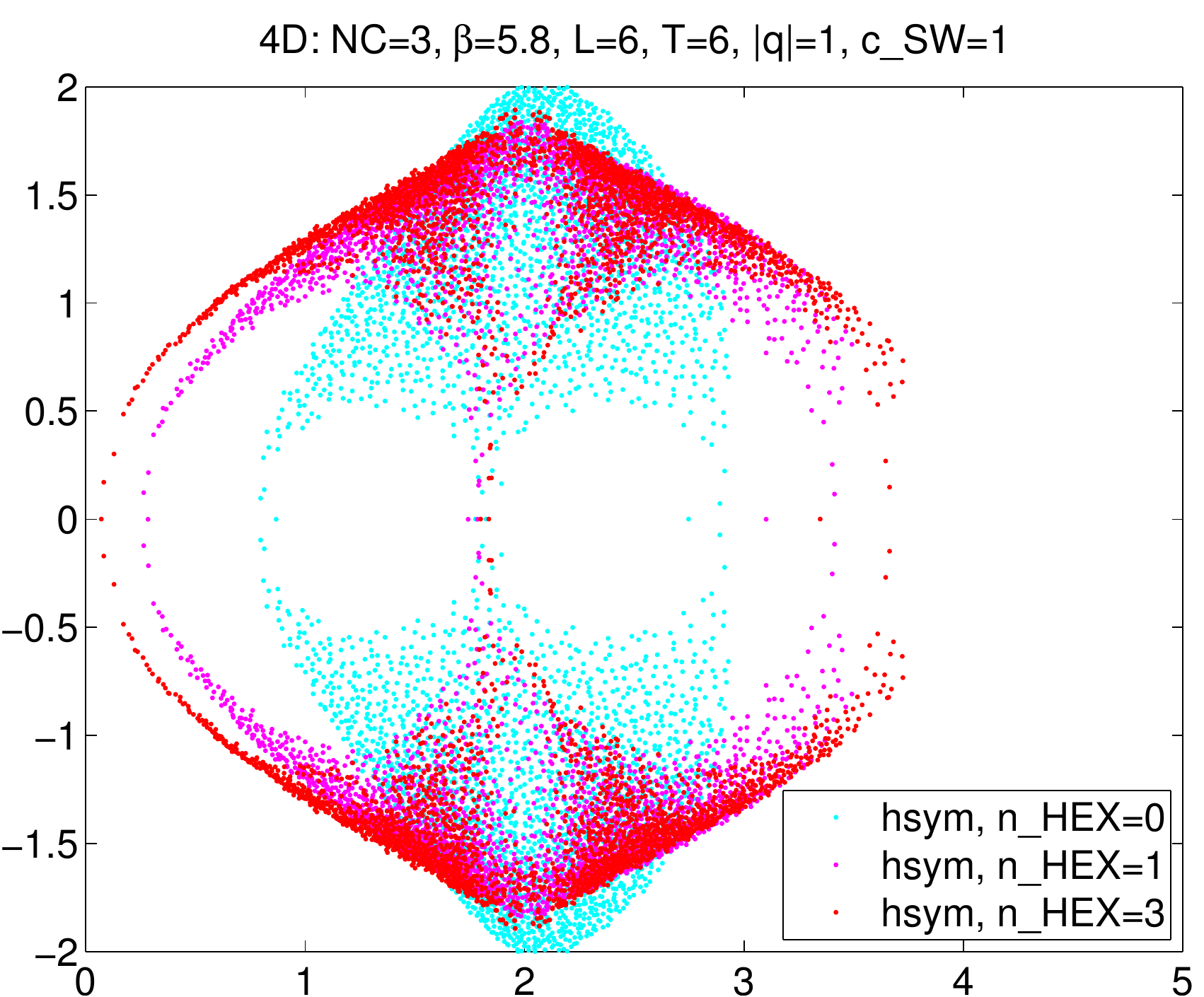}%
\includegraphics[width=0.5\textwidth]{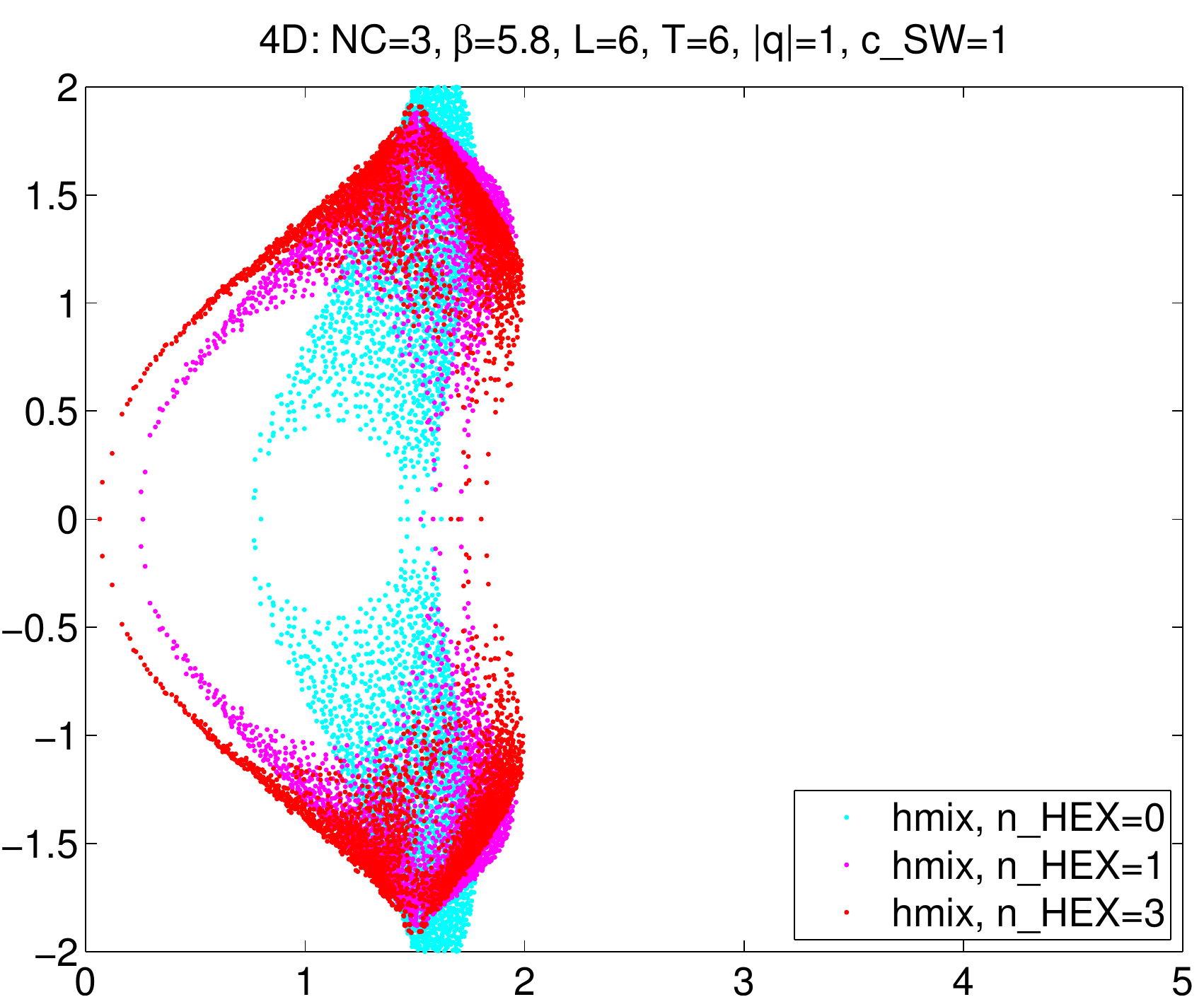}%
\caption{\label{fig5}\sl
Eigenvalue spectra of the four improved operators $(c_\mr{SW}\!=\!1)$ at
$m\!=\!0$ for up to three levels of HEX smearing -- the smearing seems
essential for the bellies to form.}
\end{figure}

\begin{figure}[!tb]
\includegraphics[width=0.5\textwidth]{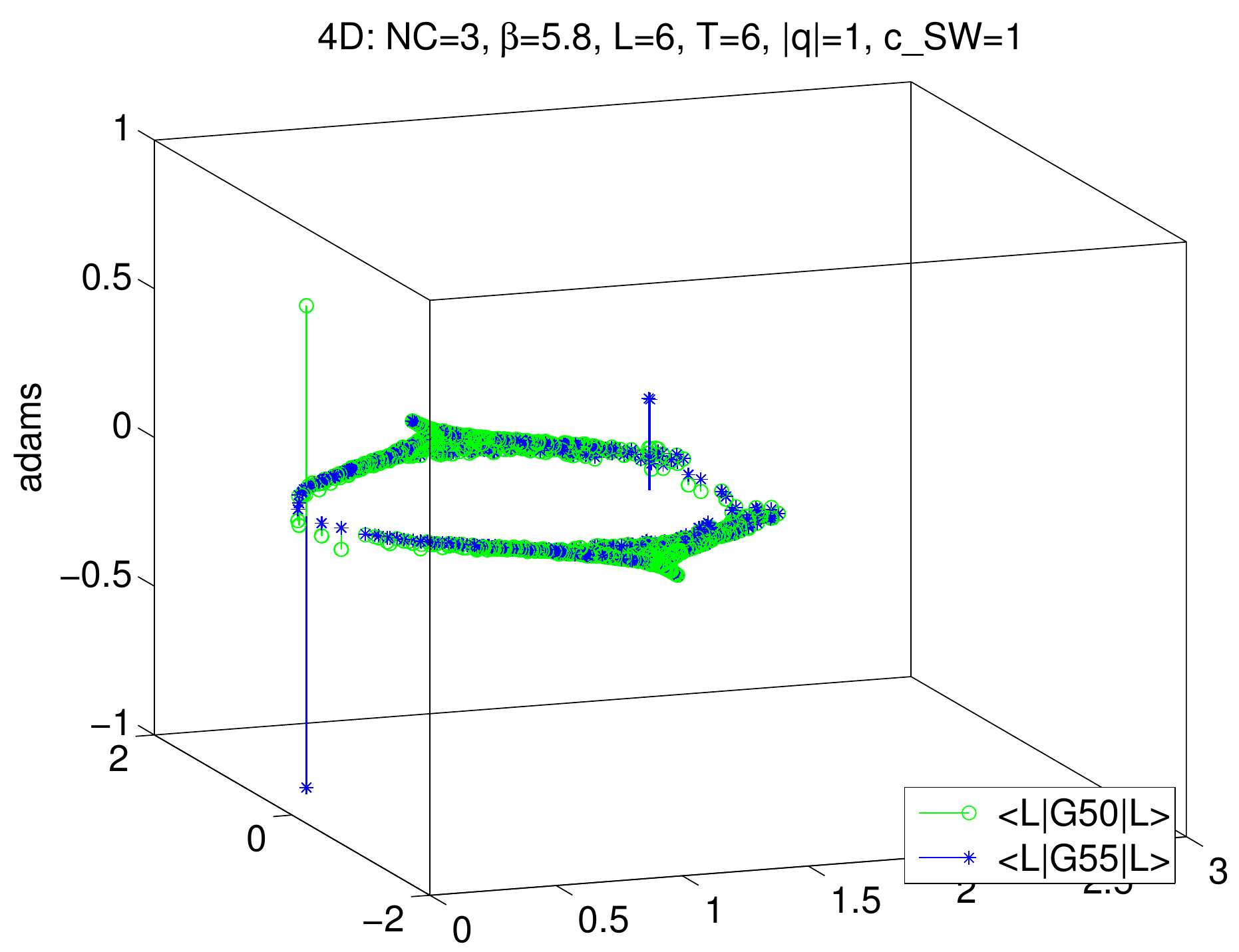}%
\includegraphics[width=0.5\textwidth]{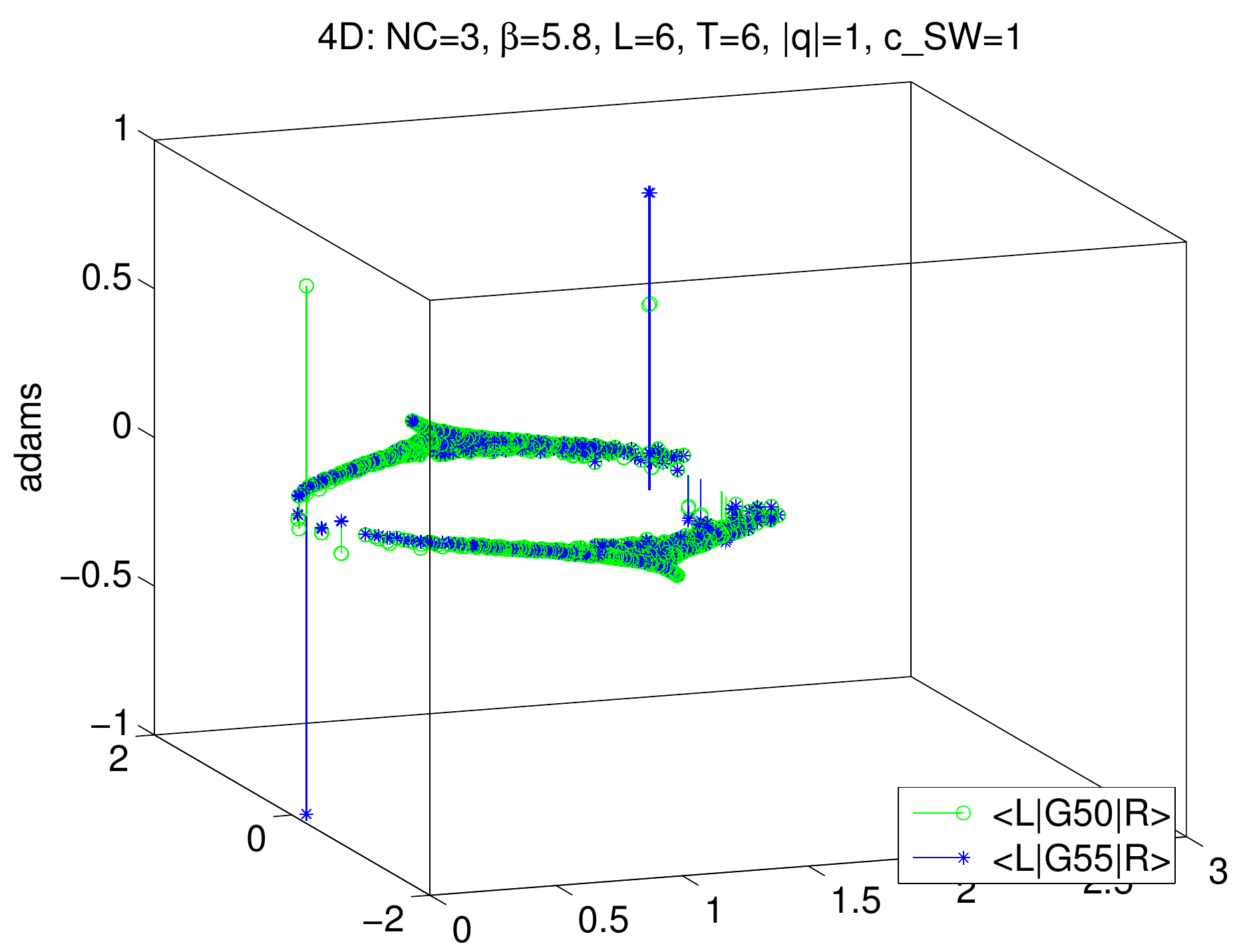}%
\\[2mm]
\includegraphics[width=0.5\textwidth]{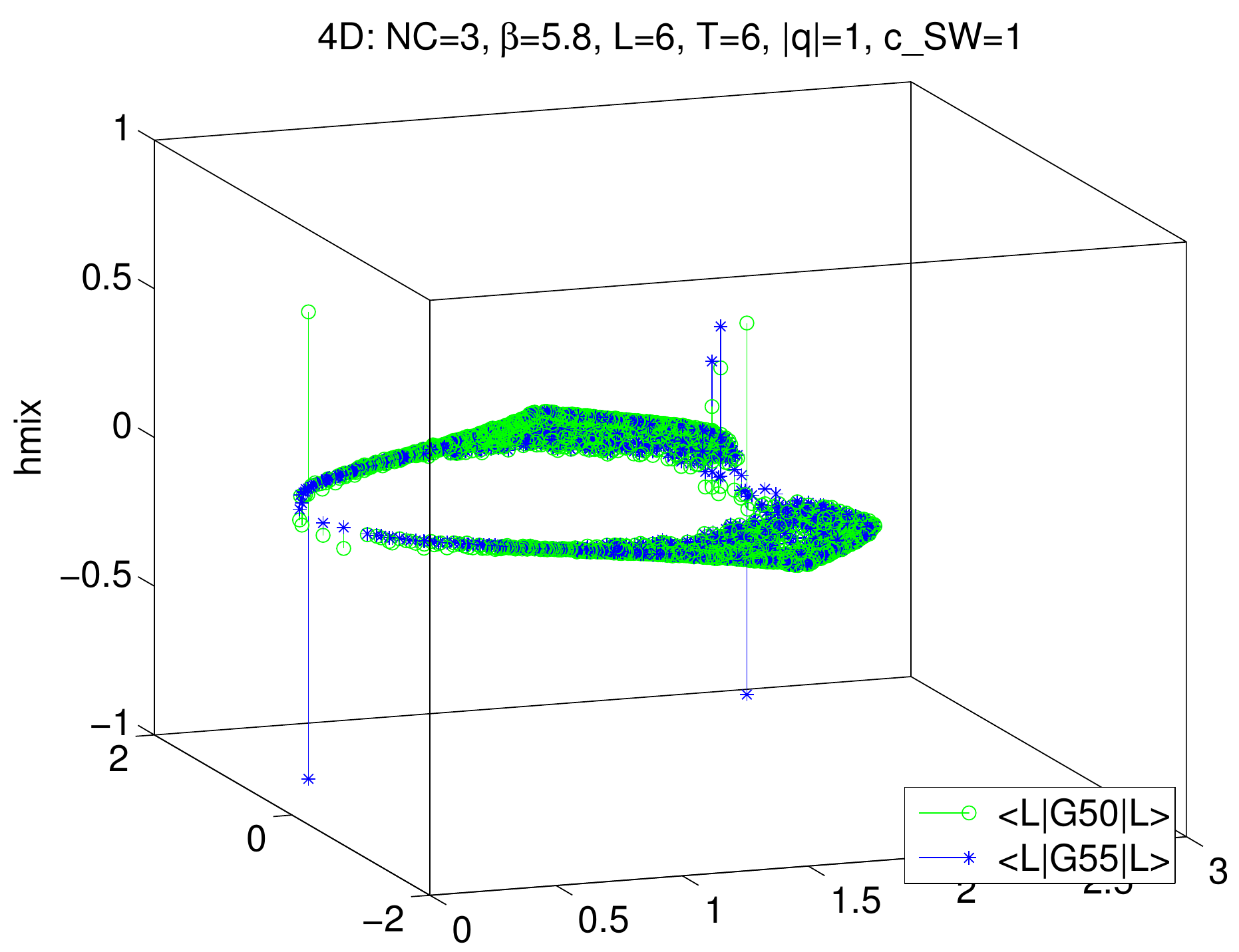}%
\includegraphics[width=0.5\textwidth]{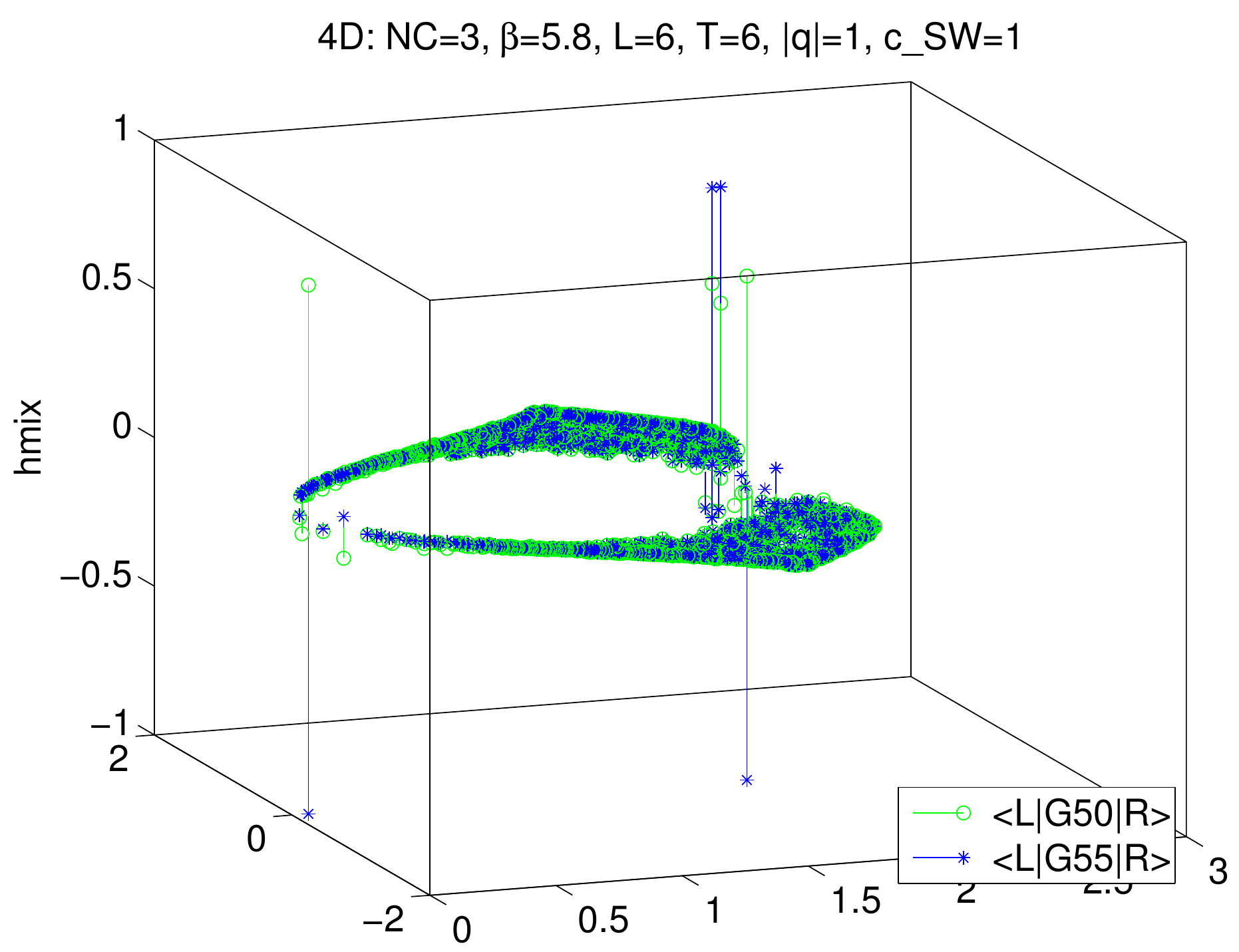}%
\caption{\label{fig6}\sl
Needle plots of the LL (left) and LR (right) chiralities of the Adams operator
(\ref{def_A}) (top) and of the mixed operator (\ref{def_Hmix}) (bottom), with
respect to $\Ga_{50}$ (green circles) or $\Ga_{55}\!=\!\ep$ (blue stars), with
Symanzik improvement $(c_\mr{SW}\!=\!1)$ and three HEX smearings.}
\end{figure}

\begin{figure}[!tb]
\includegraphics[width=0.5\textwidth]{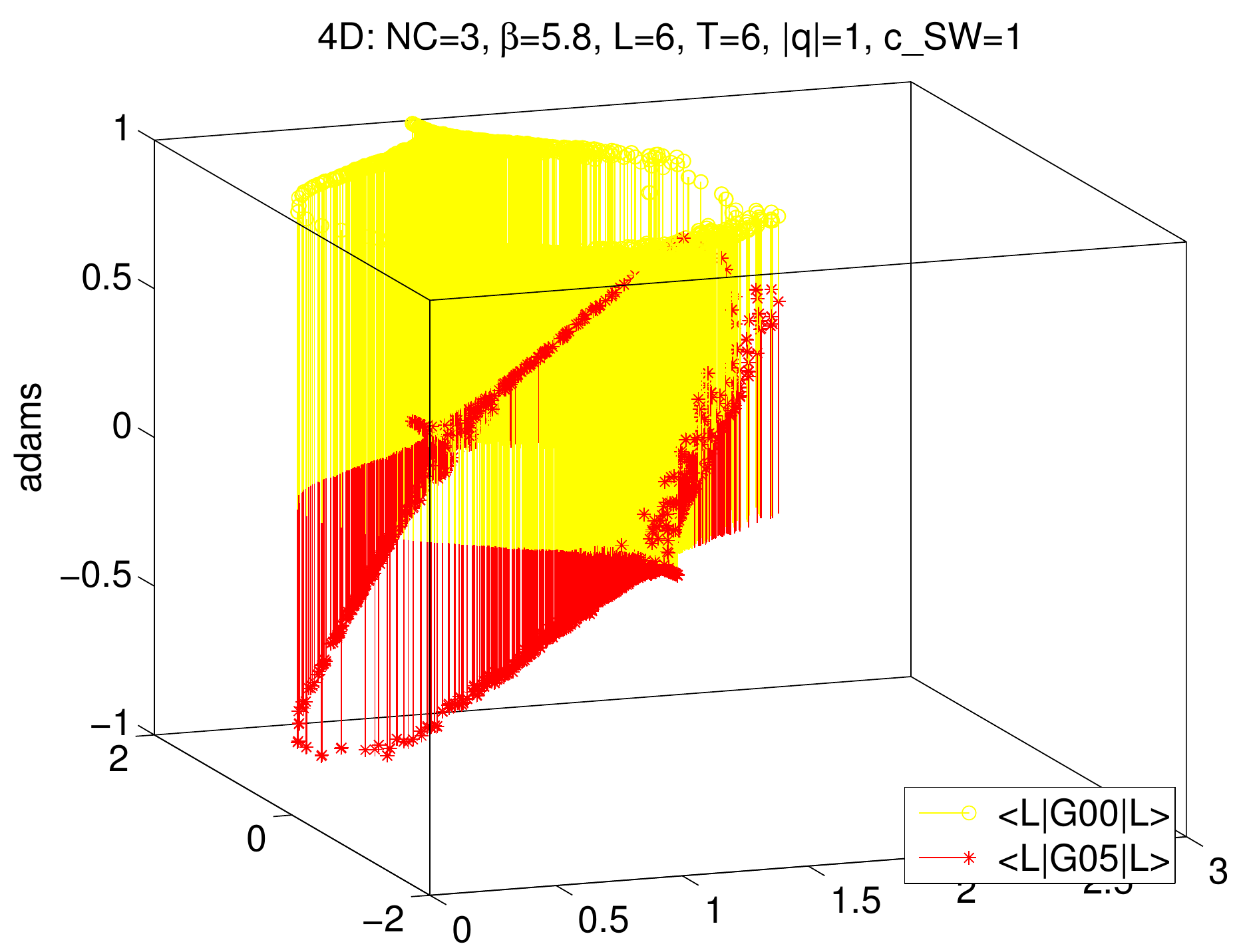}%
\includegraphics[width=0.5\textwidth]{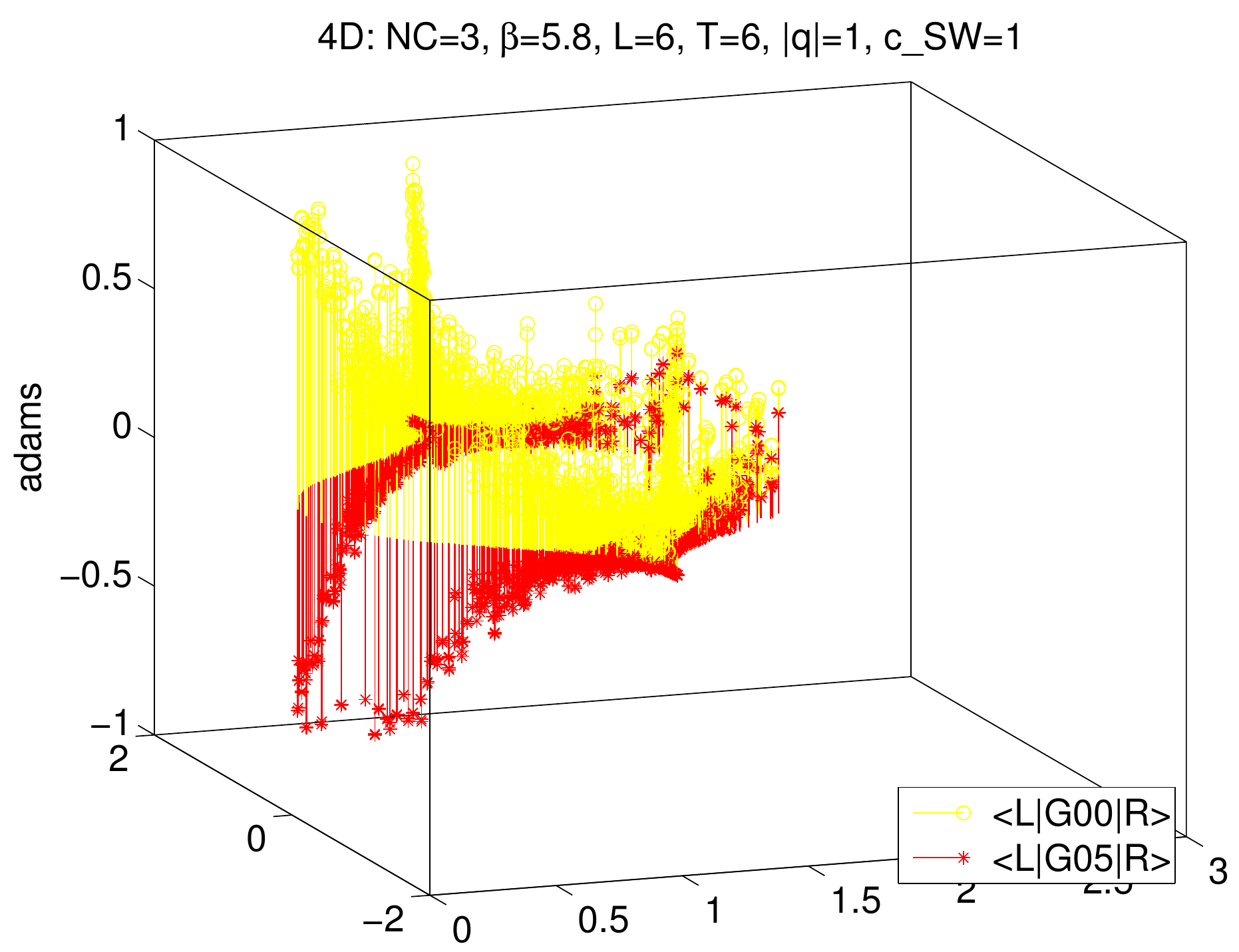}%
\\[2mm]
\includegraphics[width=0.5\textwidth]{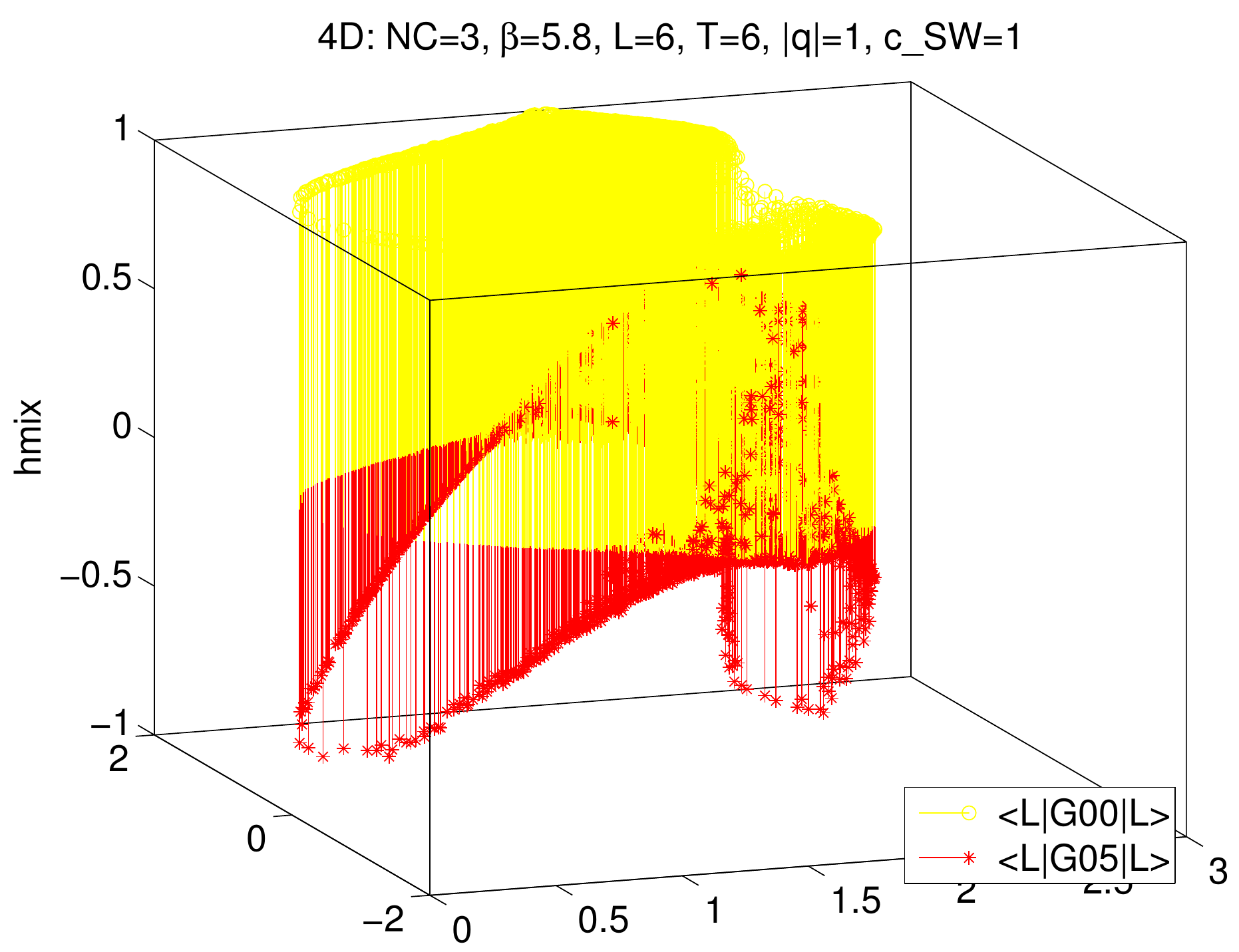}%
\includegraphics[width=0.5\textwidth]{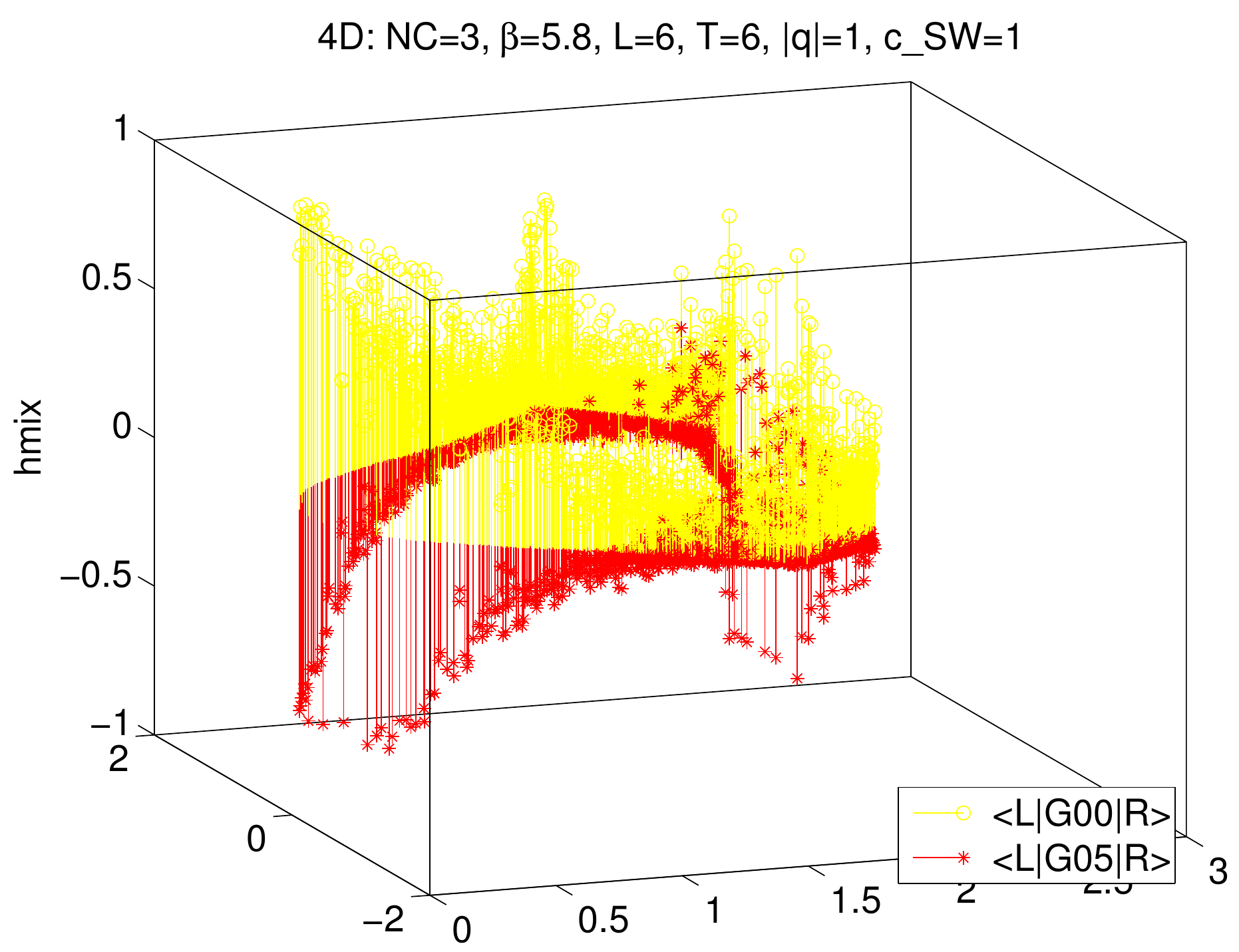}%
\caption{\label{fig7}\sl
Needle plots of the LL (left) and LR (right) overlaps $\<.|\Ga_{00}|.\>$ (yellow
circles) and taste chiralities $\<.|\Gamma_{05}|.\>$ (red stars) for the Adams
operator (\ref{def_A}) (top) and the mixed operator (\ref{def_Hmix}) (bottom),
with Symanzik improvement $(c_\mr{SW}\!=\!1)$ and three HEX smearings.}
\end{figure}

\begin{figure}[!tb]
\includegraphics[width=0.5\textwidth]{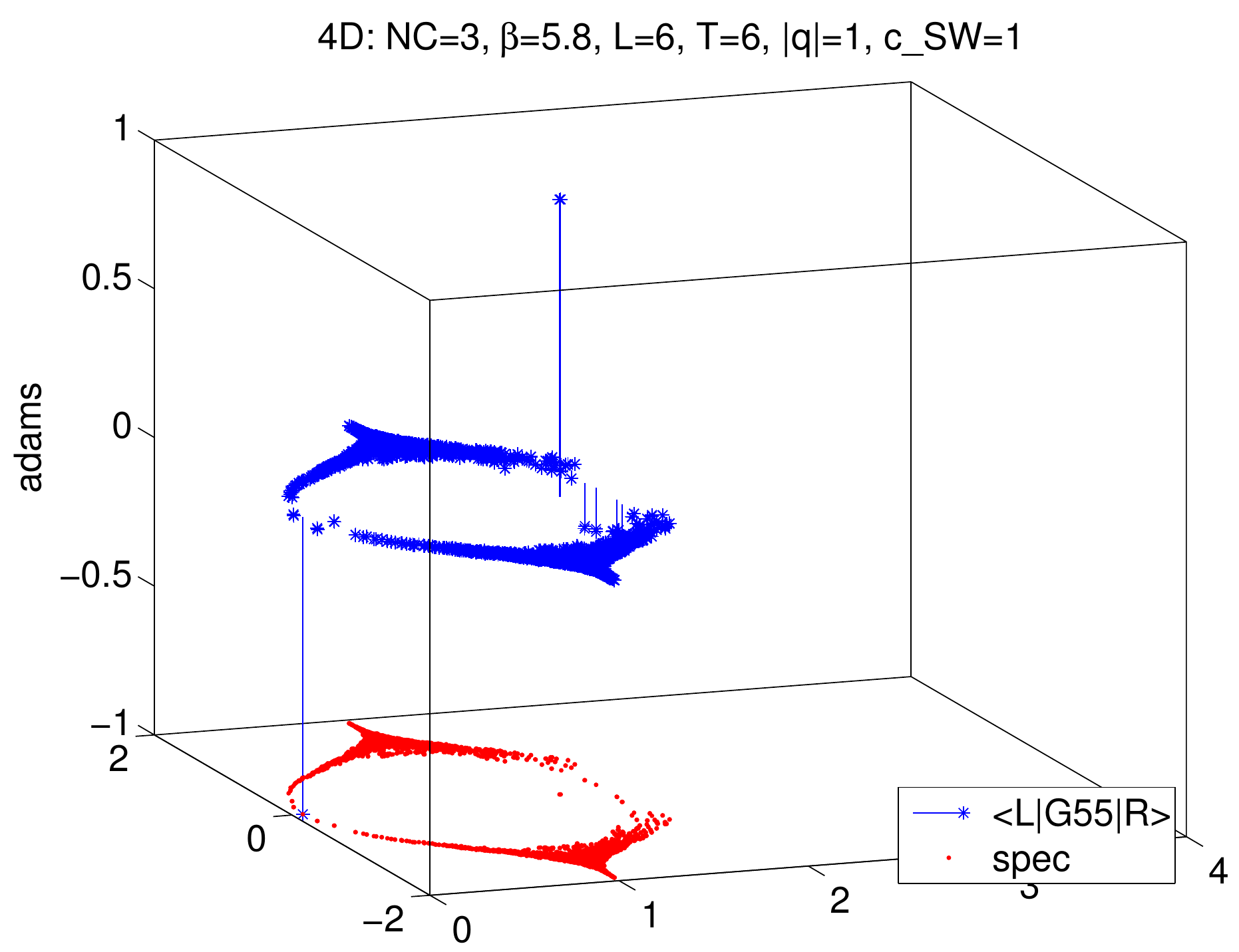}%
\includegraphics[width=0.5\textwidth]{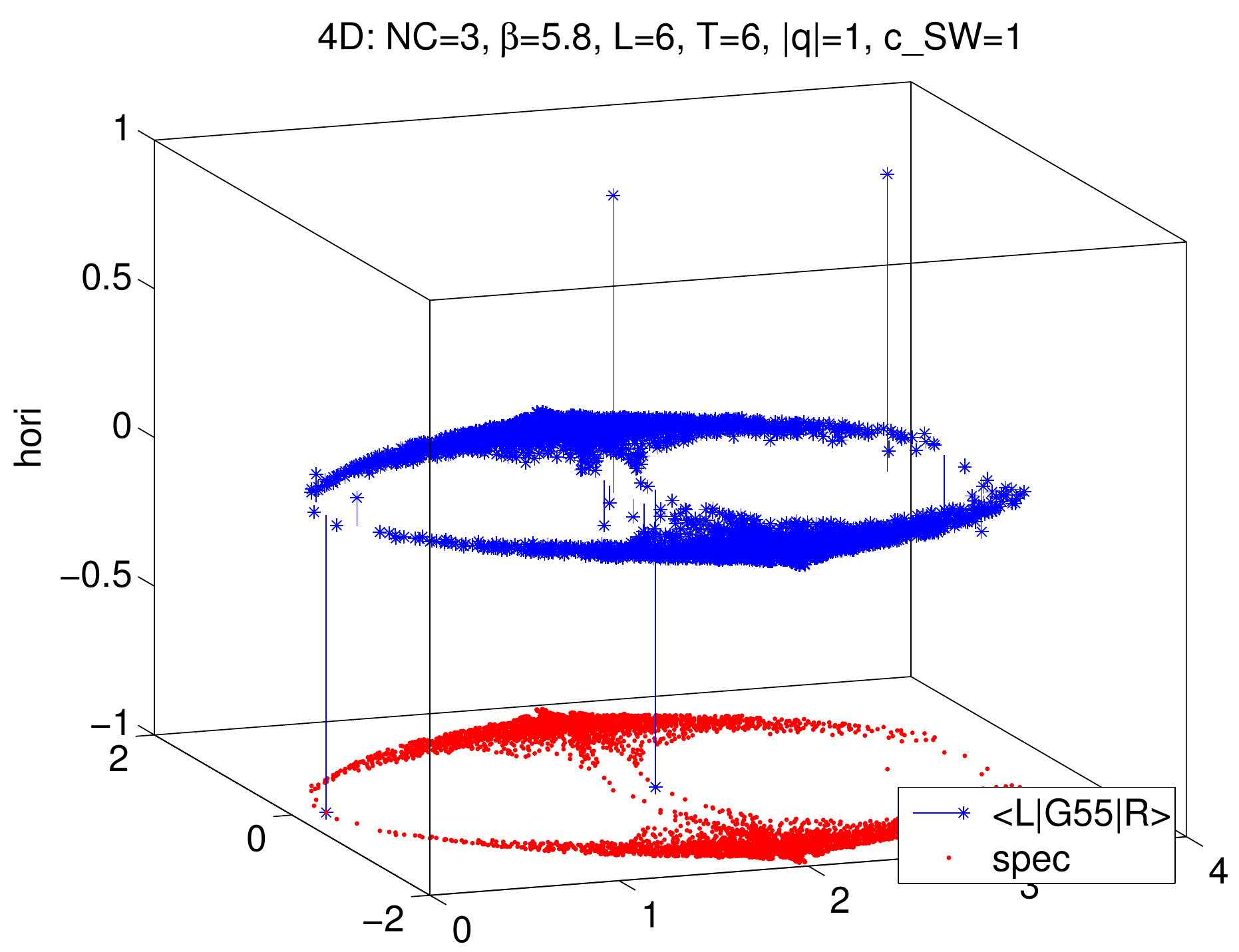}%
\\[2mm]
\includegraphics[width=0.5\textwidth]{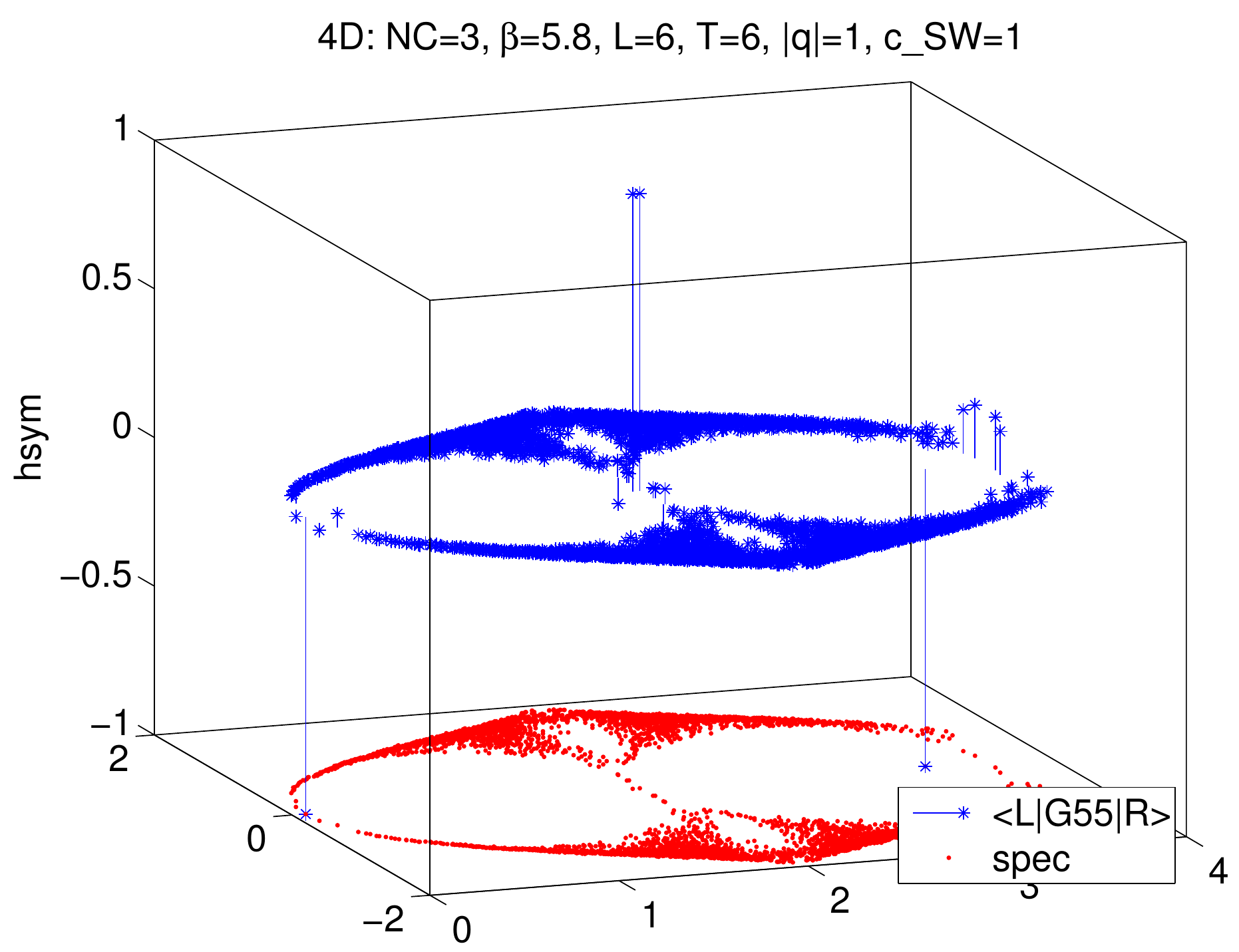}%
\includegraphics[width=0.5\textwidth]{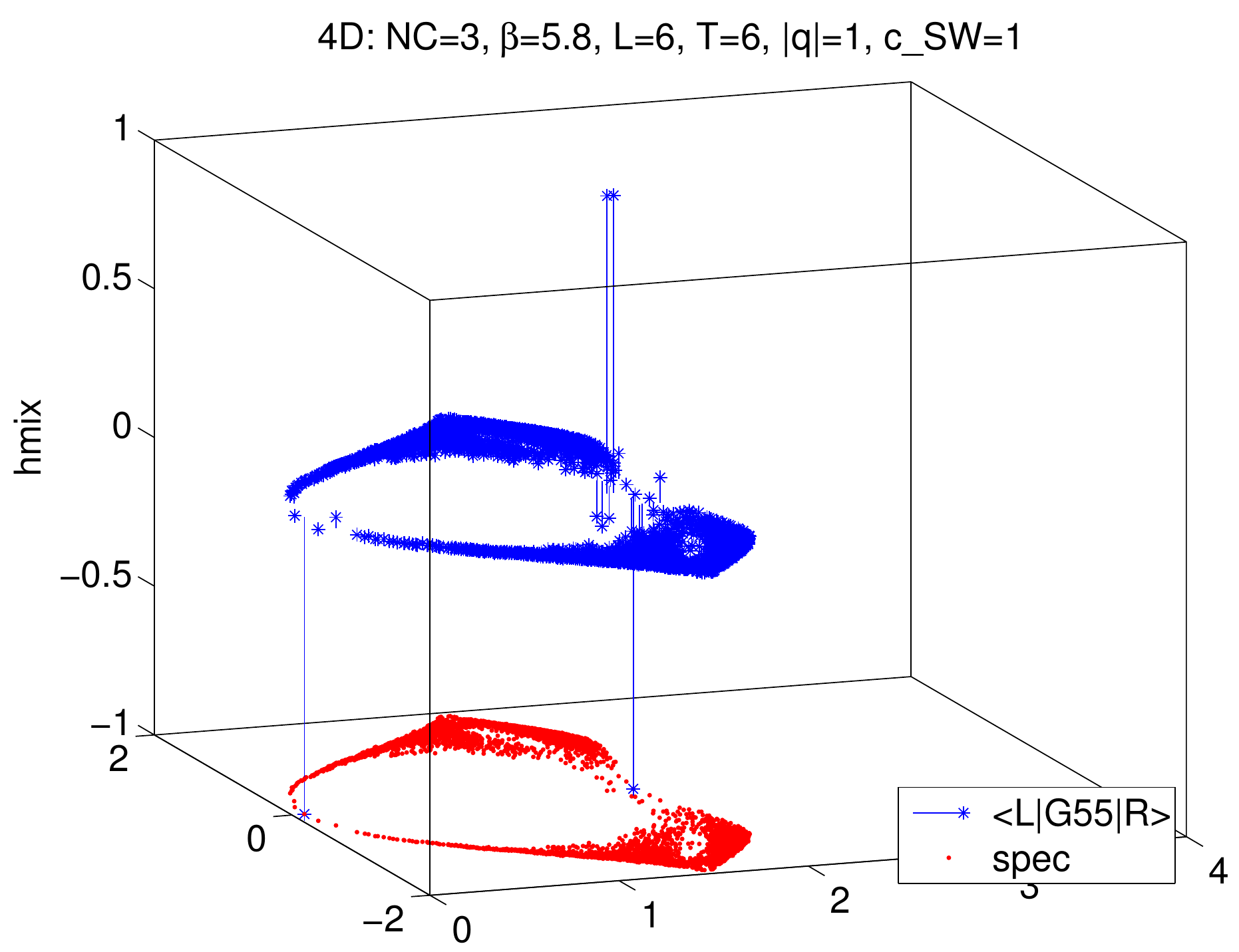}%
\caption{\label{fig8}\sl
Needle plots of the chiralities with respect to $\Ga_{55}$ of the improved
operators $(c_\mr{SW}\!=\!1)$ above their eigenvalues. The LR-definition of
$\<.|\Ga_{55}|.\>$ and three HEX smearings are used.}
\end{figure}

That link smearing proves useful for staggered actions with taste non-singlet
mass terms has already been shown in the previous section -- in
Fig.\,\ref{fig1} a clear improvement of the overall properties of the
eigenvalue spectra with an increased number of smearing steps is evident
(and consequently in Figs.\,\ref{fig2}, \ref{fig3}, \ref{fig4} only the results
for three smearing steps were displayed).

The next step is Symanzik improvement \cite{Symanzik:1983dc}.
In principle this is a program to be carried out in strict analogy to the
Wilson case -- one writes down a complete list of operators, ordered by their
mass dimensions, and eliminates those which are forbidden by symmetries or
redundant by the equation of motion.
In the Wilson case one finds that only the $d=5$ term
$\psb \ga_{\mu\nu} F_{\mu\nu} \ps$ needs to be added to the action
\cite{Sheikholeslami:1985ij}, with a tuned coefficient $c_\mr{SW}/2$ in front
to cancel all $O(a)$ cut-off effects (thus eliminating all signs of chiral
symmetry breaking to this order) \cite{Luscher:1996sc,Luscher:1996ug}.

With staggered fermions the situation is a bit more delicate, due to the
``staggered'' representation of the spinor degrees of freedom.
The point is that the operators $\Ga_\mu\equiv\et_\mu C_\mu$ and
$\Xi_\mu\equiv\ze_\mu C_\mu$ formally leave the mass dimension $d$ of a
fermion bilinear invariant, but they do involve a fermion hop [made gauge
invariant through a link $U_\mu(x)$ or $V_\mu(x)$].
In other words, the strict coincidence of the mass dimension of the operator
(as is relevant for the Symanzik analysis) and the maximum number of hops
(that affects the strength of the mixing or the ``noise'' of the operator)
is now broken.
It seems that this requires a more in-depth analysis \cite{SharpeNara}.
Here we give an incomplete and redundant list of staggered bilinears (with
``+h.c.'' implicit)
\bea
d=3: && \psb \{ 1, \Ga_\mu, \Xi_\mu,
                \Ga_\mu^2, \Ga_\mu\Xi_\mu, \Xi_\mu^2, ... \} \ps
\\
d=4: && \psb \{ \et_\mu D_\mu, \ze_\mu D_\mu,
                \Ga_\mu D_\mu, \Xi_\mu D_\mu, ... \} \ps
\\
d=5: && \psb \{ D_\mu^2, \et_\mu D_\mu^2, \ze_\mu D_\mu^2,
                \Ga_\mu D_\mu^2, \Xi_\mu D_\mu^2,
                \Ga_{\mu\nu}F_{\mu\nu}, \Xi_{\mu\nu}F_{\mu\nu}, ... \} \ps
\eea
along with a statement that most of them can be eliminated by means of the
arguments mentioned above and the identification of operators which differ only
by a factor $(1\!\otimes\!\xif)$.
The reasoning sketched in App.\,\ref{sec:appendix} suggests that only a single
operator needs to be included
\beq
-\frac{c_\mr{SW}}{4}\Big\{
\Ga_{12}F_{12}+...+\Ga_{34}F_{34}+F_{12}\Ga_{12}+...+F_{34}\Ga_{34}
\Big\}=
-\frac{c_\mr{SW}}{4}\Big\{
\sum_{\mu<\nu}\Ga_{\mu\nu}F_{\mu\nu}+\mr{h.c.}
\Big\}
\label{def_improvement}
\eeq
to be compared to
$-\frac{c_\mr{SW}}{2}\sum_{\mu<\nu}\ga_{\mu\nu}F_{\mu\nu}\de_{x,y}$ with Wilson
fermions (in either case factors are chosen such that $c_\mr{SW}=1$ at tree
level).
Note that $F_{\mu\nu}$ is site-diagonal, but not color-diagonal, whereas
$\Ga_{\mu\nu}$ is neither site- nor color-diagonal.
The symmetrization in (\ref{def_improvement}) ensures gauge-covariance (in
contradistinction to gauge-invariance in the Wilson case) and hermiticity.

Fig.\,\ref{fig5} displays the eigenvalues of the four operators
(\ref{def_A}--\ref{def_Hmix}) with the improvement term
(\ref{def_improvement}) on the same gauge configuration (with $|q|=1$) as in
Sec.\,\ref{sec:eigenvalues}.
Relative to Fig.\,\ref{fig1} one finds a clear amelioration of the behavior of
the physical (leftmost) branch of eigenvalues; it is now much thinner and the
section close to the origin is much more akin to a shifted Ginsparg-Wilson
circle.
In particular the tendency of the exactly real modes to get ``soaked into''
the belly is almost removed, except for the unsymmetrized operator
(\ref{def_Hori}).
Perhaps the most remarkable feature is that for the unsmeared operators the
improvement is hardly useful.
It takes the combination of smearing and improvement to get a profound effect
-- in strict analogy to what is observed for the Wilson case
\cite{Capitani:2006ni}.
For clarity we add that the links in the operator, in $\Ga_{\mu\nu}$ and in
$F_{\mu\nu}$ have all undergone the same kind of smearing.
Finally, let us add that it was explicitly checked that the alternative
operator $\sum_{\mu<\nu}\Xi_{\mu\nu}F_{\mu\nu}+\mr{h.c.}$ does not lead to a
similar improvement of the eigenvalues in the physical branch.

Fig.\,\ref{fig6} displays the same kind of needle plots for the $\Ga_{50}$ and
$\Ga_{55}$ chiralities of the eigenmodes of the operators (\ref{def_A}) and
(\ref{def_Hmix}) that were presented previously without improvement.
Relative to Fig.\,\ref{fig2} the most significant change is that the difference
between $\<L|.|L\>$ and $\<L|.|R\>$ in the physical branch is now less
pronounced.
This is a hint that the non-normality of the operator, when restricted to the
physical subspace, is reduced by the improvement term.

Fig.\,\ref{fig7} displays a feature that was not discussed in the unimproved
case.
The overlaps $\<L|1|L\>$ and $\<L|1|R\>$ are shown for the operators
(\ref{def_A}) and (\ref{def_Hmix}) -- the former ones are
trivially one, but the latter ones indicate how normal the operator is, and
indeed this figure is close to 1 in the physical branch.
In addition the taste chiralities are displayed; both $\<L|\Ga_{05}|L\>$
and $\<L|\Ga_{05}|R\>$ tend to assign each branch a fixed taste chirality which
is not sensitive to topology of the background (as mentioned earlier, with our
choices the taste chirality in the physical branch is $\xif=-1$).
The conclusion is that the taste chiralities in the interacting case look
similar to the free-field results presented in \cite{deForcrand:2012bm}.

Fig.\,\ref{fig8} displays the $\<L|\Ga_{55}|R\>$ chiralities of all four
operators (\ref{def_A}--\ref{def_Hmix}) with the clover term
(\ref{def_improvement}).
These chiralities are very well pronounced (i.e.\ very close to $\pm1$) and
linked to the topology of the gauge background; a cut at $\pm0.5$ is well
suited to identify the chiral modes.






\section{Results with overlap projection \label{sec:overlap}}


\begin{figure}[!tb]
\includegraphics[width=0.5\textwidth]{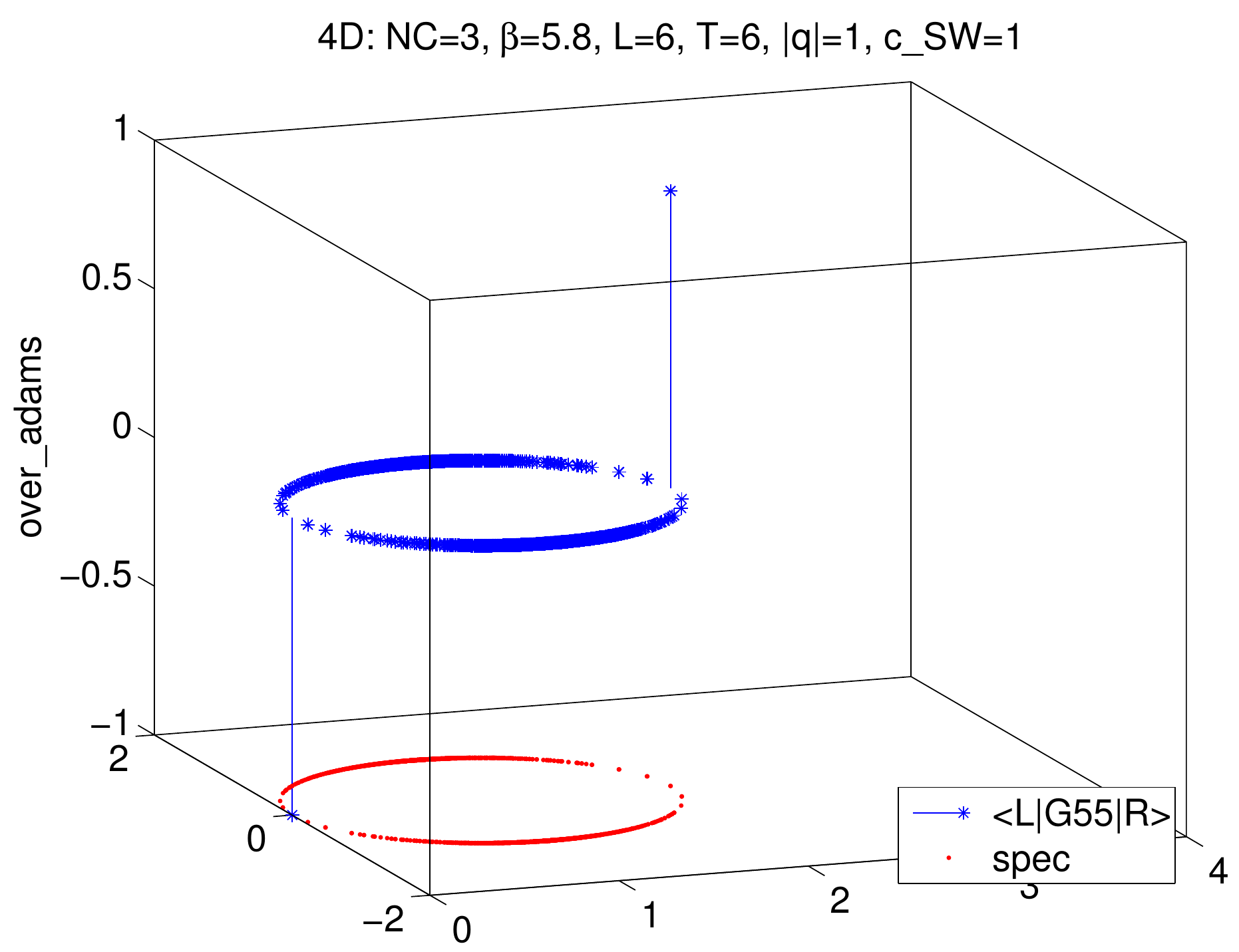}%
\includegraphics[width=0.5\textwidth]{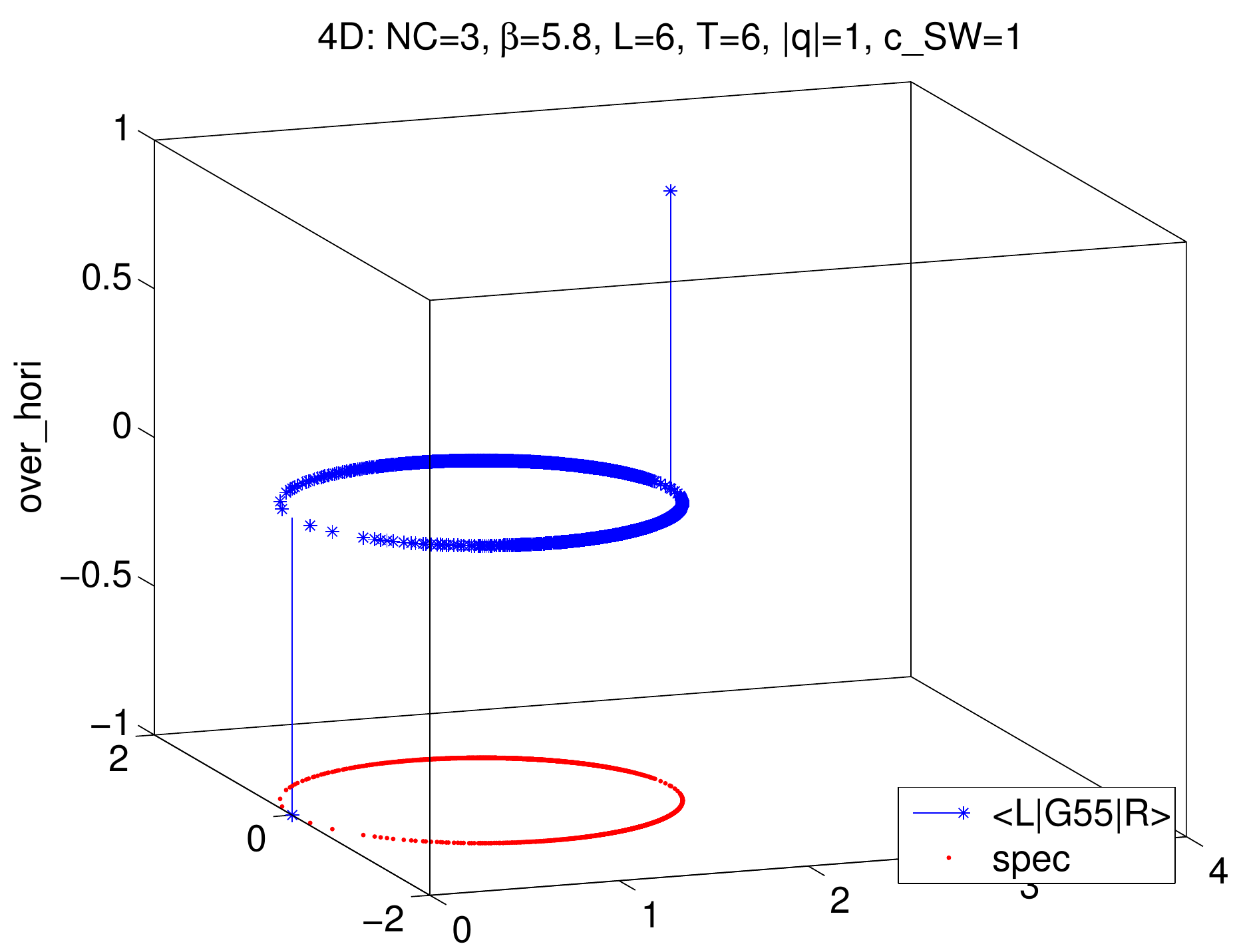}%
\\[2mm]
\includegraphics[width=0.5\textwidth]{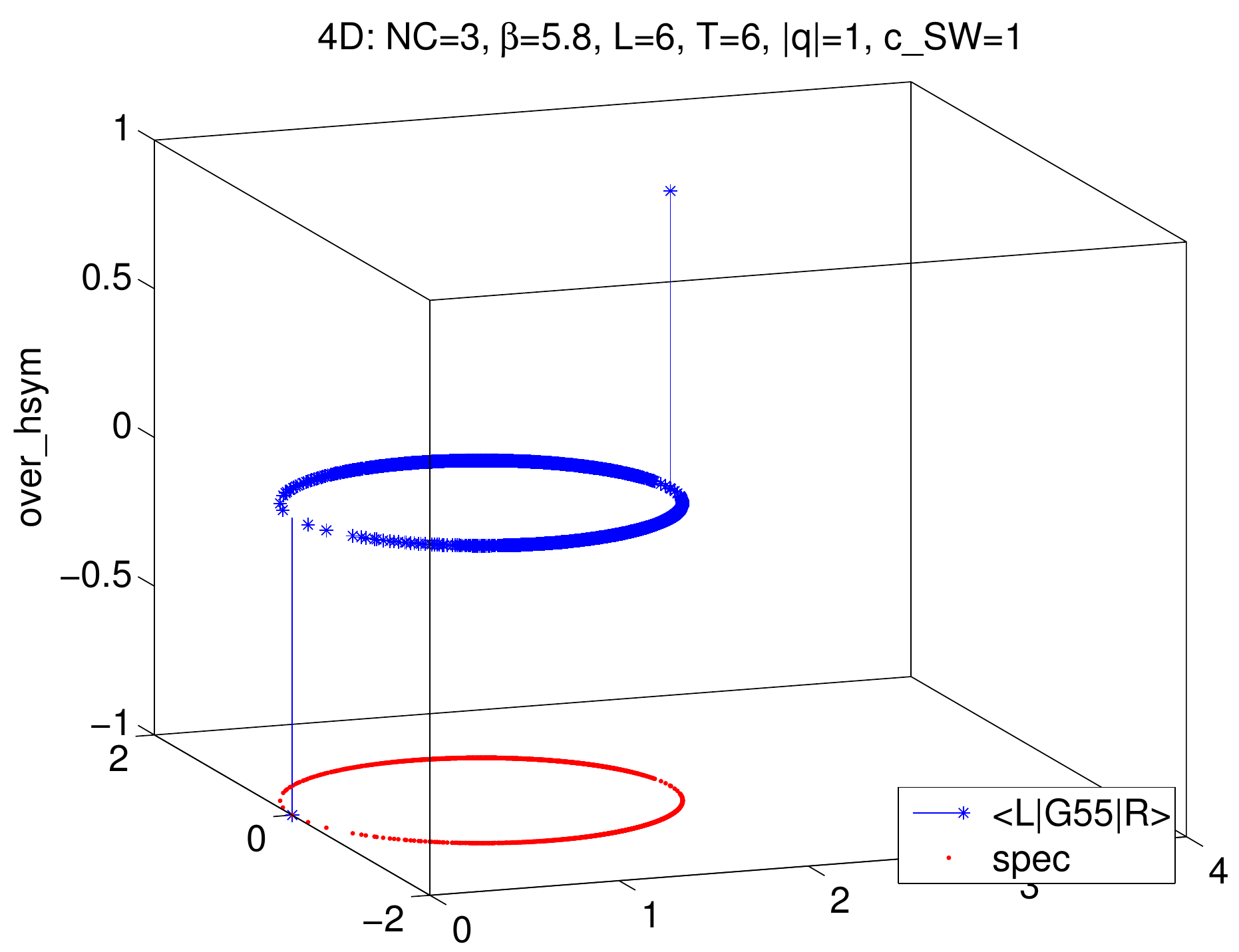}%
\includegraphics[width=0.5\textwidth]{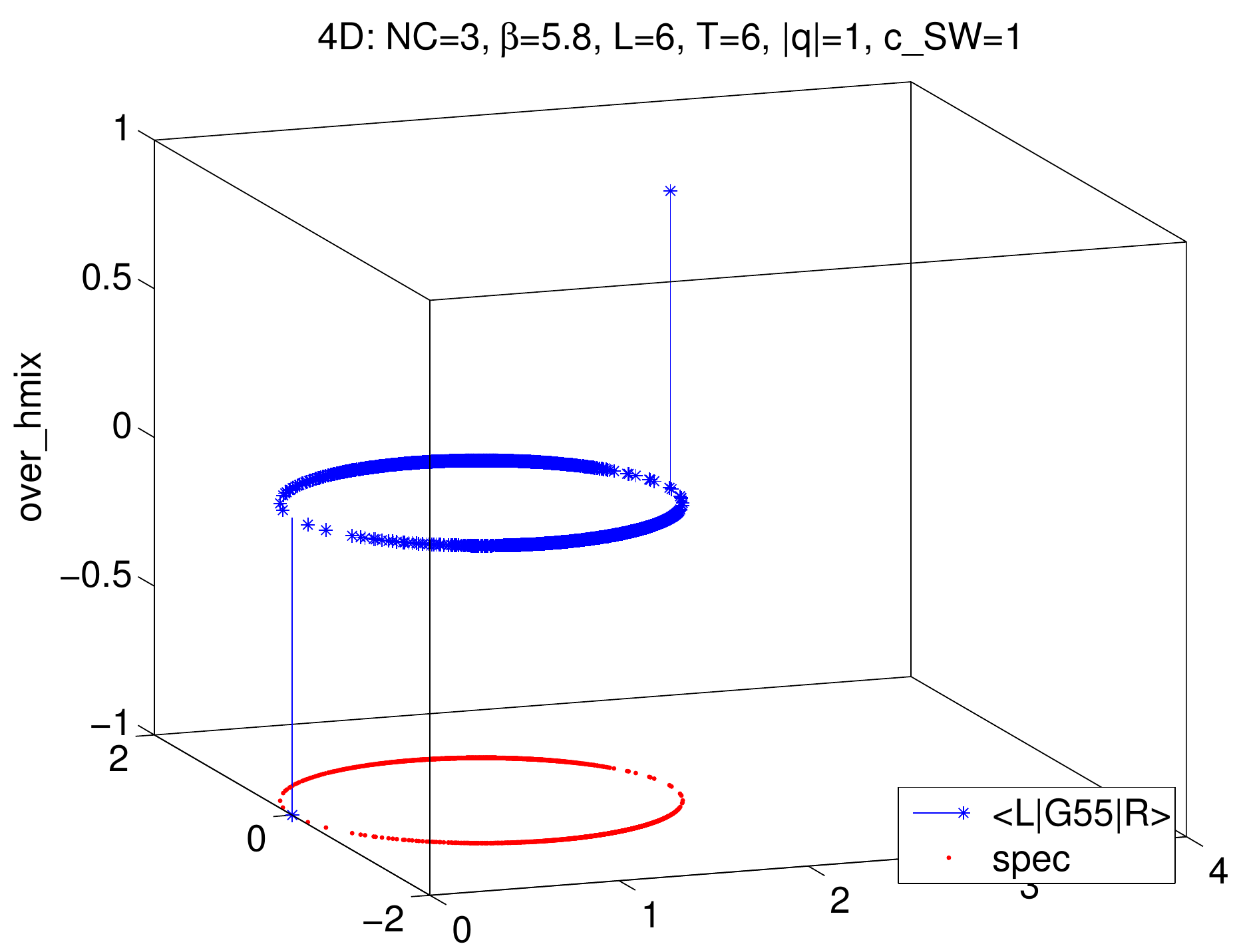}%
\caption{\label{fig9}\sl
Needle plots of the chiralities with respect to $\Ga_{55}$ of the overlap
actions with improved kernels $(c_\mr{SW}\!=\!1)$. All L/R-definitions of
$\<.|\Ga_{55}|.\>$ are equivalent; three HEX smearings are used.}
\end{figure}

The definition of the massless overlap Dirac operator takes the form
\cite{Neuberger:1997fp,Neuberger:1998wv}
\beq
\Dov=
\frac{\rh}{a}\Big(1+X(X\dag X)^{-1/2}\Big)=
\frac{\rh}{a}\Big(1+(X X\dag)^{-1/2}X\Big)
\label{def_over}
\eeq
where $X=a\Dke-\rh$ and the kernel $\Dke$ may be any undoubled fermion action
[as is the case with (\ref{def_Hori}--\ref{def_Hmix})] or a doubled one where
the tastes in the physical branch \emph{share one chirality} [as is the case
with (\ref{def_A})].
Evidently, the overlap inherits the multiplicity of the kernel operator, e.g.\
two-fold with (\ref{def_A}) as kernel.
The canonical value of the shift parameter $0<\rh<2$ is $\rh=1$, a choice which
we shall adopt in the following.
Because the eigenvalue spectra are somewhat boring (the spectra lie on the
shifted unit circle), we proceed directly to the chiralities.

Fig.\,\ref{fig9} displays the $\<.|\Ga_{55}|.\>$ chiralities of the overlap
action built from the kernels (\ref{def_A}--\ref{def_Hmix}).
These operators are normal, i.e.\ the left-eigenvectors are daggered copies of
the right-eigenvectors, viz.\ $\<L_i|=|R_i\>\dag$, and the distinction between
the four L/R-versions of $\<.|\Ga_{55}|.\>$ is now obsolete.
The overlap procedure being a projection, together with the $\Ga_{55}$
hermiticity of the kernel, ensures that the resulting $\<.|\Ga_{55}|.\>$
chiralities are exactly $\pm1$ (topological modes) or exactly $0$
(non-topological modes).
As a side-effect in the conglomerate of unphysical branches (near $x=2$) only
one type of $\Ga_{55}$ chirality survives [contrary to what a naive shift of
the corresponding modes of the kernel operator in Fig.\,\ref{fig8} would
suggest].
Thus the needle at $z=(x,y)=(2,0)$ is two-fold populated with the Adams
kernel, but one-fold in all other cases.

\begin{figure}[!tb]
\includegraphics[width=0.5\textwidth]{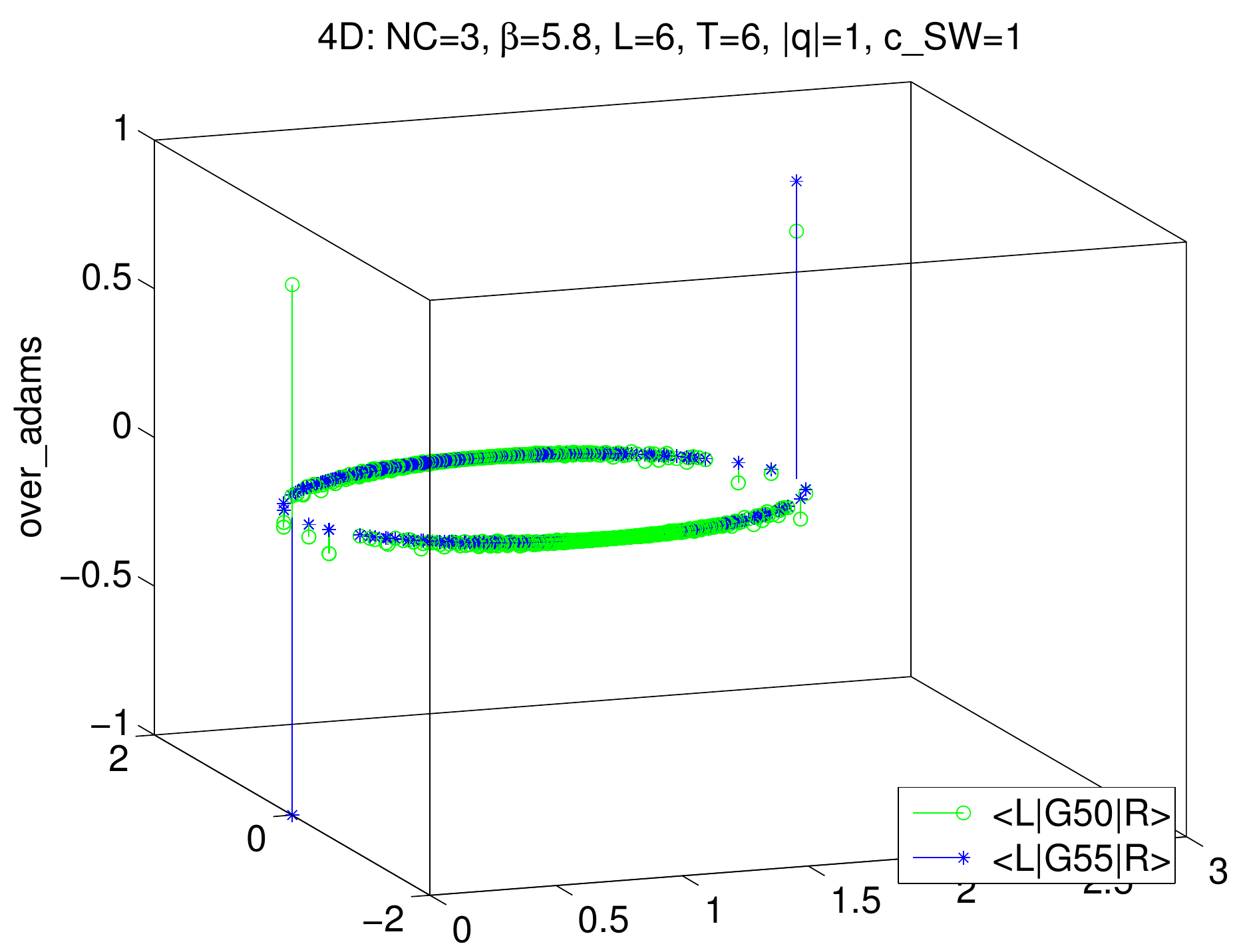}%
\includegraphics[width=0.5\textwidth]{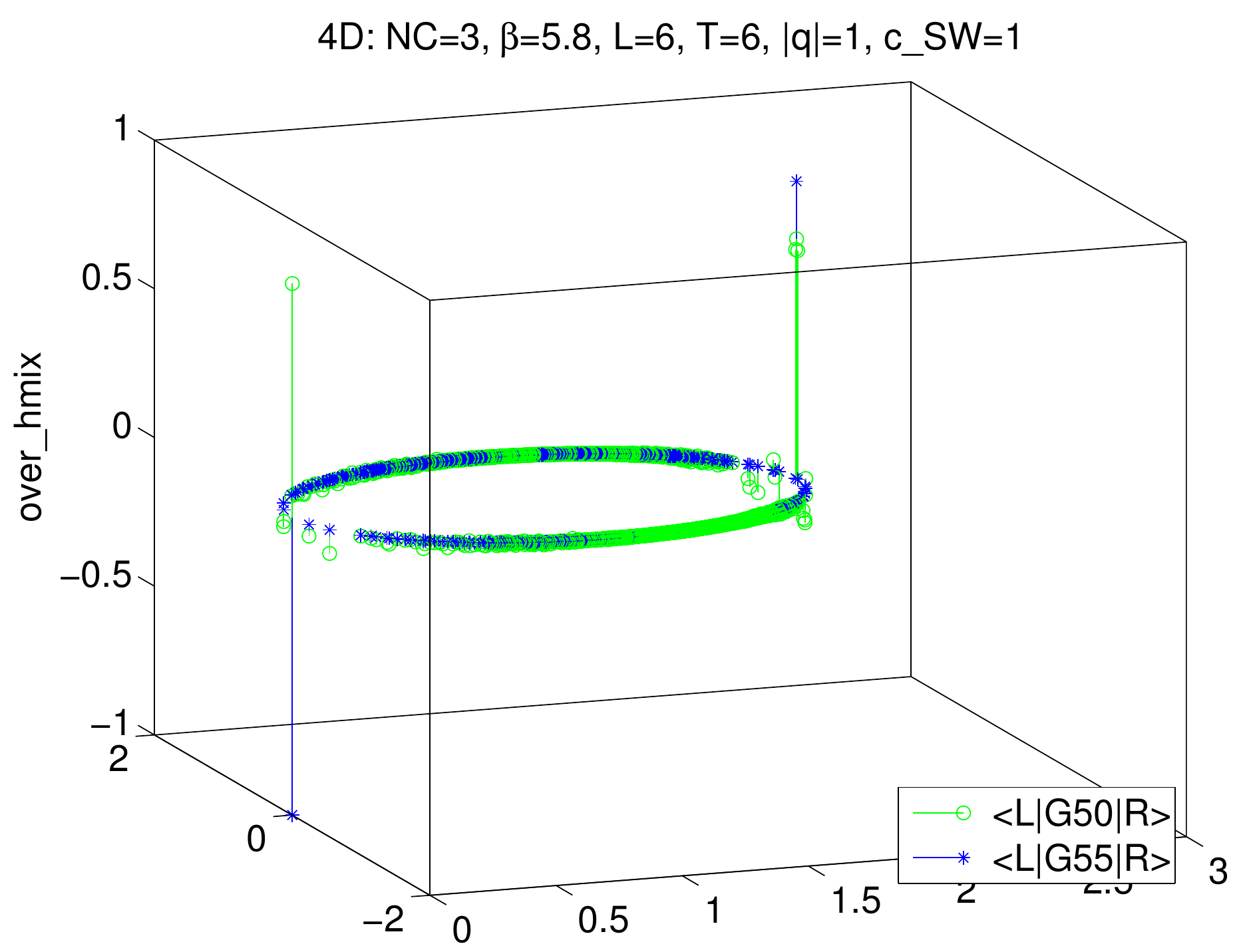}%
\caption{\label{fig10}\sl
Needle plots of the chiralities of the overlap actions based on the Adams
kernel (\ref{def_A}) (left) or the mixed kernel (\ref{def_Hmix}) (right), with
respect to $\Ga_{50}$ (green circles) or $\Ga_{55}\!=\!\ep$ (blue stars).
Either kernel uses Symanzik improvement $(c_\mr{SW}\!=\!1)$ and three HEX
smearings.}
\end{figure}

Fig.\,\ref{fig10} compares the $\<.|\Ga_{55}|.\>$ chiralities of the overlap
operators with taste-split staggered kernel to the $\<.|\Ga_{50}|.\>$
chiralities (still on the same $|q|=1$ configuration).
While the former are exactly $\pm1$ (for the topological modes) the latter are
not, in spite of the overlap projection.
As discussed before this is a consequence of the kernels being $\Ga_{55}$
hermitian but not $\Ga_{50}$ hermitian.

As a practical point let us remark that the versions with unimproved kernels
($c_\mr{SW}=0$) look superficially identical to the versions with clover
improved kernels ($c_\mr{SW}=1$) that were presented in the last two figures.
Given that the combination of clover improvement and link smearing facilitated
the separation between the physical branch and the unphysical branch(es), it is
hardly surprising that the condition number in the overlap construction is
significantly smaller in the latter case (which reduces the order of the
polynomial/rational approximation to the sign function and thus the
computational requirements).
In short, it is strongly recommended to equip the kernel operator with a clover
term and overall link smearing.

\begin{figure}[!tb]
\includegraphics[width=0.5\textwidth]{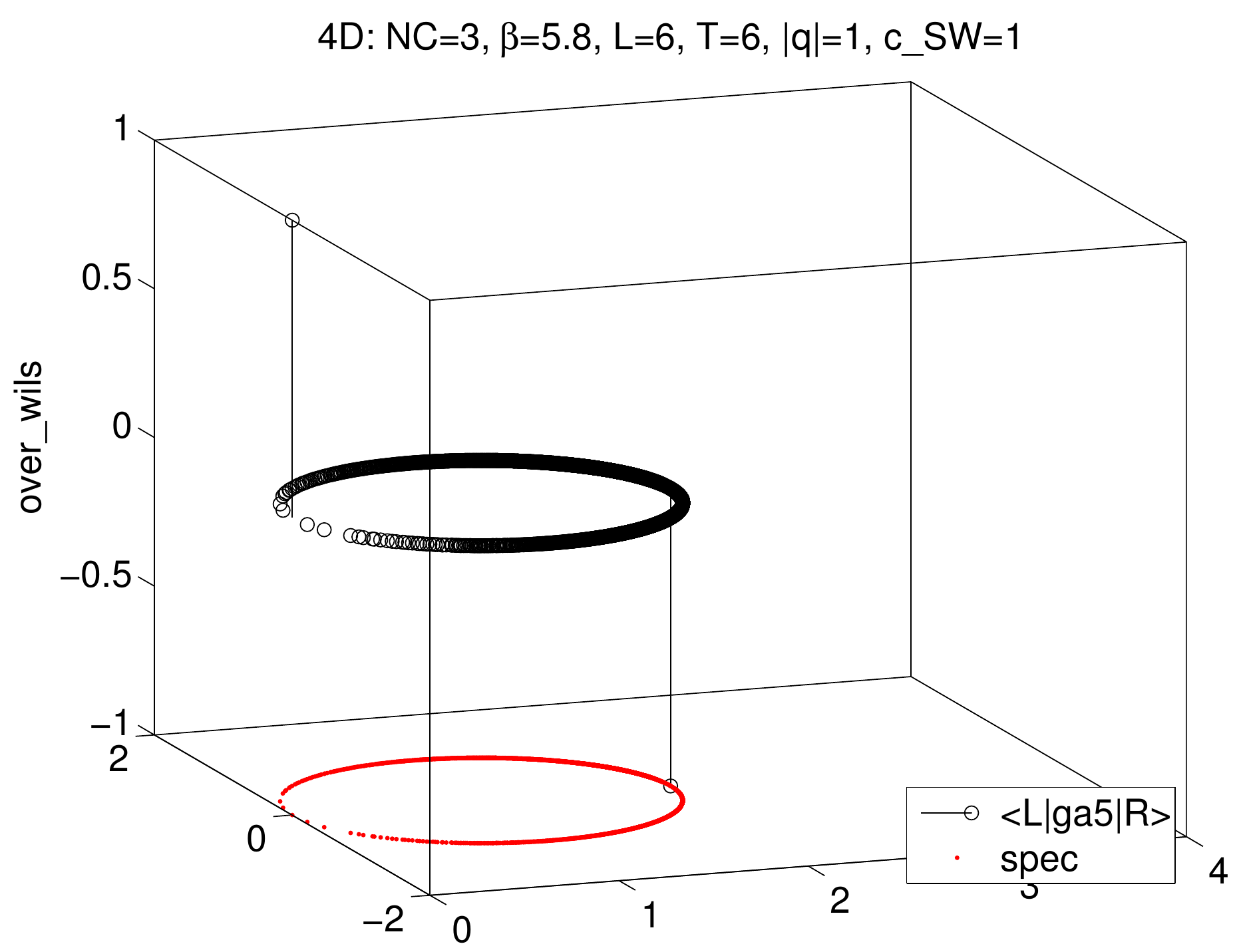}%
\includegraphics[width=0.5\textwidth]{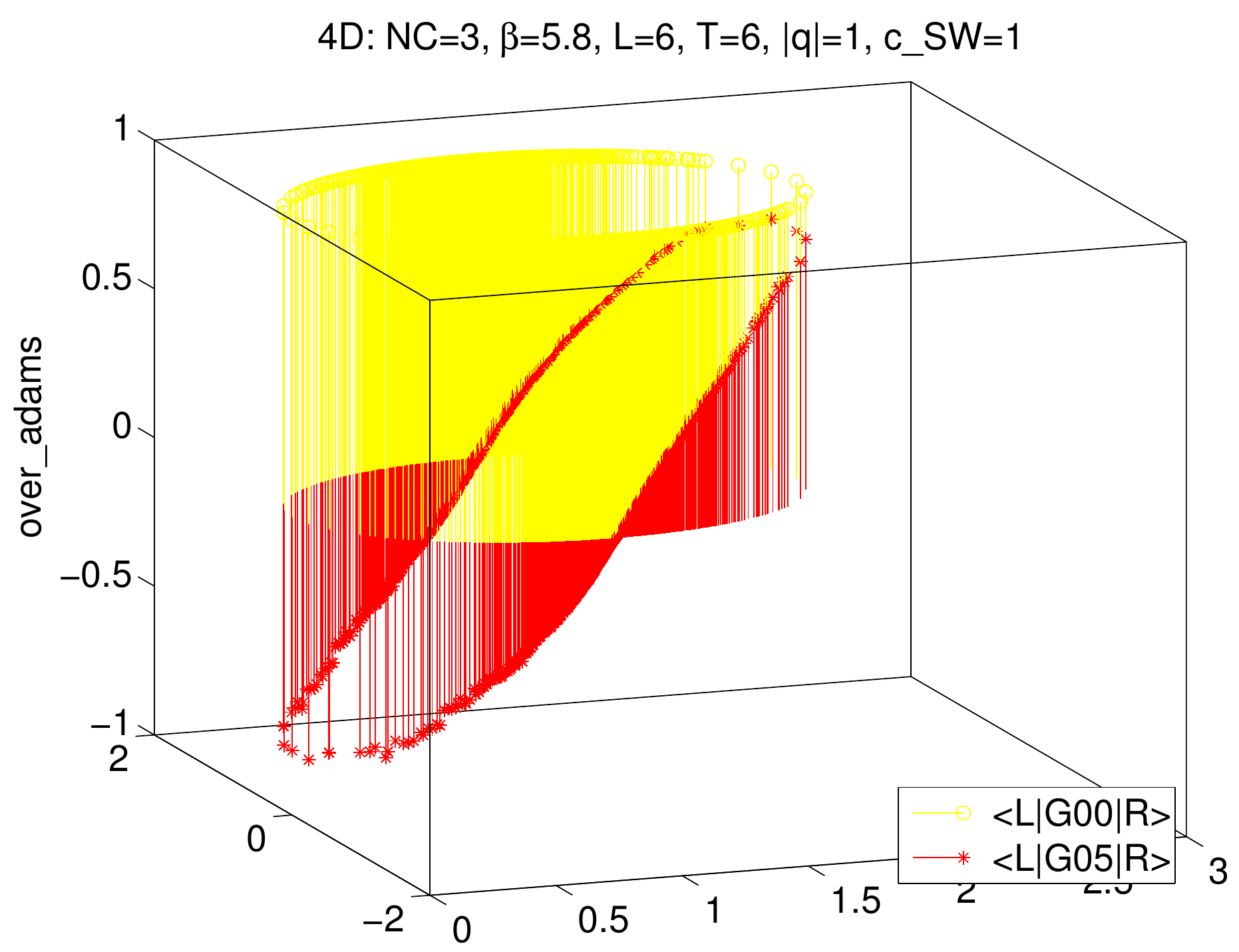}%
\caption{\label{fig11}\sl
Needle plot of the $\<.|\gaf|.\>$ chiralities of the eigenmodes of the overlap
action with clover improved ($c_\mr{SW}\!=\!1$) Wilson kernel (left) and
$\<L|\Ga_{05}|R\>$ taste-chiralities of the overlap action with clover
improved ($c_\mr{SW}\!=\!1$) Adams kernel (right).}
\end{figure}

Fig.\,\ref{fig11} contains two addenda to the overlap theme.
In the left panel the $\<.|\gaf|.\>$ chiralities of the overlap operator with
Wilson kernel are shown, and the analogy to Fig.\,\ref{fig10} is evident.
In the right panel the taste chiralities $\<.|\Ga_{05}|.\>$ of the overlap
operator with Adams kernel are shown, and the smooth pattern that would show up
in the kernel only in the $\<L|.|L\>$ version (cf.\ Fig.\,\ref{fig7}) is now
generic (as the L/R distinction is gone).


\section{Eigenvalue comparison to Wilson-type actions\label{sec:comparison}}


Since the taste-split staggered operators (\ref{def_A}--\ref{def_Hmix}) were
found to behave, on many practical issues, like Wilson fermions, it seems
instructive to compare their eigenvalue spectra (still on the same $|q|=1$
configuration) to those of the Wilson \cite{Wilson:1974sk} and Brillouin
\cite{Durr:2010ch} action.

\begin{figure}[!tb]
\includegraphics[width=0.5\textwidth]{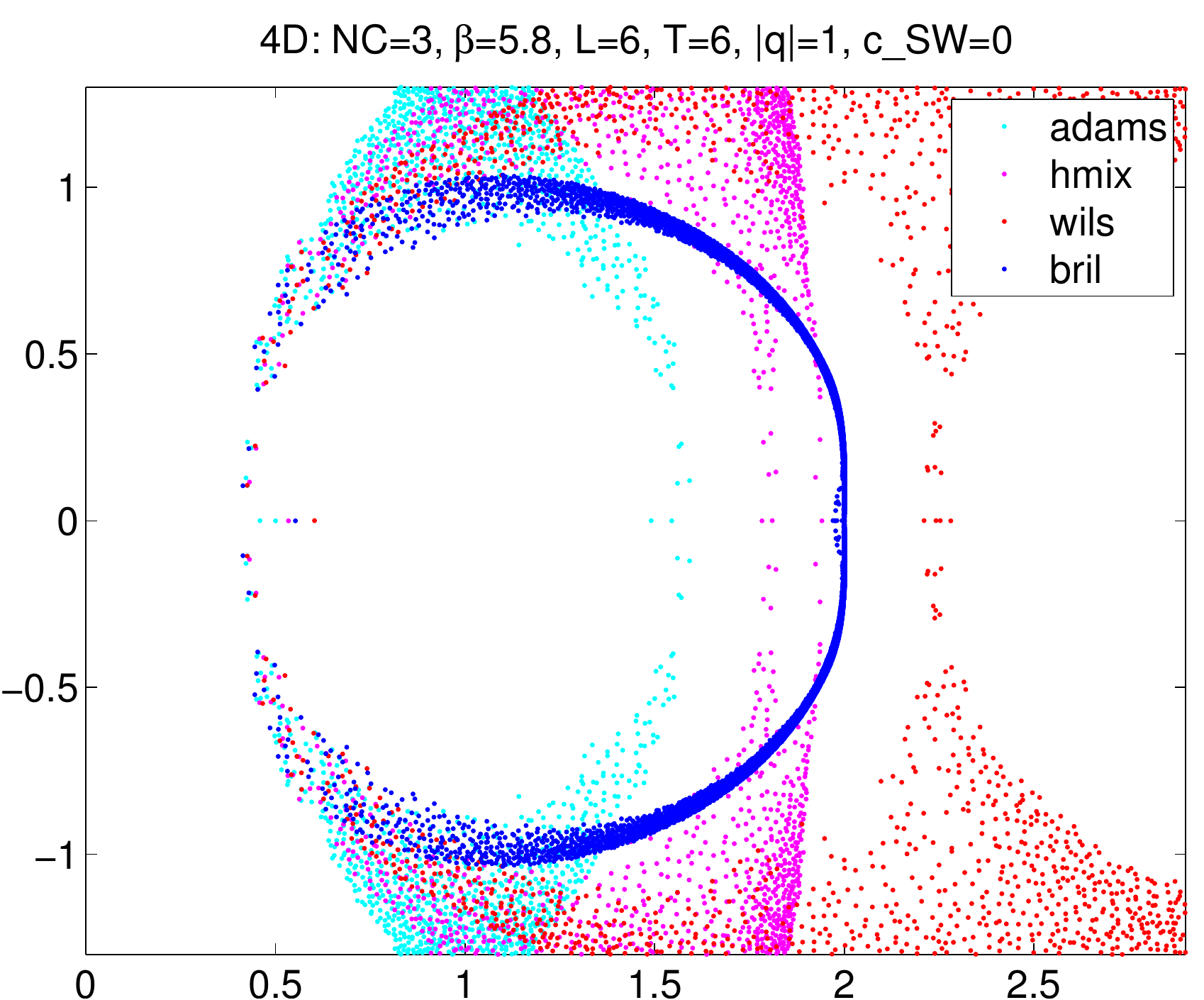}%
\includegraphics[width=0.5\textwidth]{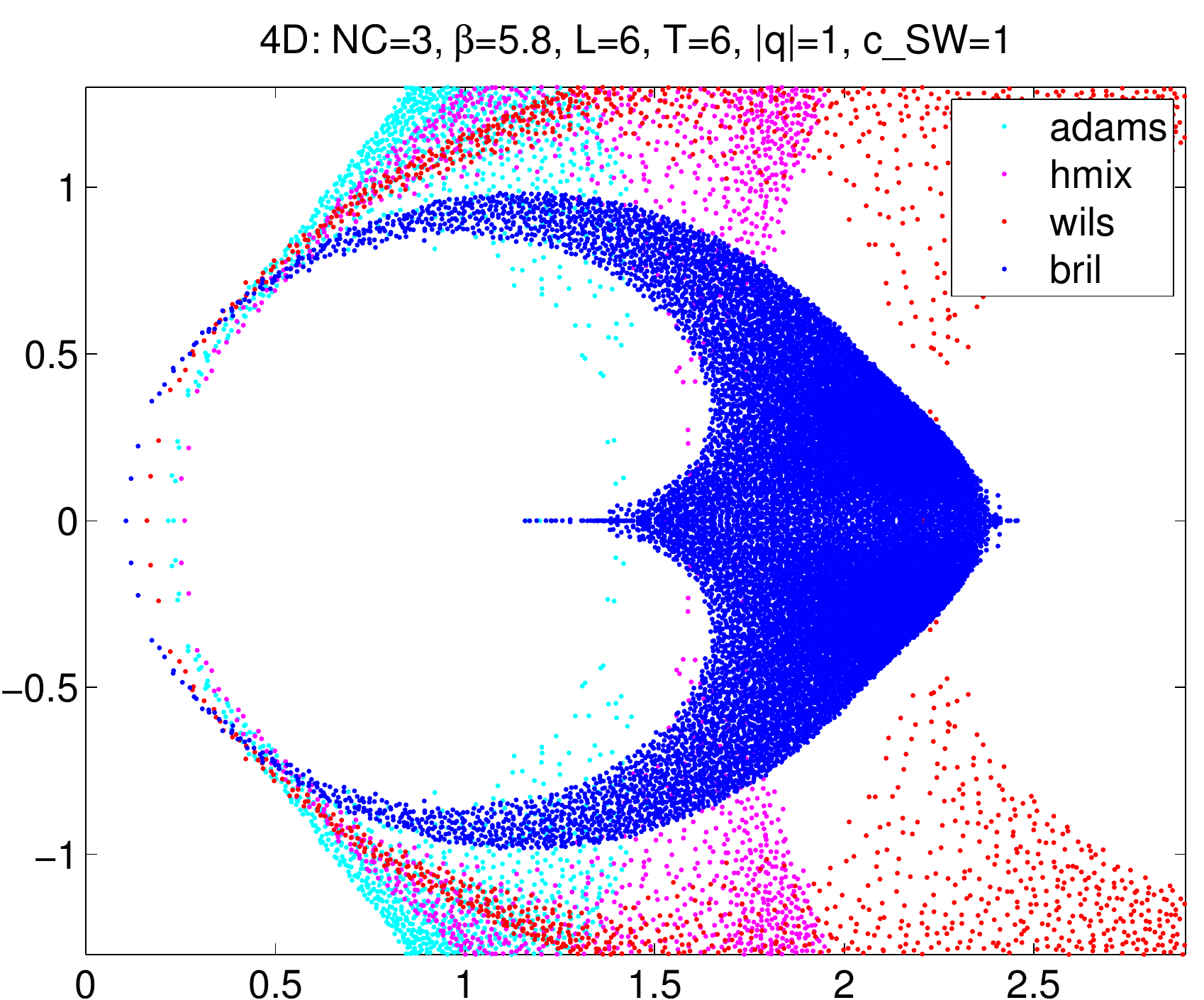}%
\\
\includegraphics[width=0.5\textwidth]{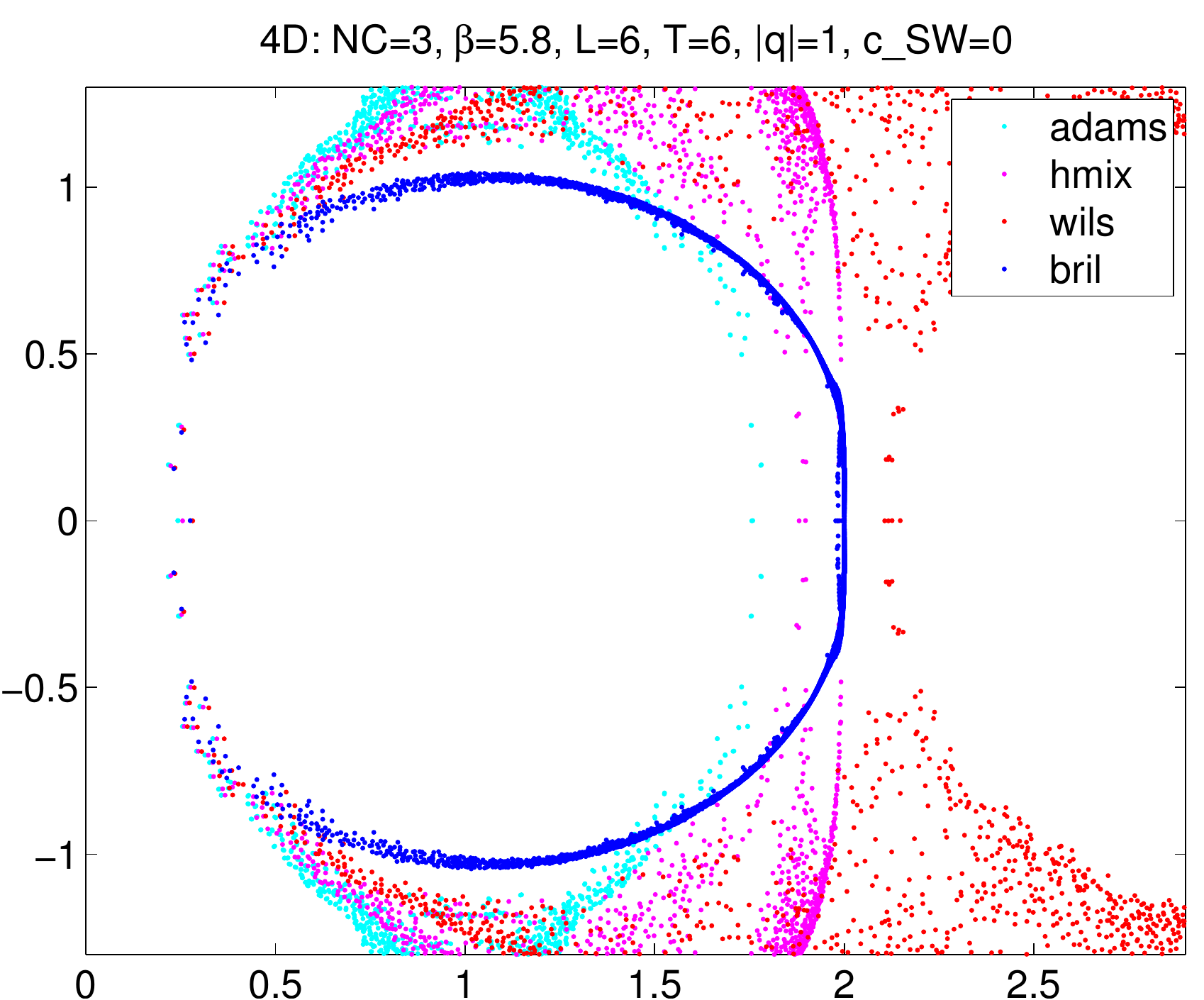}%
\includegraphics[width=0.5\textwidth]{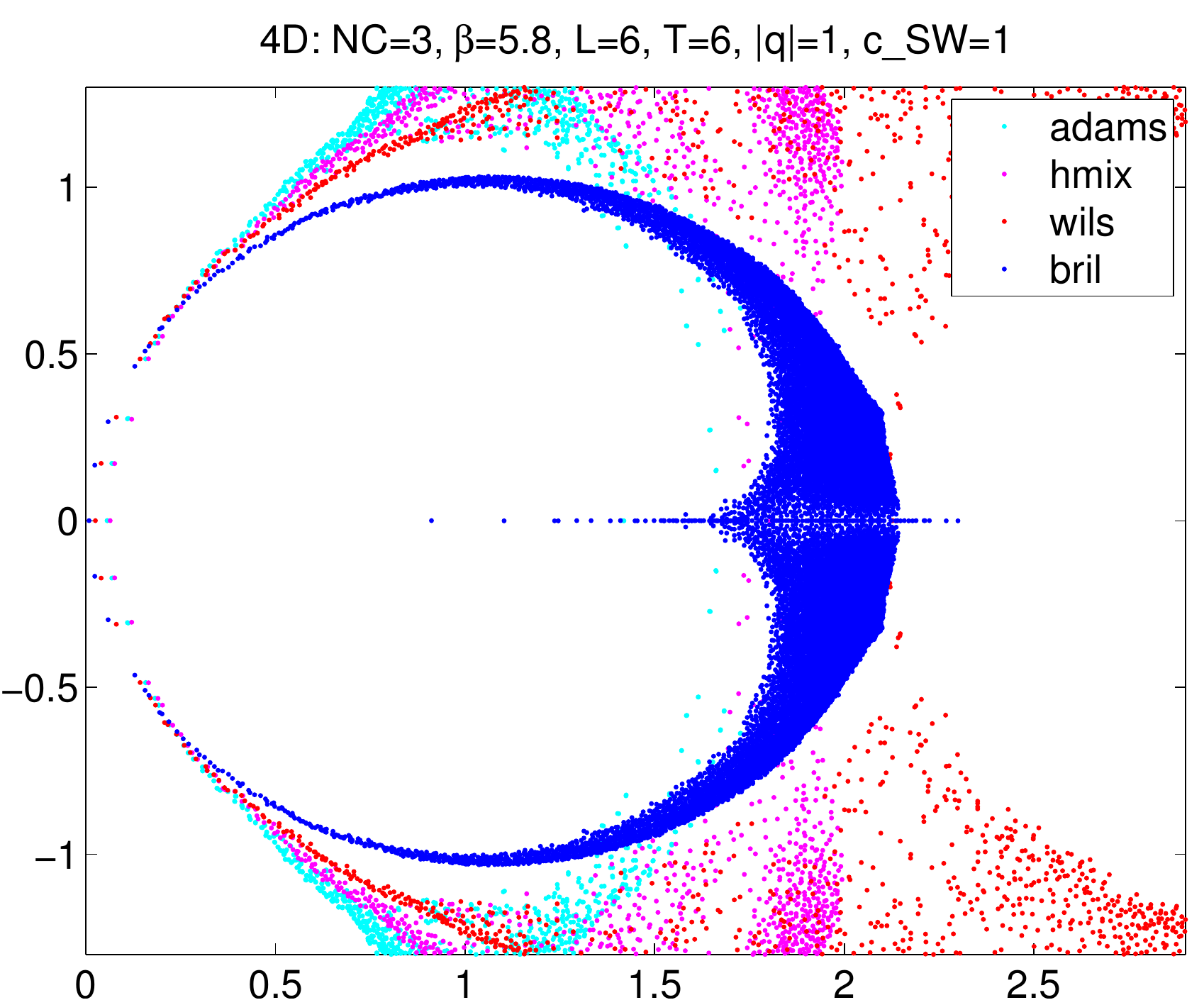}%
\caption{\label{fig12}\sl
Eigenvalue spectra of the operators (\ref{def_A}) and (\ref{def_Hmix}) and of
the Wilson and Brillouin operators on a gauge configuration with topological
charge $|q|=1$, with $c_\mr{SW}=0$ (left) and $c_\mr{SW}=1$ (right), after
1 (top) and 3 (bottom) steps of HEX smearing.}
\end{figure}

Fig.\,\ref{fig12} gives such a comparison, without improvement (left) and with
tree-level clover improvement (right).
Throughout gauge links $V_\mu(x)$ are used which have undergone 1 or 3 steps of
HEX smearing \cite{Capitani:2006ni}.
The striking feature with $c_\mr{SW}=0$ is that the lowest non-topological
modes of all operators sit essentially in the same place (with two nearly
degenerate copies in case of the Adams operator), but the tendency of the
topological modes to get ``soaked into the belly'' is different (the Adams
operator fares best, the Wilson operator fares worst on this point).
With $c_\mr{SW}=1$ the situation is just opposite; the ``soak in'' phenomenon
seems cured with all operators, but now the position of the physical branch
varies, and this time the Brillouin operator performs best.
A peculiar issue with the latter action is the ``thorn'' of eigenvalues that
grows from the doubler point 2 into the belly, once the clover term starts
spreading the 15 unphysical modes.
It is not clear whether this is an advantage (it might facilitate topology
changes) or a disadvantage (topology might be less clearly defined) of this
action -- in the latter case it might be cured by multiplying its clover term
with a factor $(1+\lap^\mr{Bri}/4)$, with symmetrization, where
$\lap^\mr{Bri}$ is defined in \cite{Durr:2010ch}.
In short these spectra underline the value of the improvement term
(\ref{def_improvement}) and suggest that --~at least for some applications~--
the actions (\ref{def_A}, \ref{def_Hmix}) might fare well.
The conceptual issue related to the rotational symmetry breaking of
(\ref{def_Hmix}) is discussed in Sec.\,\ref{sec:rotation}.

The reader might wonder whether further elements of the set of recently tried
modifications to the Wilson action (other smearings, twisted-mass term, etc.)
may prove useful in the context of taste-split staggered actions.
Regarding the link smearing it is clear that the fermion action is fairly
insensitive to the details of this procedure.
In view of dynamical fermion simulations the stout, n-HYP and HEX recipes are
most interesting, since they can be combined with the (R)HMC algorithm, but the
reader is free to select a different procedure.
A twisted-mass term $\ri\mu(\ep\!\otimes\!\ta_3)$ [based on the exact
$\ep\!=\!\ga_5\!\otimes\!\xi_5$ symmetry] would bring a lower bound on the
determinant, but it requires pairs of Hoelbling fermions.
This would be interesting if it allows to cure the problem of rotational
symmetry breaking discussed in Sec.\,\ref{sec:rotation}.
A relevant issue is how the cut-off induced isospin breaking of such a
formulation compares to the one in the Adams formulation (\ref{def_A}).
Clearly, such questions are well beyond the scope of this article.


\section{Signs of rotational symmetry breaking \label{sec:rotation}}


\begin{figure}[!tb]
\includegraphics[width=0.5\textwidth]{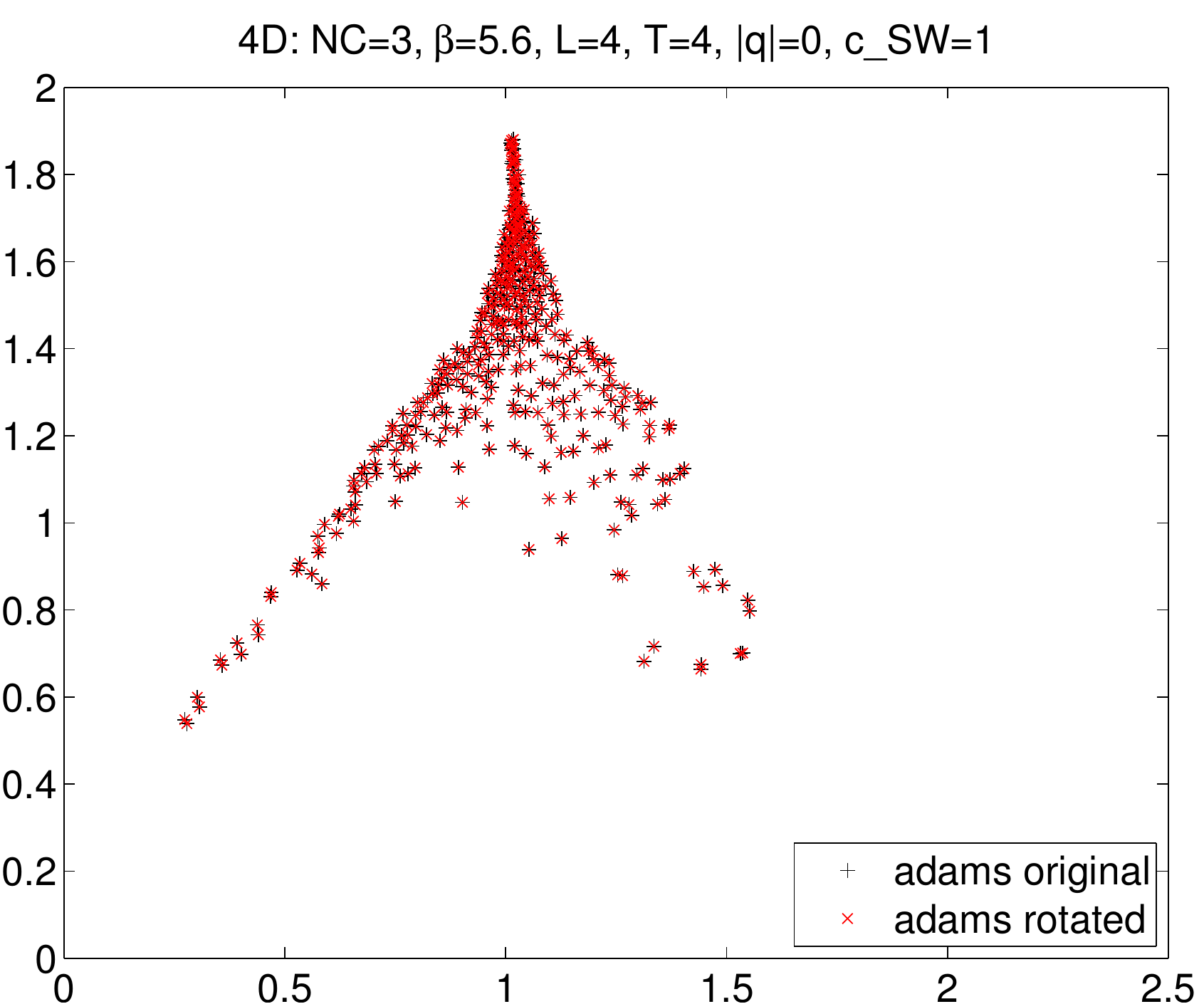}%
\includegraphics[width=0.5\textwidth]{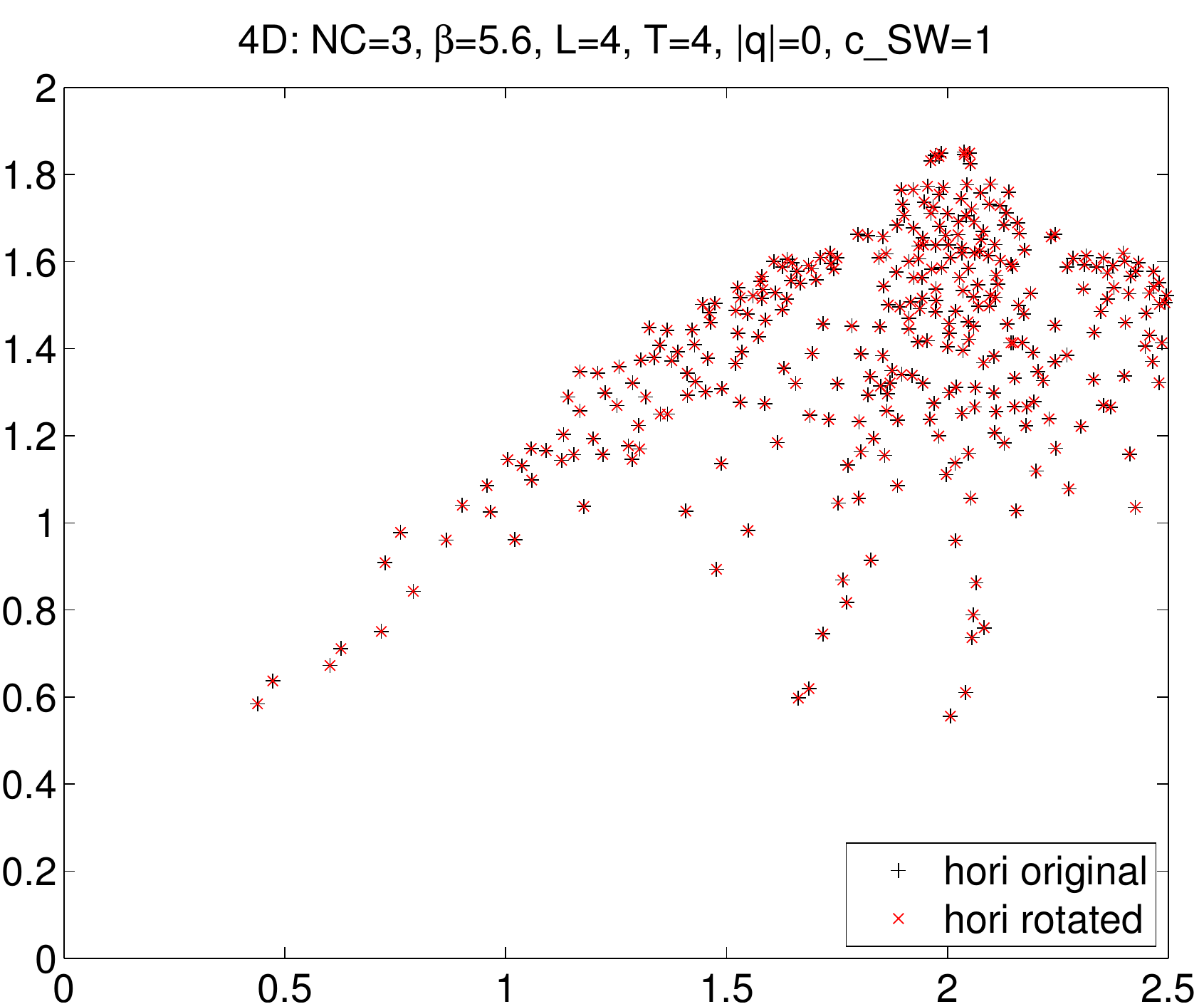}%
\\
\includegraphics[width=0.5\textwidth]{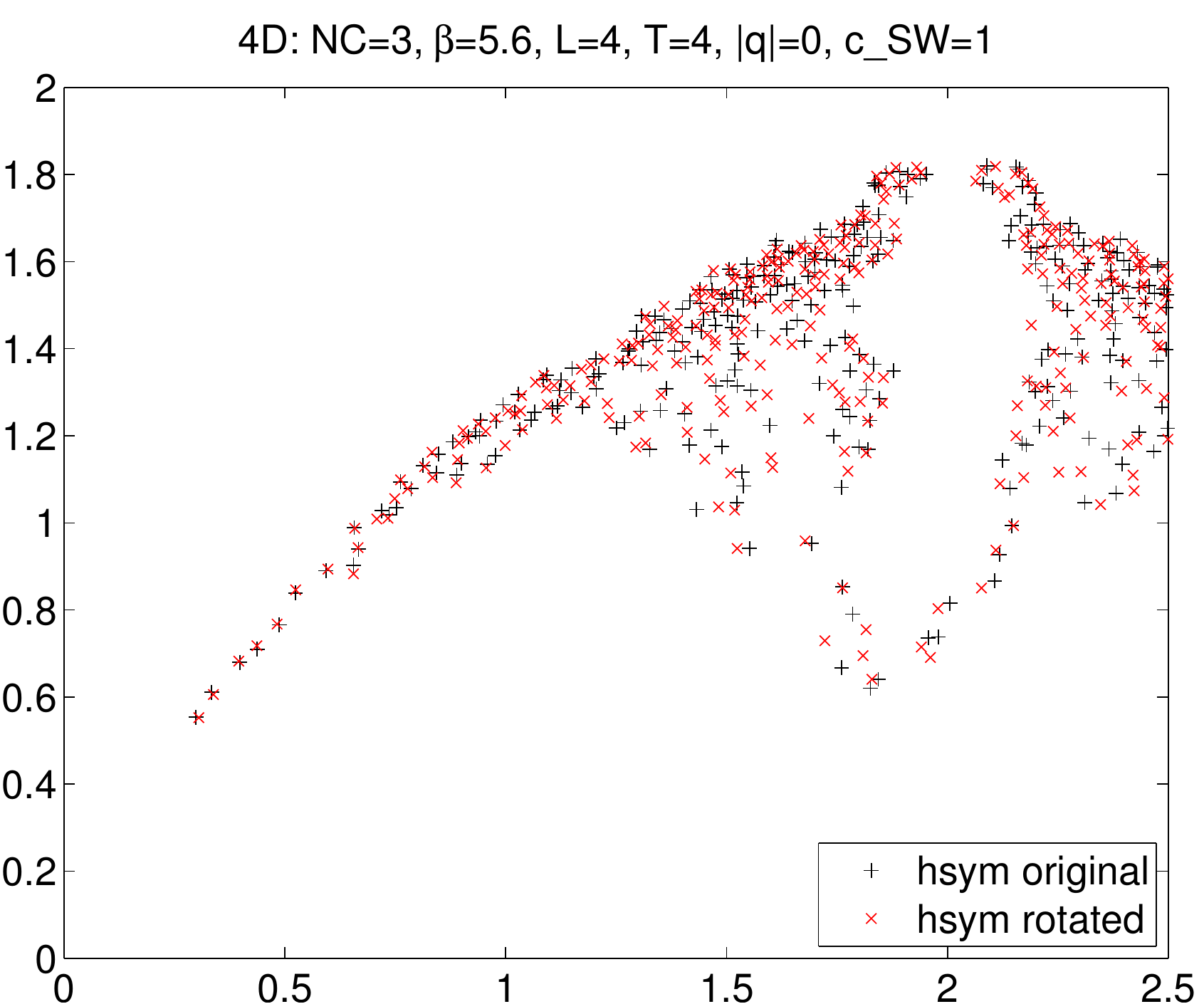}%
\includegraphics[width=0.5\textwidth]{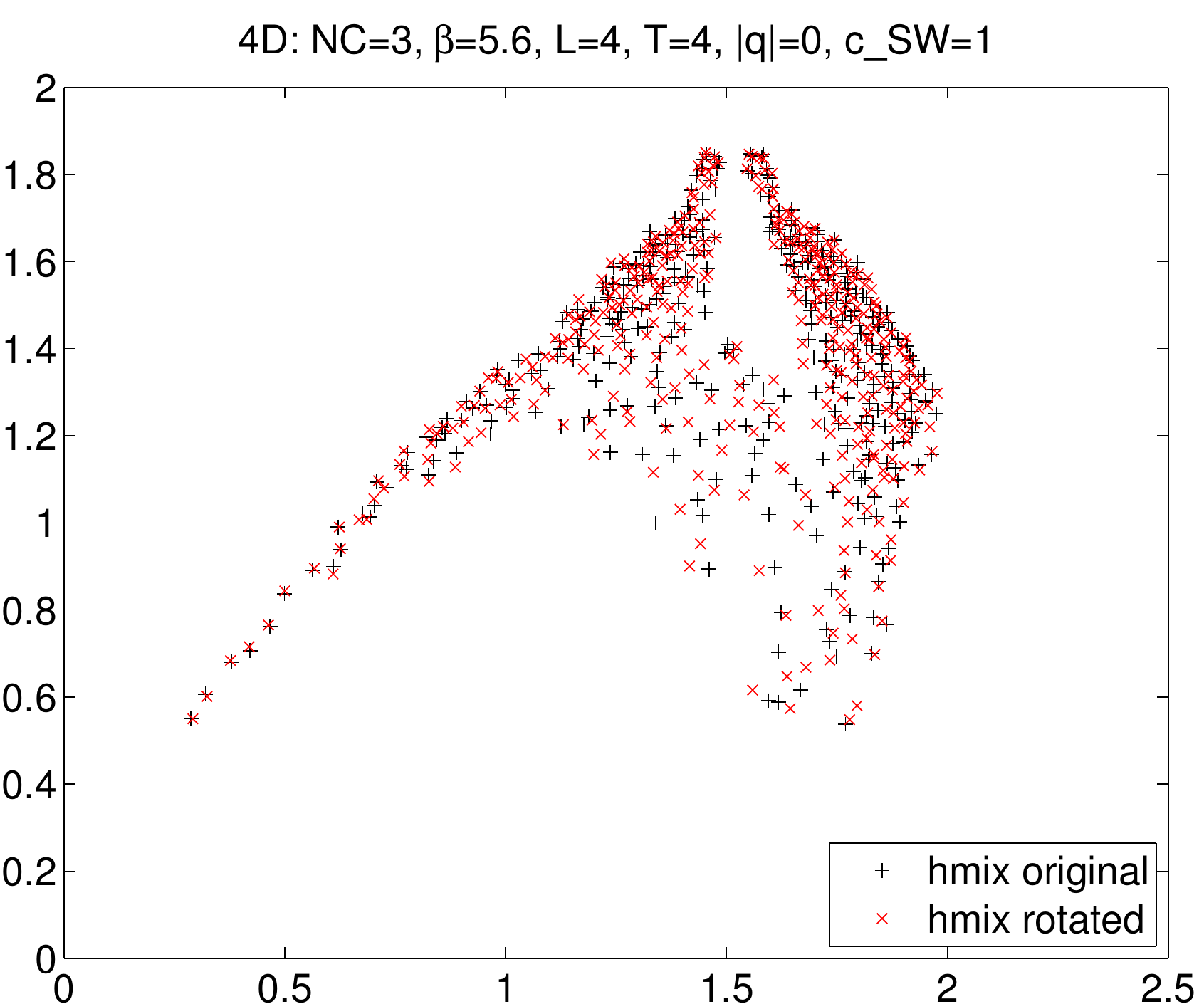}%
\caption{\label{fig13}\sl
Eigenvalue spectra of the operators (\ref{def_A},\ref{def_Hori}) (top) and
(\ref{def_Hsym},\ref{def_Hmix}) (bottom) with $c_\mr{SW}=1$, before and after
rotating the gauge background in the $(1,2)$ plane.}
\end{figure}

As mentioned in the introduction the taste-tensor mass term in the Hoelbling
operators (\ref{def_Hori}-\ref{def_Hmix}) breaks the rotational [hypercubic]
symmetry group $R_{\mu\nu}$ of the staggered action, leading to undesirable
operators generated through quantum effects \cite{SharpeNara,deForcrand:2012bm,
Misumi:2012sp,Misumi:2012eh}.
In consequence in a dynamical simulation with such an action one would expect
to see signs of anisotropy, for instance unequal expectation values
$\<W_{r\times t}\>$ of planar Wilson loops oriented in the six $(\mu,\nu)$
planes \cite{SharpeNara}.

Fig.\,\ref{fig13} displays (half of) the eigenvalue spectra of the four
operators (\ref{def_A}-\ref{def_Hmix}) on a single $4^4$ configuration before
and after rotating the gauge background in the $(1,2)$ plane.
The eigenvalue spectrum of the Adams operator (\ref{def_A}) is unaffected, as
it must be, since $R_{\mu\nu}$ is a symmetry.
Also the eigenvalue spectrum of the original Hoelbling operator (\ref{def_Hori})
with $M_\mr{Hori}=\Xi_{12}$ is unaffected, but it is worth noting that the same
statement does not hold true for other rotations [e.g.\ in the $(2,3)$ plane].
The eigenvalue spectrum of the symmetrized operator (\ref{def_Hsym}) is
affected by the rotation in the ($1,2$) plane, but interestingly the change is
predominantly due to the modes in the first doubler branch near $\mr{Re}(z)=2$;
the physical low-energy modes get barely changed.
The same statement is found to hold true for the mixed operator (\ref{def_Hmix}).
In either case it was checked that the determinant does not show an accidental
symmetry, i.e.\ it changes too.

In short we arrive at the tantalizing situation where the symmetrized Hoelbling
operator (\ref{def_Hsym}) shows ``less symmetries'' than the original variety
(\ref{def_Hori}).
The breaking is predominantly due to the UV modes (which are susceptible to\
other changes, too, e.g.\ a switching between periodic and antiperiodic
boundary conditions in one direction) and barely affects the IR properties of
the operator.
This raises a number of questions.
Could it be that the breaking disappears in the limit of infinite statistics\,?
If not, could it be that the breaking disappears on the way to the continuum\,?
Likely the proper tool to answer such questions is an operator analysis in the
Symanzik effective theory \cite{SharpeNara} and/or a perturbative analysis as
pursued for minimally doubled actions \cite{Capitani:2010nn}, but a careful
study is well beyond the scope of this article.


\section{Summary \label{sec:summary}}


The main findings of this paper may be summarized as follows:
\begin{enumerate}
\item
The discretizations (\ref{def_A}) and (\ref{def_Hsym}), as proposed by Adams
\cite{Adams:2010gx} and Hoelbling \cite{Hoelbling:2010jw}, respectively, and
the new linear combination (\ref{def_Hmix}) yield staggered fermions with
non-standard (i.e.\ taste non-singlet) mass terms (cf.\ Sec.\,\ref{sec:review}).
They are ``hybrids'' in the sense that they distribute the spinor degrees of
freedom over spacetime in the manner of staggered fermions, while technically
being close to Wilson fermions, with additive mass renormalization and the same
consequences of non-normality ($\<L|$ and $|R\>$ eigenmodes).
\item
The main technical challenge of staggered fermions --~the non-coincidence of
the exact (tasteful) $\ep\equiv\Ga_{55}$ and the non-exact (taste-singlet)
$\Ga_{50}$ chiral symmetries~-- persists.
What changes is the details of how the operator is sensitive to the topological
charge $q$ of the gauge background.
With just a scalar mass term, $\<V|\Ga_{55}|V\>$ is zero between all eigenmodes
of the operator and only $\<V|\Ga_{50}|V\>$ is sensitive to topology.
With a sufficiently large taste-tensor or taste-pseudoscalar mass term, as is
the case in (\ref{def_A}--\ref{def_Hmix}) with $r=1$, the matrix element
$\<L|\Ga_{50}|R\>$ continues to be sensitive to topology (with equal sign in
the physical and the unphysical branches), while $\<L|\Ga_{55}|R\>$ acquires an
even better sensitivity to topology (with net zero sign across all branches).
\item
The physical branches of the operators (\ref{def_A}--\ref{def_Hmix}) show a
cancellation-free sensitivity to topology with $2|q|$ or $|q|$ exactly real
modes.
But they also suffer from the same symptoms of chiral symmetry breaking with
induced $O(a)$ cut-off effects as Wilson fermions, and it seems thus natural to
try similar remedies.
In this article it is conjectured that the leading Symanzik improvement term to
these actions takes the form (\ref{def_improvement}), and the connection to the
overlap procedure suggests that the tree-level improvement coefficient is
again $c_\mr{SW}=1$.
The combination of Symanzik improvement and link-smearing is found to have a
pronounced (and beneficial) effect on the formation of a ``belly'' between the
physical and the unphysical branches in the eigenvalue plot (which, in turn, is
seen as a sign that the mixing with other operators is suppressed).
By explicitly rotating the gauge background it is confirmed that the
taste-tensor mass term in (\ref{def_Hori}-\ref{def_Hmix}) breaks the rotational
symmetry group, but this seems to be linked to UV modes which barely affect IR
physics.
\item
The kernel operators (\ref{def_A}) and (\ref{def_Hmix}), once equipped with
link smearing and a clover term, bear the promise of a cheap overlap
construction (in contrast to the mild savings that were found without these
ingredients \cite{deForcrand:2012bm}) due to a moderate spectral range over
which the sign function is to be constructed.
Accordingly, a combination of these two overlap operators might be an
attractive option for simulating QCD with $2\!+\!1\!+\!1$ active flavors.
\end{enumerate}

\bigskip

\noindent{\bf Acknowledgments}:
The author wishes to thank Christian Hoelbling and Stefan Krieg for useful
discussion.
This work was supported in part by the German SFB TRR-55.

\appendix


\section{Leading terms generated by the overlap procedure \label{sec:appendix}}


To facilitate the analysis of the necessary counterterms for $O(a)$ improvement
of the taste-split staggered actions (\ref{def_A}--\ref{def_Hmix}) let us begin
with a reflection on the situation with Wilson fermions.
The idea that for Wilson fermions the overlap procedure of
\cite{Neuberger:1997fp,Neuberger:1998wv} would automatically generate the
necessary improvement term is mentioned in the review by Niedermayer
\cite{Niedermayer:1998bi}.
Unfortunately, this account is not very verbose, and this is why we attempt
a short summary.

One essential ingredient in what follows is the continuum relation
\beq
\Dslash^2=D_\mu\ga_\mu D_\nu\ga_\nu=
\Big(\half\{\ga_\mu,\ga_\nu\}+\half[\ga_\mu,\ga_\nu]\Big)D_\mu D_\nu=
D^2+\sum_{\mu<\nu}\si_{\mu\nu}F_{\mu\nu}
\eeq
with $\si_{\mu\nu}=\frac{\ri}{2}[\ga_\mu,\ga_\nu]$ which is also dubbed
$\ga_{\mu\nu}$ in Sec.\,\ref{sec:review}.
The other ingredient is the definition (\ref{def_over}) of the overlap
procedure with $X=a\Dke-\rh$.
This definition does not rely on any special property of the kernel operator
$\Dke$.
With an argument which is $\gaf$ or $\ep$ hermitian the procedure can be recast
in a form which involves the sign function; this will be relevant for numerical
applications with (\ref{def_A}--\ref{def_Hmix}) as a kernel but not for
the analytical considerations below.

With the Wilson operator as kernel one starts from the transcription (here we
follow \cite{Ikeda:2009mv})
\bea
D_\mr{W}=\sum_\mu\Big\{\ga_\mu \nab_\mu-\frac{a}{2}\lap_\mu\Big\}&\sim&
\Dslash-\frac{a}{2}D^2+O(a^2)
\label{transcription_W}
\eea
in terms of continuum operators.
Here $\nab_\mu$ and $\lap_\mu$ denote the gauge covariant first and second
discrete derivative, respectively, the latter one to be distinguished from the
continuum $D_\mu^2$.
This implies
$X=-\rh+a\Dslash-\frac{a^2}{2}D^2+O(a^3)$ and
$X\dag=-\rh-a\Dslash-\frac{a^2}{2}D^2+O(a^3)$ and thus
\beq
X\dag X=\rh^2-(1\!-\!\rh)a^2D^2-a^2\sum_{\mu<\nu}\si_{\mu\nu}F_{\mu\nu}+O(a^3)
\eeq
in the physical branch.
Next we use the expansion $(1+z)^{-1/2}=1-\frac{1}{2}z+O(z^2)$ to arrive at
\beq
(X\dag X)^{-1/2}=\frac{1}{\rh}\Big(
1+\frac{1\!-\!\rh}{2\rh^2}a^2D^2+
\frac{a^2}{2\rh^2}\sum_{\mu<\nu}\si_{\mu\nu}F_{\mu\nu}+O(a^3)
\Big)
\eeq
and upon multiplying this with $X$ we find
\beq
X(X\dag X)^{-1/2}=
-1+\frac{a}{\rh}\Dslash-\frac{a^2}{2\rh^2}D^2
-\frac{a^2}{2\rh^2}\sum_{\mu<\nu}\si_{\mu\nu}F_{\mu\nu}+O(a^3)
\label{expression_W}
\eeq
with the consequence that the overlap operator relates to the continuum
Dirac operator like
\beq
\Dov=\Dslash
-\frac{a}{2\rh}D^2-\frac{a}{2\rh}\sum_{\mu<\nu}\si_{\mu\nu}F_{\mu\nu}+O(a^2)
\;.
\eeq
The second term may be removed by a field rotation, but the third term
indicates that $O(a)$ improvement of $D_\mr{W}$ calls for a term
$-\frac{c_\mr{SW}}{2}\sum_{\mu<\nu}\si_{\mu\nu}F_{\mu\nu}$ with $c_\mr{SW}=1$
at tree level.

The task is now to pipe the taste-split staggered operators
(\ref{def_A}--\ref{def_Hmix}) through the overlap procedure and to see which
counterterms are generated.
First we need to transcribe the staggered operator in terms of
continuum operators on the blocked lattice ($b=2a$).
For (\ref{def_A}) we find
\bea
D_\mr{A}&=&\sum_\mu\Big\{(\ga_\mu\!\otimes\!1)\nab_\mu
-\frac{b}{2}(\gaf\!\otimes\!\xi_\mu^*\xif)\lap_\mu\Big\}
+r(1\!\otimes\!\xif)+r(1\!\otimes\!1)
\nonumber
\\
&\sim&\Dslash-\frac{b}{2}\gaf D^2+O(b^2)
\label{transcription_A}
\eea
where in the second line [which is supposed to capture the effect on the
physical branch in terms of continuum operators] we use that the
violation of Lorentz symmetry takes place exclusively in the taste space.
Hence, the only relevant difference to (\ref{transcription_W}) is an additional
factor of $\gaf$.
This implies
$Y=-\rh+b\Dslash-\frac{b^2}{2}\gaf D^2+O(b^3)$ and
$Y\dag=-\rh-b\Dslash-\frac{b^2}{2}\gaf D^2+O(b^3)$ and thus
\beq
Y\dag Y=\rh^2-(1\!-\!\rh\gaf)b^2D^2-b^2\sum_{\mu<\nu}\ga_{\mu\nu}F_{\mu\nu}+O(b^3)
\eeq
in the physical branch.
With the same expansion as before we obtain
\beq
(Y\dag Y)^{-1/2}=\frac{1}{\rh}\Big(
1+\frac{1\!-\!\rh\gaf}{2\rh^2}b^2D^2+
\frac{b^2}{2\rh^2}\sum_{\mu<\nu}\ga_{\mu\nu}F_{\mu\nu}+O(b^3)
\Big)
\eeq
and upon multiplying this with $Y$ we find
\beq
Y(Y\dag Y)^{-1/2}=
-1+\frac{b}{\rh}\Dslash-\frac{b^2}{2\rh^2}D^2
-\frac{b^2}{2\rh^2}\sum_{\mu<\nu}\ga_{\mu\nu}F_{\mu\nu}+O(b^3)
\eeq
which agrees with the expression (\ref{expression_W}) which we found in the
Wilson case.
The bottom line is that in order to cancel the $O(b)$ effects we need to add a
term which acts on the physical branch like
$-\frac{c_\mr{SW}}{2}\sum_{\mu<\nu}\ga_{\mu\nu}F_{\mu\nu}$, with
$c_\mr{SW}=1$ at tree level.
To improve the actual action (\ref{def_A}) the obvious replacement is
$\ga_{\mu\nu}\to\Ga_{\mu\nu}$.
Finally, to ensure gauge covariance and hermiticity this expression needs to be
symmetrized in the manner of (\ref{def_improvement}).

None of the manipulations listed above did refer to the details of the
taste lifting term $\propto r$ in (\ref{def_A}--\ref{def_Hmix}); in fact an
essential ingredient was that this term is effectively $0$ in the physical
branch.
It follows that the operator (\ref{def_improvement}) removes the leading
cut-off effects in all taste-split staggered actions, with $c_\mr{SW}=1$ at
tree-level, regardless of the multiplicity of the physical branch.

\clearpage



\end{document}